\documentclass{iopart}

\usepackage[english]{babel}

\usepackage[letterpaper,top=2cm,bottom=2cm,left=3cm,right=3cm,marginparwidth=1.75cm]{geometry}

\usepackage{amsmath}
\usepackage{amssymb}
\usepackage{graphicx}
\usepackage{mathtools}
\usepackage[colorlinks=true, allcolors=blue]{hyperref}
\usepackage[dvipsnames,table]{xcolor}
\usepackage{empheq}
\usepackage{subcaption}
\usepackage{todonotes}
\usepackage{placeins}
\usepackage{natbib}
\usepackage{tikz}
\usetikzlibrary{decorations.text}
\usepackage{rotating}
\usepackage[outline]{contour}

\newcommand{\bhat}{\ensuremath{\boldsymbol{\hat{b}}}}
\newcommand*\bsy[1]{\mathop{}\!\boldsymbol{#1}}

\newcommand*\abs[1]{\mathop{}\!\left|#1\right|}

\begin{document}


\title{Theory of zonal flow growth and propagation in toroidal geometry}
\author{Richard Nies$^{1,2}$, Felix Parra$^{1,2}$}

\address{$^1$Department of Astrophysical Sciences, Princeton University, Princeton, NJ 08543, USA}
\address{$^2$Princeton Plasma Physics Laboratory, Princeton, NJ 08540, USA}

\eads{\mailto{richard.nies@physics.ox.ac.uk}}

\begin{abstract}
    The toroidal geometry of tokamaks and stellarators is known to play a crucial role in the linear physics of zonal flows, leading to e.g. the Rosenbluth-Hinton residual and geodesic acoustic modes. However, descriptions of the nonlinear zonal flow growth from a turbulent background typically resort to simplified models of the geometry. We present a generalised theory of the secondary instability to model the zonal flow growth from turbulent fluctuations in toroidal geometry, demonstrating that the radial magnetic drift substantially affects the nonlinear zonal flow dynamics. In particular, the toroidicity gives rise to a new branch of propagating zonal flows, the toroidal secondary mode, which is nonlinearly supported by the turbulence. We present a theory of this mode and compare the theory against gyrokinetic simulations of the secondary mode. The connection with other secondary modes -- the ion-temperature-gradient and Rogers-Dorland-Kotschenreuther secondary modes -- is also examined.
\end{abstract}

\section{Introduction}

Magnetic confinement fusion is plagued by microinstabilities which cause large turbulent heat fluxes and make it difficult to reach the high temperatures required to sustain thermonuclear fusion. Of all the microinstabilities, the most virulent is typically the ion-temperature-gradient (ITG) mode \citep{coppi_instabilities_1967}. There is however a silver lining, as ion gyroradius scale modes tend to generate strong zonal flows (ZFs) that shear apart turbulent eddies and reduce the saturation level \citep{lin_turbulent_1998, terry_suppression_2000, dimits_comparisons_2000, rogers_generation_2000}. Close to marginality, the ZFs can even lead to a nearly complete suppression of the turbulence, nonlinearly increasing the critical gradient threshold beyond which an appreciable heat flux is observed \citep{dimits_comparisons_2000}.

The linear physics of ZFs in toroidal geometry describes two separate branches: stationary (zero frequency) ZFs and the fast-oscillating geodesic acoustic modes (GAMs) \citep{winsor_geodesic_1968, conway_geodesic_2021}. Crucially, the radial magnetic drift induced by toroidicity \footnote{Theoretically, the radial magnetic drift can vanish in a toroidal `isodynamic' configuration, though the only known isodynamic toroidal equilibrium by \cite{palumbo_considerations_1968} is impractical. However, there is significant variation across toroidal devices in the size of the radial magnetic drift and therefore in the strength of the toroidal coupling between ZFs and pressure perturbations.} couples ZFs to `up-down' (in a tokamak) asymmetric pressure perturbations through the Stringer-Winsor (SW) force \citep{winsor_geodesic_1968, stringer_diffusion_1969, hassam_spontaneous_1993, hallatschek_transport_2001}. The GAM branch of ZFs stems from this SW force, with the up-down asymmetric pressure originating from the compression and expansion of the ZFs due to toroidicity. Stationary ZFs persist in a torus because they produce fluxes parallel to the magnetic field lines that compensate for the compression and expansion. The stationary ZFs are weakened by toroidicity: for typical initial conditions, ZFs decay to the Rosenbluth-Hinton residual \citep{rosenbluth_poloidal_1998, hinton_dynamics_1999}. We note that stationary ZFs and GAMs have been observed in multiple tokamak and stellarator experiments, see e.g. \cite{fujisawa_identification_2004, gupta_detection_2006, fujisawa_review_2009, de_meijere_complete_2014, kobayashi_quantification_2018}.

While the linear physics of ZFs in tori is well known, the nonlinear mechanisms for ZF growth and saturation are less well understood. As the ZF advection is within flux surfaces, the ZFs cannot linearly extract free-energy from the background radial gradients and must instead be nonlinearly driven by the drift-wave (DW) turbulence. 

Most studies of ZF drive have focused on the role of the nonlinear Reynolds and diamagnetic stresses, generally employing reduced models that ignore the effects of toroidicity. The nonlinear stresses can cause ZF growth through a modulational instability mechanism, whereby a linearly unstable `primary' mode (e.g. an ITG mode) becomes itself nonlinearly unstable to `secondary' perturbations. The paradigmatic work of \cite{rogers_generation_2000} employed this secondary mode approach to describe the nonlinear growth of ZFs, leading to a purely growing mode which we call the Rogers-Dorland-Kotschenreuther (RDK) secondary mode in this article. More generally, the ZF drive by nonlinear stresses underlying the RDK secondary mode has been studied e.g. in quasilinear models of the ZF-DW system's evolution \citep{diamond_dynamics_2001, parker_zonal_2013, zhu_wave-kinetic_2021}, in studies of Z-pinch turbulence \citep{ivanov_zonally_2020, ivanov_dimits_2022}, and through various generalisations to include kinetic effects \citep{plunk_gyrokinetic_2007}, a finite DW oscillation frequency \citep{chen_excitation_2000}, and variation of the flux surface compression along the magnetic field \citep{plunk_nonlinear_2017}. 

Another less studied mechanism for ZF drive hinges on the SW force. Up-down asymmetric pressure perturbations can be nonlinearly generated by DWs, and in turn the SW force produced by these asymmetries will drive ZFs. In previous work, the drive of ZFs by the SW force has been studied with the up-down pressure asymmetry assumed \textit{ab initio} \citep{hassam_spontaneous_1993, lee_kinetic_2023} or it was taken to result from zonal flow shearing in conjunction with magnetic shear \citep{hallatschek_transport_2001, itoh_excitation_2005}.

\begin{figure}
     \centering
     \begin{subfigure}[t]{0.49\columnwidth}
         \centering
          \includegraphics[width=\textwidth, trim={0.85cm 0.6cm 0.4cm 0.6cm},clip]{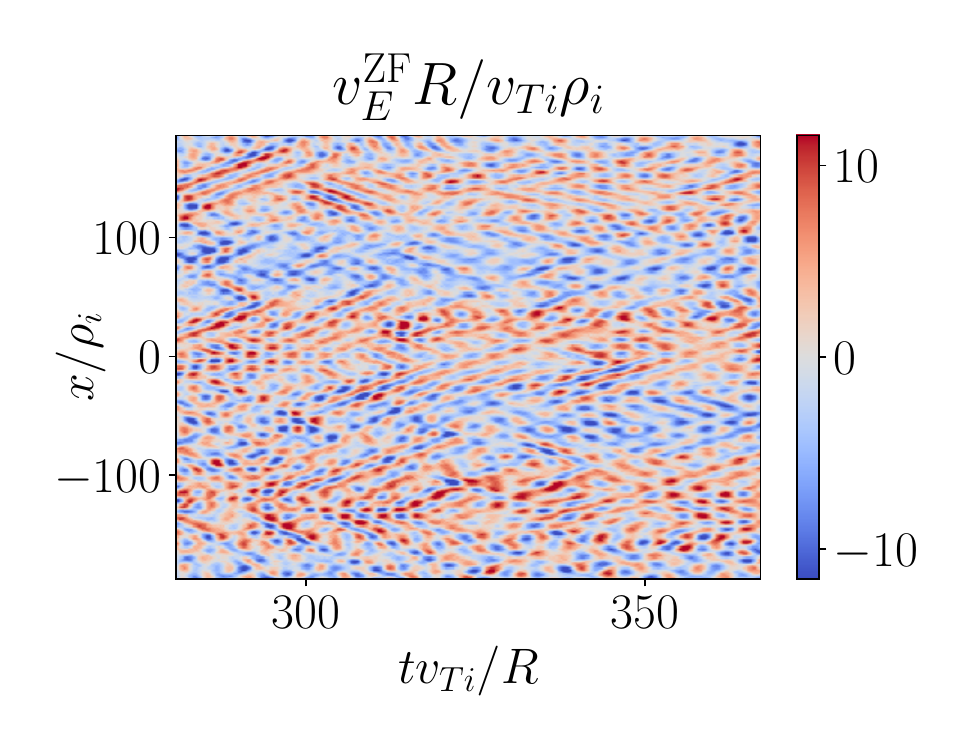}
     \end{subfigure}
     \begin{subfigure}[t]{0.49\columnwidth}
         \centering
          \includegraphics[width=\textwidth, trim={0.85cm 0.6cm 0.4cm 0.6cm},clip]{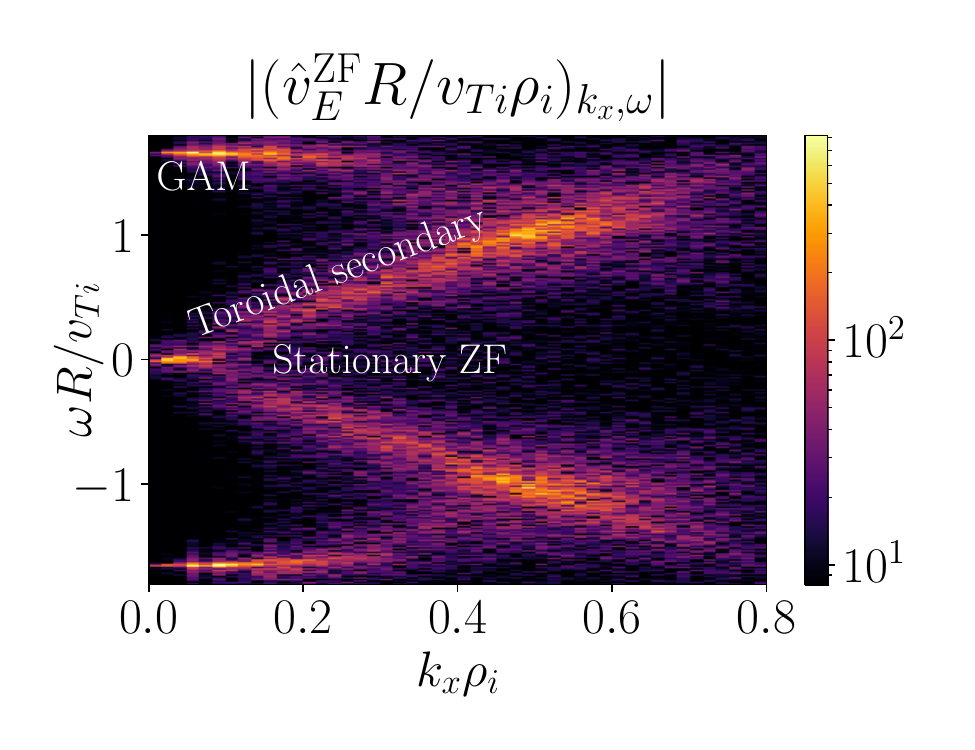}
     \end{subfigure}
    \caption{Zonal flow velocity $v_E^\mathrm{ZF}$ (see Section~\ref{sec:review} for definition; other quantities in the figure follow the standard conventions and are also defined in Section~\ref{sec:review}) in real space (left) and Fourier space (right) from a nonlinear gyrokinetic simulation of tokamak ion-temperature-gradient turbulence with safety factor $q=4.2$. There are stationary ZFs and geodesic acoustic modes (GAMs) at large scales, and propagating ZFs at smaller radial scales corresponding to toroidal secondary modes. General information about the simulations shown in this study may be found in Section~\ref{sec:GK_sims}. We note that a similar plot for a different set of parameters was previously presented in \cite{nies_saturation_2024}.}
    \label{fig:ZF_real_Fourier}
\end{figure}

In this work, we develop a generalised theory of secondary modes including toroidal geometric effects neglected in previous studies, in particular the radial magnetic drift induced by toroidicity and the ensuing SW force. The theory describes the toroidal secondary mode (TSM) \citep{nies_saturation_2024}, a new branch of zonal flows which grows and propagates radially due to the SW force. Here, the up-down pressure asymmetry is generated through a combination of zonal flow shearing and the advection by the radial magnetic drift. The TSM is crucial to understand the ZF behaviour in simulations of tokamak ITG turbulence, see e.g. Figure~\ref{fig:ZF_real_Fourier}, where the TSM explains the ZFs at short radial wavelengths -- while the stationary ZFs and GAMs are found at long radial wavelengths. Moreover, the TSM was shown in previous work \citep{nies_saturation_2024} to contribute to the saturation of strongly-driven ITG turbulence in tokamaks. 

The generalised theory of the secondary mode presented in this work also includes the ZF drive by nonlinear stresses. Therefore, the RDK secondary mode is recovered from the theory in the limit of strong nonlinear drive by the primary, in which case the linear coupling through the SW force becomes subdominant. Furthermore, the inclusion of the radial magnetic drift in the theory allows for the existence of ITG secondary modes (ISMs) driven unstable by the temperature gradient in the primary drive. The properties of the RDK secondary mode and the ISM will be discussed, as well as the connections between these modes and the TSM.

The paper is structured as follows: in Section~\ref{sec:review}, we first review the theory of gyrokinetics and the secondary mode model (Sections ~\ref{sec:GK} and ~\ref{sec:secondary}, respectively). We provide details of the numerical simulations (both of the turbulence and of the secondary mode) in Section~\ref{sec:GK_sims}. The theory of secondary modes in toroidal geometry is presented in Section~\ref{sec:secondary_modes_toroidal_geo}, starting in Section~\ref{sec:theory_ISM_TSM} with the derivation of dispersion relations for the local (on a flux surface) ISMs and for the global (on a flux surface) TSM and RDK secondary mode. The ISM and the TSM are discussed in more detail in Sections~\ref{sec:ISM} and \ref{sec:TSM_R}, respectively. The strongly-driven non-resonant limit of the secondary modes is studied in Section~\ref{sec:TSM_NR}. We evaluate the importance of various physical effects not considered in the theory of the generalised secondary mode in Section~\ref{sec:robustness_TSM}, namely the effects of binormal drifts and magnetic shear on the TSM in Section~\ref{sec:vy_shat} and the inclusion of parallel streaming in Section~\ref{sec:streaming_TSM}. We summarise and discuss our results in Section~\ref{sec:conclusions}. The limit of the secondary mode theory for strong primary drive and long perpendicular wavelengths is presented in \ref{sec:strongly_driven}, and the RDK secondary mode dispersion relation for arbitrary radial wavelengths is derived in \ref{sec:RDK_SW}.

\section{Gyrokinetic secondary mode theory} \label{sec:review}

We begin by reviewing in Section~\ref{sec:GK} the theory of gyrokinetics used to describe microinstabilities in strongly magnetised plasmas. The gyrokinetic secondary mode model will then be discussed in Section~\ref{sec:secondary}, and specifics of the gyrokinetic simulations performed in this work will be presented in Section~\ref{sec:GK_sims}.

\subsection{Gyrokinetics} \label{sec:GK}

In the strongly magnetised plasmas typical of magnetic confinement fusion, the characteristic timescales of the microinstabilities that drive turbulence are much longer than the period of the Larmor gyration. One may therefore average over the fast particle gyromotion and describe the evolution of rings of charge instead of that of particles. In the core, one may further assume small-scale fluctuations compared to the Maxwellian background $F_{M\sigma}$, i.e. the distribution function $f_\sigma$ for any species $\sigma$ satisfies $\abs{\delta f_\sigma}=\abs{f_\sigma - F_{M\sigma}} \ll F_{M\sigma}$. For the collisionless and electrostatic limits considered in this study, the distribution of gyrocentres $g_\sigma = \langle \delta f_\sigma \rangle_{\bsy{R}_\sigma}$ may be shown to evolve according to the gyrokinetic (GK) equation \citep{catto_linearized_1978, frieman_nonlinear_1982}
\begin{equation} \label{eq:gyrokinetic_eq_realspace}
    \partial_t g_\sigma + \left(v_\parallel \bhat + \bsy{\tilde v}_{M\sigma}  \right) \cdot \nabla\left(  g_\sigma +  \frac{Z_\sigma e \langle \varphi \rangle_{\bsy{R}_\sigma} }{T_\sigma} F_{M\sigma}\right) + \langle \bsy{v}_E \rangle_{\bsy{R}_\sigma} \cdot \nabla (F_{M\sigma} + g_\sigma) = 0,
\end{equation}
where the derivatives are taken at constant particle energy $E_\sigma = m_\sigma(v_\parallel^2 + v_\perp^2)/2$ and magnetic moment $\mu_\sigma = m_\sigma v_\perp^2 / 2 B$, with the parallel and perpendicular velocities being $v_\parallel = \bsy{v}\cdot \bhat$ and $v_\perp= |{\bsy{v} \times \bhat}| $, respectively. The gradients are performed with respect to the gyrocentre position $\bsy{R}_\sigma = \bsy{r} - \bhat \times \bsy{v_\perp} / \Omega_\sigma$, shifted from the real space position $\bsy{r}$ by the gyroradius vector $\bhat \times \bsy{v}_\perp/\Omega_\sigma$. Here, $\bhat = \bsy{B}/B$ is the unit magnetic field vector, $B=|\bsy{B}|$ is the magnetic field strength, and $Z_\sigma e$, $m_\sigma$, $v_{T\sigma}$, $T_\sigma = m_\sigma v_{T\sigma}^2/2$, and $\Omega_\sigma = Z_\sigma e B / m_\sigma$ are the charge, mass, thermal speed, temperature, and Larmor gyration frequency of species $\sigma$, respectively. Angular brackets $ \langle \cdot \rangle_{\bsy{r}}$ and $\langle \cdot \rangle_{\bsy{R}_\sigma}$ denote gyro-averages at fixed real space position $\bsy{r}$ and gyrocentre position $\bsy{R}_\sigma$, respectively. Furthermore, $\bsy{v}_E = \bhat\times\nabla\varphi/B$ is the $\bsy{E}\times\bsy{B}$ velocity and $\bsy{\tilde v}_{M\sigma}$ is the magnetic drift velocity. The tilde on $\bsy{\tilde v}_{M\sigma}$ is meant to make it distinct from the quantity $v_{Mx}$ that we will introduce later in equation \eqref{eq:v_Mx}. The magnetic drift contains both the curvature and $\nabla B$ drifts and may be expressed as
\begin{equation} \label{eq:v_M}
    \bsy{\tilde v}_{M\sigma} = \rho_\sigma v_{T \sigma} \left( \frac{ v_\parallel^2 + v_\perp^2/2}{v_{T \sigma}^2} \frac{\bsy{B}\times \nabla B }{B^2} + \frac{v_\parallel^2}{v_{T \sigma}^2} \mu_0 \frac{\bsy{B}\times \nabla p }{B^3}  \right),
\end{equation}
where  $\rho_\sigma = v_{T\sigma}/\Omega_\sigma$ is the gyroradius of a thermal particle and $\mu_0$ is the vacuum permeability. The curvature drift contribution was simplified in \eqref{eq:v_M} by assuming ideal magnetohydrostatic force balance $(\nabla\times\bsy{B})\times\bsy{B}=\mu_0 \nabla p$, with $\nabla p$ the background pressure gradient. The electrostatic potential fluctuation $\varphi$ in the GK equation \eqref{eq:gyrokinetic_eq_realspace} must be determined self-consistently from quasineutrality
\begin{equation} \label{eq:quasineutrality_general}
    0 = \sum_\sigma Z_\sigma \int\mathrm{d}^3 v\, \delta f_\sigma = \sum_\sigma Z_\sigma \int\mathrm{d}^3 v\; \langle g_\sigma \rangle_{\bsy{r}} - \sum_\sigma  \frac{Z_\sigma^2 e n_\sigma}{T_\sigma} \left( 1 - \hat\Gamma_{0\sigma} \right) \varphi,
\end{equation}
where 
\begin{equation} \label{eq:Gamma0_hat}
   \hat\Gamma_{0\sigma} \varphi \equiv \frac{1}{n_\sigma} \int\mathrm{d}^3 v \, F_{M\sigma} \langle \langle \varphi \rangle_{\bsy{R}_\sigma} \rangle_{\bsy{r}}.
\end{equation}
The charge density in the quasineutrality equation \eqref{eq:quasineutrality_general} has contributions from the density of gyrocentres $\int\mathrm{d}^3 v\; \langle g_\sigma \rangle_{\bsy{r}}$ and from the polarisation of gyro-orbits, which is proportional to $\left( 1 - \hat\Gamma_{0\sigma} \right) \varphi$.

We consider ion-scale fluctuations, with typical spatial variation perpendicular to the magnetic field on the ion gyroradius scale, $\abs{\nabla_\perp  \ln g_\sigma} \sim \rho_i^{-1}$, and typical temporal variation on an ion transit time, $\partial_t \ln g_\sigma \sim v_{Ti}/R$. Here, $R$ is the major radius of the tokamak or stellarator under consideration and is taken to be the system scale length. While the typical scale length of the turbulence in the directions perpendicular to the magnetic field is on the ion gyroradius scale $\rho_i$, its parallel extent is on the much larger system scale $R$, with $ R \gg \rho_i$ in a strongly magnetised plasma. Therefore, the turbulence may be modelled in a perpendicularly narrow flux tube centered on a chosen flux surface with enclosed toroidal flux $\psi=\psi_0$ and on a chosen field line $\alpha=\alpha_0$. The field line label $\alpha$ is defined such that $\bsy{B}=\nabla \psi\times\nabla \alpha$. A set of coordinates $(x,y,\theta)$ is used to parametrise the flux tube, with $x$ labelling flux surfaces, $y$ labelling field lines within a flux surface, and the distance along the magnetic field chosen here to be parametrised by the poloidal angle $\theta$ without loss of generality. The coordinates $x$ and $y$ vary on the ion gyroradius scale. Therefore, all background (non-fluctuating) quantities are evaluated at $(\psi_0, \alpha_0$) and are independent of $x$ and $y$ to leading order in $\rho_i/R \ll 1$. As a consequence, statistical periodicity may be assumed to hold in $x$ and $y$ for the fluctuating quantities, so that the fluctuating gyrocentre distribution and electrostatic potential may be expanded in Fourier harmonics
\begin{equation}
    g_\sigma = \sum_{k_x, k_y} \hat g_\sigma (k_x, k_y, \theta, v_\parallel, v_\perp) e^{i(k_x X_\sigma + k_y Y_\sigma)}, \qquad \varphi = \sum_{k_x, k_y} \hat \varphi (k_x, k_y, \theta) e^{i(k_x x + k_y y)}.
\end{equation}

Due to their comparatively fast propagation speed, the electrons may be modelled by a modified adiabatic response
\begin{equation} \label{eq:modified_adiab}
    \delta f_e = \frac{e}{T_e} \left( \varphi - \langle \varphi \rangle_\psi \right) F_{Me}.
\end{equation}
The substraction of the flux-surface averaged potential $\langle \varphi \rangle_\psi$ stems from the electrons' fast motion along magnetic field lines ($v_{Te} \sim v_{Ti} \sqrt{m_i/m_e} \gg v_{Ti}$) and small gyroradii ($\rho_e \sim \rho_i \sqrt{m_e/m_i} \ll \rho_i$). Due to these differences with ions, electrons can only respond to in-surface variations of the electrostatic potential. Here, the flux-surface average is defined as
\begin{equation} \label{eq:def_FSA}
    \langle f \rangle_\psi = \int\frac{\mathrm{d}\theta}{\bsy{B}\cdot\nabla\theta} \,f^\mathrm{Z} \bigg/ \int \frac{\mathrm{d}\theta}{\bsy{B}\cdot\nabla\theta} \equiv \langle \langle f \rangle_y \rangle_\theta,
\end{equation}
where the zonal and nonzonal components of any quantity $f$ are defined as $f^\mathrm{Z} = \langle f \rangle_y = \sum_{k_x} \hat f(k_x, k_y=0, \theta) e^{i k_x x}$ and $f^\mathrm{NZ} = f-f^\mathrm{Z}$, respectively. The $\theta$-integral in \eqref{eq:def_FSA} extends over the length of the flux tube, $\theta \in [-\pi N_\mathrm{turns}, \pi N_\mathrm{turns}]$, where $N_\mathrm{turns}$ is generally chosen to be an integer for tokamak simulations but may be a non-integer real for stellarator simulations.

Using the modified adiabatic electron response \eqref{eq:modified_adiab} and assuming a single ion species $i$ for simplicity, the quasineutrality equation \eqref{eq:quasineutrality_general} simplifies to
\begin{equation}\label{eq:quasineutrality}
    \left( 1 - \hat\Gamma_{0i} \right) \varphi + \tau \left( \varphi - \langle \varphi \rangle_\psi \right) = \frac{T_i}{Z_i e n_i} \int\mathrm{d}^3 v\, \langle g_i \rangle_{\bsy{r}},
\end{equation} 
where $n_i$ is the background ion density and $\tau = T_i/Z_i T_e$ is the ratio of ion and electron temperatures weighted by the atomic number.

The modified adiabatic electron response crucially reduces the ZF inertia at long perpendicular wavelengths, where $\hat\Gamma_{0i} \approx 1$. This may be seen explicitly by taking the time derivative and flux-surface average of \eqref{eq:quasineutrality} and inserting \eqref{eq:gyrokinetic_eq_realspace}, which leads to the vorticity equation
\begin{equation} \label{eq:vorticity}
    \left\langle \left( 1 - \hat\Gamma_{0i} \right) \partial_t \varphi \right\rangle_\psi = -\frac{T_i}{Z_i e n_i} \left\langle \int\mathrm{d}^3 v\, \left\langle \langle \bsy{v}_E \rangle_{\bsy{R}_i} \cdot \nabla  g_i + \tilde v_{Mx} \partial_x \left( g_i +  \frac{Z_i e \langle \varphi \rangle_{\bsy{R}_i} }{T_i} F_{Mi} \right) \right\rangle_{\bsy{r}} \right\rangle_\psi.
\end{equation}
The vorticity equation describes the time evolution of the ZF velocity due to nonlinear stresses and due to the Stringer-Winsor force induced by the ion radial magnetic drift velocity
\begin{equation} \label{eq:v_Mx}
    \tilde v_{Mx} \equiv \tilde{\bsy{v}}_{M i} \cdot \nabla x = \rho_i v_{Ti} \frac{\bsy{B}\times \nabla B \cdot \nabla x}{B^2} \frac{ v_\parallel^2 + v_\perp^2/2}{v_{Ti}^2}  \equiv v_{Mx}\frac{ v_\parallel^2 + v_\perp^2/2}{v_{Ti}^2},
\end{equation}
where we used the fact that the background pressure $p=p(\psi)$ in \eqref{eq:v_M} is a flux function. Note that \eqref{eq:v_Mx} implicitly defines the velocity-independent quantity $v_{Mx}$ that will turn out to be convenient below. 

Since the $\theta$-variation of the zonal potential $\varphi^\mathrm{Z}$ is generally small due to the electron response in the quasineutrality equation \eqref{eq:quasineutrality}, we refer to $v_E^\mathrm{ZF} = \langle \bsy{v}_E \cdot \nabla y \rangle_\psi$ as the zonal flow (ZF). We note that the zonal $\bsy{E}\times\bsy{B}$ flow is perpendicular to both $\mathbf{B}$ and $\nabla\psi$; therefore, it generally has both a large poloidal and a small toroidal component.

\subsection{Secondary model} \label{sec:secondary}

In this study, the growth of ZFs is modelled by describing the dynamics of small-amplitude secondary modes over a primary background frozen in time. The ion gyrocentre distribution and the fluctuating potential are expanded as $g_i=g_i^P + g_i^S$ and $\varphi= \varphi^P + \varphi^S$ with $\abs{g_i^S}\ll \abs{g_i^P}$ and $\abs{\varphi^S} \ll \abs{\varphi^P}$, allowing the GK equation \eqref{eq:gyrokinetic_eq_realspace} to be linearised to
\begin{equation} \label{eq:gyrokinetic_eq_realspace_secondary}
        \partial_t g_i^S + \left(v_\parallel \bhat + \bsy{\tilde v}_{Mi}  \right) \cdot \nabla\left(  g_i^S +  \frac{Z_i e \langle \varphi^S \rangle_{\bsy{R}_i} }{T_i} F_{Mi}\right) + \langle \bsy{v}_E^S \rangle_{\bsy{R}_i} \cdot \nabla (F_{Mi} + g_i^P) +  \langle \bsy{v}_E^P \rangle_{\bsy{R}_i} \cdot \nabla g_i^S = 0.
\end{equation}
The secondary potential $\varphi^S$ is self-consistently determined by quasineutrality \eqref{eq:quasineutrality}
\begin{equation}\label{eq:quasineutrality_secondary}
        \left( 1 - \hat\Gamma_{0i} \right) \varphi^S + \tau \left( \varphi^S - \langle \varphi^S \rangle_\psi \right) = \frac{T_i}{Z_i e n_i} \int\mathrm{d}^3 v\, \langle g_i^S \rangle_{\bsy{r}}.
\end{equation}
The primary drive is assumed to be purely nonzonal, i.e. $\langle g_i^P \rangle_y=0$, as is the case for linearly unstable modes in gyrokinetics. In contrast, the secondary distribution $g_i^S$ has both zonal and nonzonal components. The purely growing RDK secondary mode was derived in \citet{rogers_generation_2000} by neglecting the linear terms $v_\parallel \bhat \cdot \nabla, \bsy{\tilde v}_{Mi}\cdot \nabla$, and $\langle \bsy{v}_E^S \rangle_{\bsy{R}_i} \cdot \nabla F_{Mi}$ in \eqref{eq:gyrokinetic_eq_realspace_secondary}. Furthermore, the primary drive was assumed to be a streamer, i.e. $\partial_x g_i^P = 0$, because \cite{rogers_generation_2000} investigated fast growing secondary modes `catching up' to a linearly unstable primary mode, and the streamers are generally the most unstable modes. The streamer primary drive assumption $\partial_x g_i^P=0$ can be relaxed, as shown in \ref{sec:strongly_driven} where the secondary mode theory is rederived allowing for radial variation of the primary drive, in the strongly driven and long perpendicular wavelength limit.

In the main text of this paper, we will consider a frozen streamer primary drive ($\partial_x g_i^P=0, \partial_t g_i^P = 0$) for simplicity. Aside from the scenario of a secondary mode catching up to a primary mode, this assumption may also be motivated by a separation of scales between the small-scale fast-oscillating secondary modes and a primary turbulent background slowly varying in $x$ and $t$. This scale separation is indeed satisfied in strongly-driven ITG turbulence \citep{nies_saturation_2024}, whence the secondary model is apt to describe the ZF behaviour in nonlinear turbulence simulations.

Motivated by the small radial scale of the toroidal secondary modes shown in Figure~\ref{fig:ZF_real_Fourier}, we will further consider in the theory large binormal scales compared to the small radial wavelength $\abs{ \partial_y \ln g_i^S} \ll \abs{ \partial_x \ln g_i^S}$ (the case of comparable binormal and radial scales is studied numerically in Section~\ref{sec:vy_shat}). We thus neglect in \eqref{eq:gyrokinetic_eq_realspace_secondary} the magnetic drift advection in the binormal direction and the diamagnetic frequency contribution $\langle \bsy{v}_E^S \rangle_{\bsy{R}_i} \cdot \nabla F_{Mi} \propto \partial_y \langle \varphi^S \rangle_{\bsy{R}_i}$. Note that the field-line labelling coordinate is $y \propto \alpha = \zeta - q \theta + \nu(\theta, \zeta)$, with $\nu$ a single-valued function, $q=q(x)$ the safety factor measuring the average pitch of magnetic field lines on a flux surface, and $\theta$ and $\zeta$ a set of arbitrary poloidal and toroidal angles, respectively. Therefore, the gradient of $y$
\begin{equation} \label{eq:nabla_y}
    \nabla y \propto (1+\partial_\zeta \nu)\nabla \zeta  + (-q + \partial_\theta \nu) \nabla\theta - \theta \,\nabla x \frac{\mathrm{d}q}{\mathrm{d}x}
\end{equation}
has a secularly increasing component along the field line due to magnetic shear $\hat s \equiv \mathrm{d}\ln q /\mathrm{d}\ln x$. The limit of negligible $\bsy{\tilde v}_{Mi}\cdot \nabla y$ thus also requires $\abs{ \partial_y \ln g_i^S \hat s \theta}  \ll \abs{ \partial_x \ln g_i^S}$, i.e. we exclude modes that are very extended along the magnetic field in our analysis. The Fourier modes in $k_x$ of the secondary mode then become eigenmodes of the system. Had we kept the magnetic shear in our derivation, we would have had to use a boundary condition in $\theta$ that is consistent with the secular term in \eqref{eq:nabla_y} -- see Section~\ref{sec:GK_sims}. This boundary condition couples the Fourier modes in $k_x$.

With our assumptions, the gyro-averages simplify to Bessel functions
\begin{equation}  \label{eq:J0}
    J_{0i} = J_{0}\left( \sqrt{2 b_i} \,\frac{v_\perp}{v_{Ti}} \right),
\end{equation}
with the factor $b_i$ (not to be confused with the unit magnetic field vector $\bhat$) measuring the strength of the finite Larmor Radius (FLR) effects,
\begin{equation}
    b_i = \frac{k_x^2 \abs{\nabla x}^2 \rho_i^2}{2}.
\end{equation}

If we further consider exponentially growing solutions, an equation for the secondary eigenmodes $g_i^S = \hat g_i^S e^{i(k_x X_i - \omega t)}, \varphi^S = \hat \varphi^S e^{i(k_x x - \omega t)}$ is readily derived from \eqref{eq:gyrokinetic_eq_realspace_secondary},
\begin{equation} \label{eq:gyrokinetic_eq_fourierspace_secondary}
        0 = - \omega \hat g_i^S +  \left(- i v_\parallel \bhat \cdot \nabla +  k_x \tilde{v}_{Mx}\right) \left( \hat g_i^S + \frac{Z_i e \hat \varphi^S}{T_i} J_{0i} F_{Mi} \right) -  k_x \tilde{v}_g^P \frac{Z_i e \hat \varphi^S}{T_i} J_{0i} F_{Mi} + k_x v_{Ex}^P \hat g_i^S.
\end{equation}
Here, we defined the radial $\bsy{E}\times\bsy{B}$ velocity associated with the primary,
\begin{equation} \label{eq:vEx_P}
    v_{Ex}^P = \bsy{v}_E^P \cdot \nabla x = - \frac{\bhat \times \nabla x \cdot \nabla y}{B} \partial_y \varphi^P
\end{equation}
and a normalised binormal derivative of the primary distribution function (with velocity units)
\begin{equation} \label{eq:omegaP_g_E}
    \tilde{v}_g^P = -\frac{T_i}{Z_i e} \frac{\bhat \times \nabla x \cdot \nabla y}{B} \frac{\partial_y g_i^P}{ F_{Mi}}.
\end{equation}
Because the gyro-averages have reduced to Bessel functions, the integral \eqref{eq:Gamma0_hat} simplifies to $\hat \Gamma_{0i} = \Gamma_{0i} \equiv I_0(b_i) e^{-b_i}$ with $I_0$ the modified Bessel function of the first kind. The quasineutrality equation \eqref{eq:quasineutrality_secondary} in Fourier space then becomes
\begin{equation}\label{eq:quasineutrality_secondary_fourier}
        \left( 1 - \Gamma_{0i} \right) \hat \varphi^S + \tau \left( \hat \varphi^S - \langle \hat \varphi^S \rangle_\psi \right) = \frac{T_i}{Z_i e n_i} \int\mathrm{d}^3 v\, J_{0i} \hat g_i^S.
\end{equation}

\subsection{Gyrokinetic simulations} \label{sec:GK_sims}

All simulations presented in this study use the code \texttt{stella} \citep{barnes_stella_2019} to model a hydrogenic plasma ($Z_i=1$) on the flux surface at half radius $r(\psi_0)=a/2$ of the Cyclone Base Case, a tokamak with flux surfaces of circular cross-section and minor radius $a=0.36 R$. The radial coordinate is $x=r-r(\psi_0)$ and the binormal coordinate is $y =  \alpha r(\psi_0)/q(\psi_0)$. The electrons follow the modified adiabatic response \eqref{eq:modified_adiab} and the ratio of ion to electron temperatures is set to $\tau=1$. The flux tube is chosen to extend over a single poloidal turn, $N_\mathrm{turns}=1$. The safety factor $q$ is frequently varied across simulations and is thus specified in the captions of the respective figures. In all simulations but those of Figure~\ref{fig:scan_kyP}, the magnetic shear $\hat s \equiv \mathrm{d}\ln q/\mathrm{d}\ln x$ measuring the radial variation of the field line pitch is set to $\hat s = 0.8$.

Two types of GK simulations are presented in this study. First, the zonal flows from fully nonlinear GK simulations of ITG turbulence are studied in Figures~\ref{fig:ZF_real_Fourier} and \ref{fig:NL_spectra_scan_q}. In these cases, a background ion temperature gradient $R/L_{Ti} = -R \,\mathrm{d} \ln T_i / \mathrm{d} r = 13.9$ and background density gradient $R/L_{n} = -R \,\mathrm{d} \ln n_i / \mathrm{d} r =  2.2$ are assumed. The twist-and-shift boundary condition \citep{beer_fieldaligned_1995} is used at the ends of the flux-tube domain,
\begin{equation} \label{eq:BC_twist-shift}
    \hat g_\sigma(k_x, k_y, \theta=\pi N_\mathrm{turns}) = \hat g_\sigma(k_x + 2\pi N_\mathrm{turns} \hat s k_y, k_y, \theta=-\pi N_\mathrm{turns}),
\end{equation}
such that different radial wavenumbers are coupled through the magnetic shear. The box size in the binormal and radial directions is $L_y = 2\pi/\min(k_y) = 377 \rho_i$ and $L_x = 2\pi/\min(k_x) = 375 \rho_i$, respectively. The simulations include perpendicular wavenumbers up to $k_y \rho_i = 1.42$ and $k_x \rho_i = 2.13$, with a parallel resolution of $N_\theta = 32$. The velocity-space resolutions are $N_{v_\parallel}= 32$ for the parallel velocity grid and $N_\mu = 8$ for the magnetic moment grid. Parallel velocities between $-3 v_{Ti}$ and $3 v_{Ti}$ are considered, and the maximum of $\mu$ is chosen such that the maximum perpendicular velocity is $3 v_{Ti}$ at the minimum value of $B$.

In all other GK simulations (Figures~\ref{fig:summary_regions_GK},~\ref{fig:validation_D_TSM},~\ref{fig:forbidden_region},~\ref{fig:TSM_phase},~\ref{fig:TSM_mechanism},~\ref{fig:non_resonant_omega_AP},~\ref{fig:etaprp_etapar},~\ref{fig:scan_kyP},~and~\ref{fig:scan_qinp}), the linearised secondary system of equations \eqref{eq:gyrokinetic_eq_realspace_secondary} and \eqref{eq:quasineutrality_secondary} is solved to obtain the complex frequency of exponentially growing secondary modes at a specified radial wavenumber $k_x = k_x^S \neq 0$. In these simulations, the $(k_x =k_x^P = 0, k_y=k_y^P$) Fourier mode is enforced to be constant in time to represent the streamer primary mode, i.e. a slowly growing linear mode or a slowly evolving turbulent background. For simplicity, the primary drive is thus always chosen in simulations to vary sinusoidally in the binormal coordinate $y$ (in contrast, the theory of Section~\ref{sec:secondary_modes_toroidal_geo} allows for arbitrary $y$-variation). Unless stated otherwise, the GK simulations of the secondary are performed with binormal resolution $N_y = 32$, parallel resolution $N_\theta = 96$, parallel velocity resolution $N_{v_\parallel} = 72$, and magnetic moment resolution $N_\mu = 16$. 

We characterise the strength of the primary drive through the normalised amplitude
\begin{equation} \label{eq:A_P}
    \mathcal{A}^P\equiv \sqrt{\left\langle \left( \frac{v_{Ex}^{P}}{v_{Ti}}\frac{R}{\rho_i} \right)^2 \right\rangle_\psi},
\end{equation}
which provides a measure of the relative strengths of the primary $\bsy{E}\times\bsy{B}$ flow velocity and the radial magnetic drift velocity $v_{Mx} \sim \rho_i v_{Ti}/R$. In \cite{nies_saturation_2024}, GK simulations of the secondary mode were shown for an ITG eigenmode taken as the primary drive. To allow for a direct comparison between GK simulations and the secondary mode theory that we will present below, the primary drive is here set to be a bi-Maxwellian with density $n^P$, parallel temperature $T_\parallel^P$, and perpendicular temperature $T_\perp^P$, i.e.
\begin{equation} \label{eq:model_gP_M}
	g_i^P = F_{Mi} \left[ \frac{n^P}{n_i}  + \frac{T_\parallel^P}{T_i}\left( \frac{v_\parallel^2}{v_{Ti}^2} - \frac{1}{2}\right) +  \frac{T_\perp^P}{T_i}\left( \frac{v_\perp^2}{v_{Ti}^2} - 1\right) \right].
\end{equation}
We choose to consider parallel and perpendicular primary temperatures separately because they contribute differently to the growth of various secondary modes. In particular, the RDK secondary modes will be shown to depend only on $T_\perp^P$ as they rely on finite Larmor radius effects and the gyroradius size depends only on the perpendicular particle energy, see also \eqref{eq:J0}. In contrast, the TSM instability mechanism hinges on the radial magnetic drift \eqref{eq:v_Mx}, which depends on both parallel and perpendicular particle energy. 

For the bi-Maxwellian primary distribution \eqref{eq:model_gP_M}, equation \eqref{eq:omegaP_g_E} simplifies to
\begin{equation} \label{eq:v_gP}
    \tilde{v}_g^P = \tau v_{Ex}^P \left[ 1 + \eta_\parallel^P\left( \frac{v_\parallel^2}{v_{Ti}^2} - \frac{1}{2}\right) +  \eta_\perp^P \left( \frac{v_\perp^2}{v_{Ti}^2} - 1\right) \right].
\end{equation}
Here, we defined the dimensionless ratios of the primary parallel and perpendicular temperature gradients to the density gradient,
\begin{equation}\label{eq:eta_par_prp}
    \eta_{\parallel}^P = \frac{n_i}{T_i} \frac{\partial_y T_{\parallel}^P}{\partial_y  n^P}, \qquad \eta_{\perp}^P = \frac{n_i}{T_i} \frac{\partial_y T_{\perp}^P}{\partial_y  n^P},
\end{equation}
and used quasineutrality \eqref{eq:quasineutrality} in the limit of long binormal wavelengths to relate the primary density and potential,
\begin{equation}
    -\frac{T_i}{Z_i e} \frac{\bhat \times \nabla x \cdot \nabla y}{B} \partial_y \ln n_i^P = \tau v_{Ex}^P.
\end{equation}
Arbitrary variations in $y$ and $\theta$ of $v_{Ex}^P$, $\eta_{\parallel}^P$, and $\eta_{\perp}^P$ are allowed in the theory. In the simulations, $\eta_{\parallel}^P$ and $\eta_{\perp}^P$ are taken to be constants and the primary drive's variation on the flux surface is captured entirely by $v_{Ex}^P(y,\theta)$, with the exception of the simulations in Figure~\ref{fig:validation_D_TSM_scan_phase} where a phase shift in $y$ between the density and temperature moments of the primary distribution function is considered. The variation of the primary drive $v_{Ex}^P(y,\theta)$ along the magnetic field for each simulation is specified in the corresponding figures.

The primary mode's binormal wavenumber $k_y^P \rho_i = 10^{-3}$ is chosen to be small in all simulations (except those of Figure~\ref{fig:scan_kyP}) to ensure the binormal derivatives and magnetic shear are negligible -- see discussion around \eqref{eq:nabla_y}. When $k_y \rho_i$ is sufficiently small, the dynamics at different radial wavenumbers are effectively decoupled, and we therefore impose the phase-shift-periodic boundary condition \citep{st-onge_phase-shift-periodic_2022, volcokas_ultra_2023}
\begin{equation} \label{eq:BC_phase-shift-periodic}
    \hat g_\sigma(k_x, k_y, \theta=\pi N_\mathrm{turns}) = \exp\left( i k_y L_y \Theta \right) \hat g_\sigma(k_x, k_y, \theta=-\pi N_\mathrm{turns})
\end{equation}
in the GK simulations of the secondary mode, with $L_y = 2\pi/\min(k_y)$ the size of the binormal domain. We note that the zonal modes ($k_y = 0$) always satisfy periodic boundary conditions, as may be seen from \eqref{eq:BC_twist-shift} and \eqref{eq:BC_phase-shift-periodic}. The phase-shift-periodic boundary condition \eqref{eq:BC_phase-shift-periodic} models a flux surface with zero magnetic shear. The flux surface is rational if $\Theta \in \mathbb{Z}$ and irrational if $\Theta \in \mathbb{R} \setminus \mathbb{Z}$. In practice, this boundary condition is chosen because it does not couple different $k_x$ Fourier modes. The choice of $\Theta$ will dictate the degree of decorrelation at the ends of the flux tube; e.g., small values of $\Theta$ allow modes to remain coherent after wrapping around the flux tube. We will show that most results of interest are not affected by this boundary condition because the choice of boundary conditions is important for the secondary modes only when the parallel streaming contribution to \eqref{eq:gyrokinetic_eq_fourierspace_secondary} is relevant to the dynamics; some discussion of these effects will be provided in Section~\ref{sec:streaming_TSM}.

\section{Secondary modes in toroidal geometry}\label{sec:secondary_modes_toroidal_geo}

\begin{figure}
    \centering
    \begin{subfigure}[t]{0.6\columnwidth}
         \centering
         \includegraphics[width=\textwidth, trim={0.2cm 0.8cm 0.4cm 0.8cm},clip]{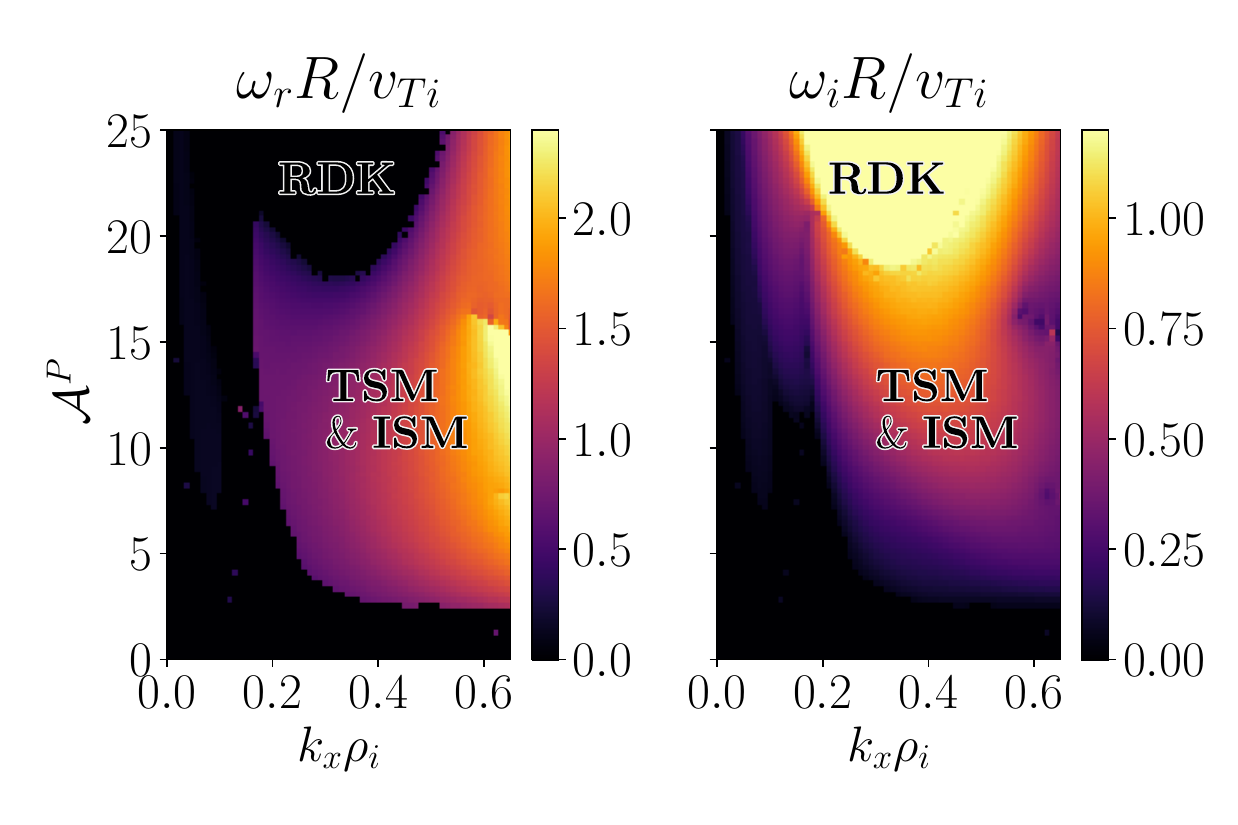}
         \caption{Secondary mode real frequency (left) and growth rate (right) from gyrokinetic simulations. The primary distribution function is given by \eqref{eq:model_gP_M} with $\eta_\parallel^P=\eta_\perp^P=2$ and varies along the magnetic field as $g_i^P \propto e^{-(\theta/\pi)^2}$. The safety factor is set to $q=2.8$. The parallel boundary condition is the phase-shift-periodic boundary condition \eqref{eq:BC_phase-shift-periodic} with $\Theta=0.288572618$. Due to computational cost, the simulations are performed with a reduced resolution $\{N_y, N_\theta, N_{v_\parallel}, N_\mu\} = \{16, 48, 32, 8\}$.}
        \label{fig:summary_regions_GK}
    \end{subfigure}
    \begin{subfigure}[t]{0.391\columnwidth}

\begin{tikzpicture}
        \node[anchor=south west,inner sep=0] (image) at (0,0) {\includegraphics[width=\textwidth, trim={0.45cm 0.6cm 0cm 0cm},clip]{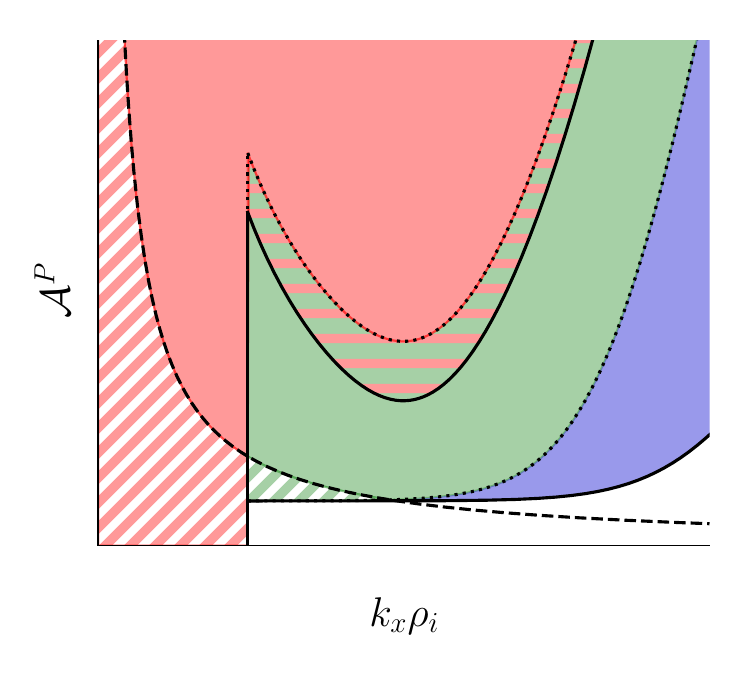}};
        \begin{scope}[x={(image.south east)},y={(image.north west)}]

        \node[anchor=center, font=\normalsize] at (0.46,0.89) {\contour{white}{\textbf{RDK}} (\ref{sec:TSM_NR}, \ref{sec:strongly_driven})};
        
        \node[anchor=center, font=\normalsize] at (0.51,0.43) {(\ref{sec:TSM_NR})};

        \node[anchor=center, font=\normalsize] at (0.55,0.3) {\contour{white}{\textbf{TSM}} (\ref{sec:TSM_R})};

        \node[anchor=center, rotate=-78, font=\normalsize] at (0.21,0.6) {$\omega_\parallel \sim k_x \tilde v_g^P$};
        
        \node[anchor=center, rotate=-60, font=\normalsize] at (0.37,0.7) {$\omega_\mathrm{GAM} \sim$};
        
        \node[anchor=center, rotate=-45, font=\normalsize] at (0.48,0.55) {$k_x \tilde v_g^P$};

        \node[anchor=center, font=\normalsize] at (0.3,0.12) {Landau};
        \node[anchor=center, font=\normalsize] at (0.3,0.05) {(\ref{sec:streaming_TSM})};

        \node[anchor=center, font=\normalsize] at (0.88,0.45) {\contour{white}{\textbf{ISM}}};
        \node[anchor=center, font=\normalsize] at (0.88,0.375) {(\ref{sec:ISM})};

    \end{scope}
\end{tikzpicture}
        \centering
        \caption{Approximate regions of Rogers-Dorland-Kotschenreuther (RDK) secondary mode \citep{rogers_generation_2000}, toroidal secondary mode (TSM) \citep{nies_saturation_2024}, and ITG secondary mode (ISM).  The parallel streaming frequency is $\omega_\parallel \sim v_\parallel \hat{b}\cdot\nabla \sim v_{Ti}/qR$ and the geodesic acoustic mode (GAM) frequency is $\omega_\mathrm{GAM} \sim v_{Mx}/\rho_i \sim v_{Ti}/R$ \eqref{eq:omega_GAM}.}
        \label{fig:summary_regions_overview}
    \end{subfigure}
   \caption{Illustrative case of secondary modes in the space of the primary amplitude $\mathcal{A}^P$ and secondary radial wavenumber $k_x$ (a), and qualitative picture (b). The TSM ($\omega_r \neq 0$) is found at short radial wavelengths $k_x \gtrsim \omega_\parallel / v_{Mx}$ as it is Landau damped at long radial wavelengths (Section~\ref{sec:streaming_TSM}), and above a primary amplitude threshold $\mathcal{A}^P \gtrsim 1$ as it is otherwise subject to kinetic damping by the radial magnetic drift (Section~\ref{sec:TSM_R}). The RDK secondary mode ($\omega_r = 0$) is found at large primary amplitudes, $k_x \tilde v_{g}^P \gtrsim \omega_\mathrm{GAM}$ for short wavelengths $k_x \gtrsim \omega_\parallel / v_{Mx}$ (Section~\ref{sec:TSM_NR}), and $k_x \tilde v_{g}^P \gtrsim \omega_\parallel$ for long wavelengths $k_x \lesssim \omega_\parallel / v_{Mx}$ (Section~\ref{sec:streaming_TSM}). The horizontally dashed region in (b) is where the non-resonant limit of the toroidal secondary mode exists (Section~\ref{sec:TSM_NR}), between the TSM and RDK regions. The ISMs, discussed in Section~\ref{sec:ISM}, are dominant at short wavelengths $k_x \gtrsim \omega_\parallel / v_{Mx}$ for weak primary drive $\mathcal{A}^P \sim 1$, and at sub-Larmor scales $k_x \rho_i \gtrsim 1$. The diagonally hashed regions in (b) have weak primary drive $k_x \tilde v_g^P \lesssim \omega_\parallel$; the comparatively fast parallel streaming rate then makes the choice of parallel boundary condition important and the assumptions made in our secondary mode theory may become inappropriate (Section~\ref{sec:streaming_TSM}). Note that the size and shape of the regions can vary depending on the parameters used, e.g. the dependence of the primary drive $k_x \tilde v_g^P$ on velocity, $y$, and $\theta$, or the magnetic geometry.}  
   \label{fig:summary_regions}
\end{figure}
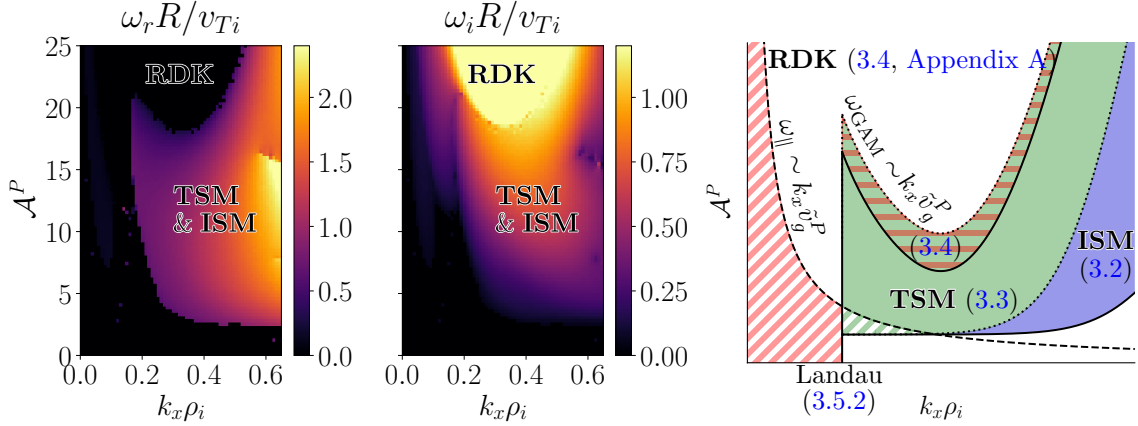

An overview of the secondary modes discussed in this study is shown in Figure~\ref{fig:summary_regions} for varying radial wavenumber $k_x$ and primary amplitude $\mathcal{A}^P$. In addition to the purely growing RDK secondary mode at large $\mathcal{A}^P$, growing TSMs and ISMs with finite $\omega_r$ are observed at smaller $\mathcal{A}^P$. As discussed in Section~\ref{sec:GK_sims}, we consider the primary drive to be given by the bi-Maxwellian \eqref{eq:model_gP_M} to allow comparisons between the secondary mode theory and the GK simulations. The bi-Maxwellian primary seems to appropriately capture the behaviour of secondary modes: indeed, Figure~\ref{fig:summary_regions_GK} is qualitatively similar to Figure~2 in \cite{nies_saturation_2024}, which uses a linear ITG mode as the primary mode driving the secondary and also exhibits TSMs.

After deriving the theory of the various secondary modes in Section~\ref{sec:theory_ISM_TSM}, the ISM and TSM will be discussed in Sections~\ref{sec:ISM} and \ref{sec:TSM_R}, respectively. Then, the non-resonant limit of the secondary modes will be considered in Section~\ref{sec:TSM_NR}, including the RDK limit (see also \ref{sec:strongly_driven}). Finally, the effects of finite binormal wavenumbers and parallel streaming will be studied in Section~\ref{sec:robustness_TSM}, including a discussion of the Landau damping threshold for the TSM (see Figure~\ref{fig:summary_regions_overview}) in Section~\ref{sec:streaming_TSM}.

\subsection{Theory of ITG and toroidal secondary modes} \label{sec:theory_ISM_TSM}

We consider the limit where the effects of parallel streaming are negligible, e.g. by considering a tokamak with a large safety factor, as $v_\parallel \bhat \cdot \nabla \sim v_{Ti} / qR$. The required size of $q$ to make the parallel streaming sufficiently small depends on the primary amplitude and the radial wavenumber under consideration, as will be discussed further in Section~\ref{sec:streaming_TSM}. From here on, we use the symbol $\omega_\parallel$ to refer to the typical size of the parallel streaming rate, $\omega_\parallel \sim v_\parallel \bhat\cdot\nabla$. In the absence of parallel streaming, the GK equation for the secondary eigenmodes \eqref{eq:gyrokinetic_eq_fourierspace_secondary} simplifies to an algebraic equation
\begin{equation} \label{eq:gyrokinetic_eq_fourierspace_secondary_no_streaming}
        0 = - \omega \hat g_i^S +  k_x \tilde{v}_{Mx} \left( \hat g_i^S + \frac{Z_i e \hat \varphi^S}{T_i} J_{0i} F_{Mi} \right) -  k_x \tilde{v}_g^P \frac{Z_i e \hat \varphi^S}{T_i} J_{0i} F_{Mi} + k_x v_{Ex}^P \hat g_i^S,
\end{equation}
with solution
\begin{equation} \label{eq:gS_TSM}
	\hat g_i^S = \frac{k_x \tilde v_{Mx} - k_x \tilde{v}_g^P}{\omega - k_x v_{Ex}^P - k_x \tilde v_{Mx}} J_{0i} F_{Mi} \frac{Z_i e \hat \varphi^S}{T_i}.
\end{equation}
Inserting \eqref{eq:gS_TSM} into the quasineutrality equation \eqref{eq:quasineutrality_secondary_fourier} and rearranging, we obtain
\begin{equation}\label{eq:quasineutrality_GSec}
    \tau \langle\hat \varphi^S\rangle_\psi = \hat \varphi^S \left( 1-\Gamma_{0i} + \tau - \mathcal{N} \right),
\end{equation}
where we defined the ion response function
\begin{equation} \label{eq:curlyN}
    \mathcal{N} = \frac{1}{n_i} \int\mathrm{d}^3 v\, J_{0i}^2 F_{Mi} \frac{k_x \tilde v_{Mx} - k_x \tilde{v}_g^P}{\omega - k_x v_{Ex}^P - k_x \tilde v_{Mx}}.
\end{equation}

Equation \eqref{eq:quasineutrality_GSec} has two classes of solutions. First, when the zonal flow component of the secondary mode vanishes ($\langle \hat \varphi^S \rangle_\psi / \mathrm{max}(\hat\varphi^S)=0$), there are local modes on the flux surface (localised around particular values of $y$ and $\theta$) where
\begin{equation} \label{eq:D_ISM}
    0 = \mathcal{D}^\mathrm{ISM}(y, \theta) \equiv \left( 1-\Gamma_{0i} + \tau - \mathcal{N} \right) / \tau.
\end{equation}
These modes may be identified as local ISMs driven unstable by the gradients in the primary drive. They will be discussed further in Section~\ref{sec:ISM}. 

The second class of solutions to \eqref{eq:quasineutrality_GSec} corresponds to secondary modes with nonzero zonal flow component $\langle \hat \varphi^S \rangle_\psi \neq 0$, which will be our principal focus. The corresponding dispersion relation is derived from \eqref{eq:quasineutrality_GSec} and the consistency condition $1=\langle \hat \varphi^S / \langle \hat \varphi^S \rangle_\psi \rangle_\psi$,
\begin{equation} \label{eq:D_TSM}
    0 = \mathcal{D} \equiv 1-\left\langle \frac{1}{\mathcal{D}^\mathrm{ISM}} \right\rangle_\psi = 1 - \left\langle \frac{\tau}{ 1-\Gamma_{0i} + \tau - \mathcal{N}} \right\rangle_\psi.
\end{equation}
These modes are global on the flux surface: once the mode frequency has been obtained by solving \eqref{eq:D_TSM}, the shape of the eigenfunction in $y$ and $\theta$ is given by \eqref{eq:quasineutrality_GSec}. As we will show explicitly below (see Section~\ref{sec:TSM_NR}), the dispersion relation \eqref{eq:D_TSM} includes the TSM, the RDK secondary mode, and the GAM.

To evaluate the dispersion relations numerically, we follow the procedure of \cite{terry_kinetic_1982} and \cite{biglari_toroidal_1989} and express the resonant denominator in \eqref{eq:curlyN} as
\begin{equation}
    \left(\omega - k_x v_{Ex}^P - k_x \tilde v_{Mx}\right)^{-1} = -i \int_0^\infty \mathrm{d}\lambda \, \exp\left[ i \lambda (\omega - k_x v_{Ex}^P - k_x \tilde v_{Mx}) \right],
\end{equation}
where we assumed exponentially growing secondary modes ($\omega_i > 0$) so that the $\lambda$-integral converges at infinity. If we further assume a bi-Maxwellian primary distribution function \eqref{eq:model_gP_M}, the parallel and perpendicular velocity integrals in \eqref{eq:curlyN} may be evaluated to give
\begin{align} \label{eq:curlyN_lambdaint}
    \mathcal{N} & = -2i \int_0^\infty \mathrm{d}\lambda \, \frac{e^{i \lambda(\omega - k_x v_{Ex}^P)}}{\left(1 + i \lambda k_x v_{Mx} \right)^{1/2}\left(2 + i \lambda k_x v_{Mx} \right)} \Bigg[ \left( \Gamma_{0i}(\tilde b_i) + \tilde b_i \Gamma_{0i}'(\tilde b_i) \right) \frac{k_x v_{Mx} - 2\tau k_x v_{Ex}^P \eta_\perp^P}{2+i \lambda k_x v_{Mx}} \nonumber \\
    & + \Gamma_{0i}(\tilde b_i) \left( \tau k_x v_{Ex}^P \left( \frac{\eta_\parallel^P}{2} + \eta_\perp^P - 1 \right) + \frac{k_x v_{Mx}-\tau k_x v_{Ex}^P \eta_\parallel^P}{2 (1 + i \lambda k_x v_{Mx})} \right) \Bigg],
\end{align}
where $\Gamma_{0i}'(\tilde b_i) = \mathrm{d}\Gamma_{0i}(\tilde b_i)/\mathrm{d}\tilde b_i$ and
\begin{equation}
    \tilde b_i = \frac{2 b_i}{2 + i \lambda k_x v_{Mx}}.
\end{equation}
The integrand in \eqref{eq:curlyN_lambdaint} decays exponentially at large $\lambda$ as $\omega_i > 0$ and the integral may be evaluated efficiently numerically.

\begin{figure}
    \begin{subfigure}[t]{\columnwidth}
            \centering
            \includegraphics[width=\textwidth, trim={0.5cm 0.1cm 0.5cm 0cm}, clip]{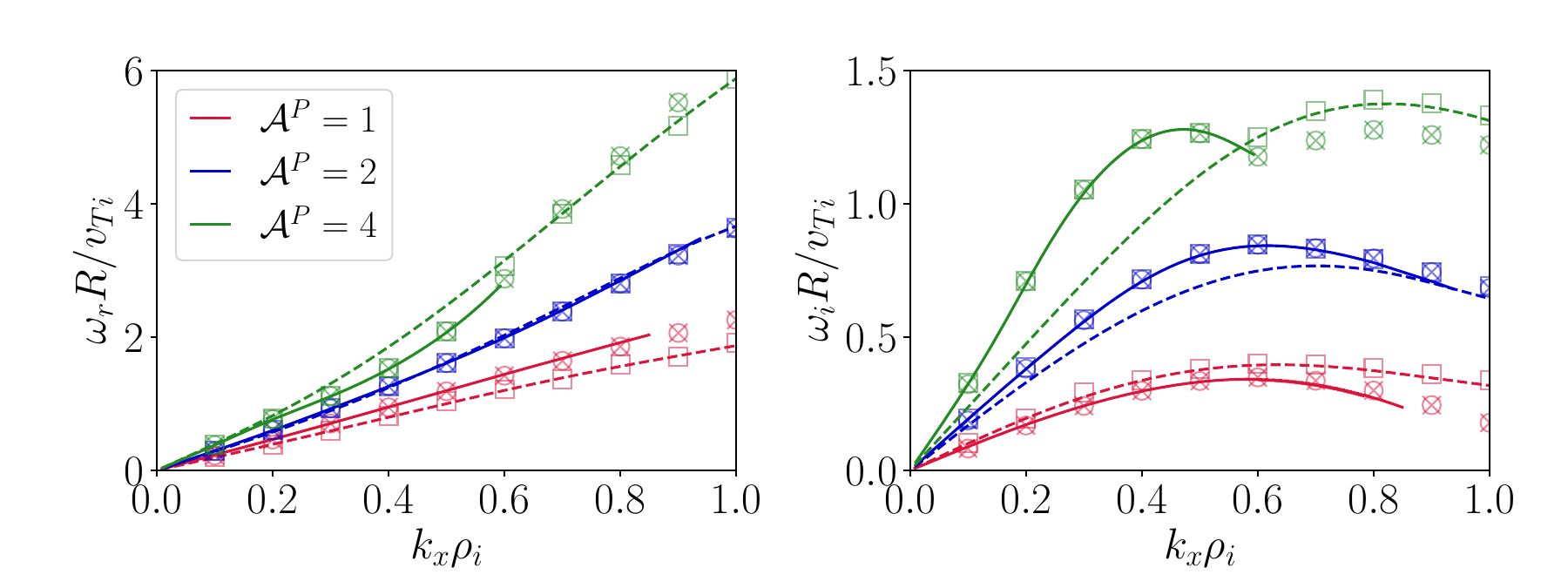}
        \caption{Varying secondary mode wavenumber $k_x \rho_i$ at fixed $\delta = 0$.}
        \label{fig:validation_D_TSM_scan_kx}
    \end{subfigure}
    \begin{subfigure}[t]{\columnwidth}
            \centering
            \includegraphics[width=\textwidth, trim={0.5cm 0.1cm 0.5cm 0cm}, clip]{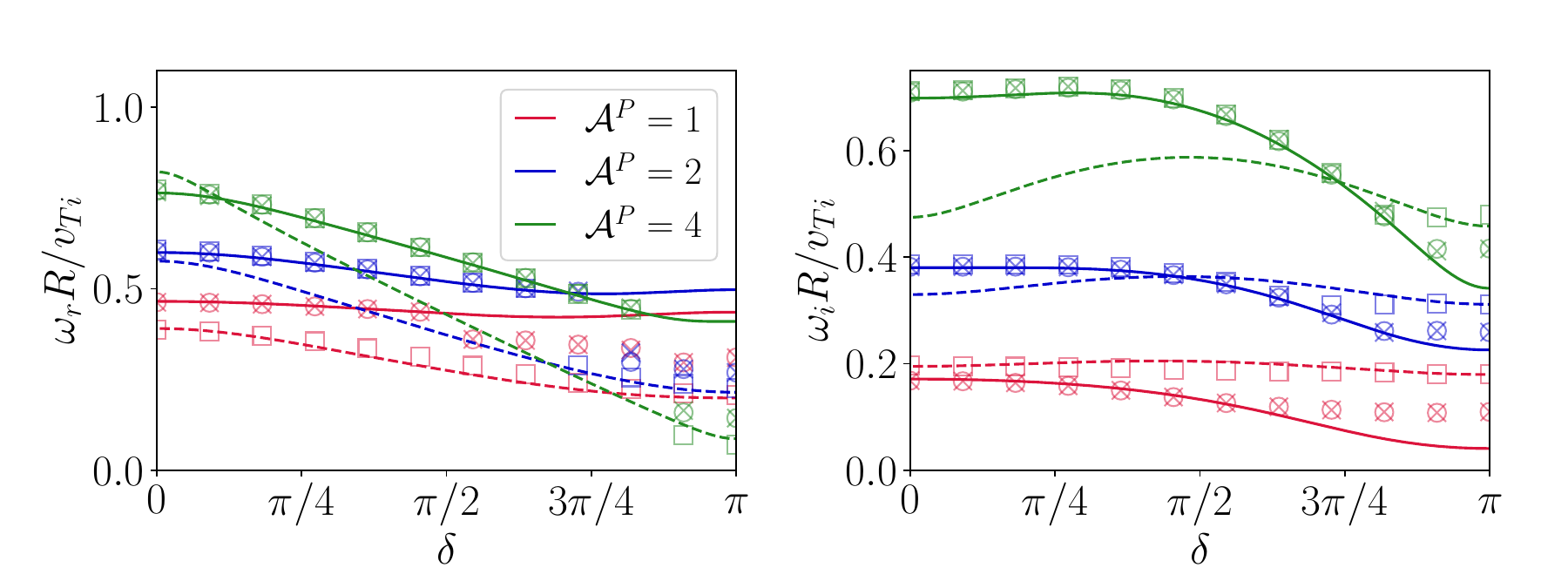}
        \caption{Varying phase shift $\delta$ between $(T_\parallel^P, T_\perp^P)$ and $n^P$ at fixed $k_x \rho_i = 0.2$.}
        \label{fig:validation_D_TSM_scan_phase}
    \end{subfigure}
    \caption{Secondary mode frequency (left) and growth rate (right) of the fastest growing modes as a function of $k_x \rho_i$ (a) and phase shift $\delta$ (b) (see definition of $\delta$ below in this caption). We calculate the frequency and growth rate from the generalised secondary mode dispersion relation \eqref{eq:D_TSM} (solid lines), the ISM dispersion relation \eqref{eq:D_ISM} (dashed lines), and from gyrokinetic simulations of the secondary mode (markers). Different primary amplitude values $\mathcal{A}^P$ (colors) are examined. The primary drive has the form \eqref{eq:model_gP_M} with $v_{Ex}^P \propto \sin(k_y^P y)$ and $\eta_\perp^P = \eta_\parallel^P = 3 \sin(k_y^P y + \delta) / \sin(k_y^P y)$. The primary drive varies along the magnetic field as $g_i^P \propto e^{-(\theta/\pi)^2}$. The gyrokinetic simulations of the secondary mode have a safety factor value $q=20$ and employ the phase-shift-periodic boundary condition \eqref{eq:BC_phase-shift-periodic} with $\Theta = 0$ (circles) and $\Theta = 0.288572618$ (crosses); the squares correspond to gyrokinetic simulations where the parallel streaming and binormal magnetic drift were artificially switched off in the simulations. The three markers overlap in those cases where the secondary mode is unaffected by the choice of boundary condition and by the inclusion of in-surface derivatives.}
    \label{fig:validation_D_TSM}
\end{figure}

The dispersion relations \eqref{eq:D_ISM} and \eqref{eq:D_TSM} are validated against gyrokinetic simulations in Figure~\ref{fig:validation_D_TSM}. Excellent agreement is obtained between the theory and the gyrokinetic simulations for both the secondary mode frequencies and growth rates. We note that in-surface derivatives (parallel streaming and binormal drifts), which were neglected when deriving \eqref{eq:gyrokinetic_eq_fourierspace_secondary_no_streaming}, affect the ISM frequency and growth rate because the solutions of \eqref{eq:D_ISM} correspond to modes localised to a point on the flux surface and hence have large derivatives with respect to $\theta$ and $y$. The ISM frequency and growth rate are exactly recovered only when artificially removing the parallel streaming and binormal drift in the GK simulation, see square markers in Figure~\ref{fig:validation_D_TSM}. In contrast, the TSM's global (on the flux surface) character means the mode frequency is correctly captured even for finite (but small) values of parallel streaming and binormal drifts; the effects of large $k_y^P$ and small $q$ on the TSM will be discussed further in Section~\ref{sec:robustness_TSM}. When the ISM growth rate sufficiently exceeds that of the TSM, the GK simulations with finite parallel streaming and binormal drifts capture an ISM with reduced growth rate, as may be seen for $\mathcal{A}^P=4, k_x\rho_i \gtrsim 0.6$ in Figure~\ref{fig:validation_D_TSM_scan_kx} and for $\delta \approx \pi$ in Figure~\ref{fig:validation_D_TSM_scan_phase}.

The primary amplitude is seen in Figure~\ref{fig:validation_D_TSM_scan_kx} to strongly affect the growth rate of the TSM, which has a finite $\mathcal{A}^P$ threshold for instability. In contrast, increasing $\mathcal{A}^P$ causes smaller relative changes to the real frequency of the TSM, which approximately satisfies $\omega_r \approx 2 k_x \rho_i v_{Ti}/R$, in agreement with Figure~\ref{fig:ZF_real_Fourier}. The smaller relative changes in $\omega_r$ compared to $\omega_i$ with increasing $\mathcal{A}^P$ may also be observed in the TSM region of Figure~\ref{fig:summary_regions_GK}. 

Moreover, as shown in Figure~\ref{fig:validation_D_TSM_scan_phase}, the frequency and growth rate of the TSM vary smoothly with the phase shift $\delta$ between the density and temperature of the primary drive. The TSM growth rate reaches its maximum for $\delta =0$ and its minimum for $\delta=\pi$, corresponding to the primary density and temperature gradient being parallel and anti-parallel, respectively. While the phase shift $\delta$ is of crucial importance for the primary modes (as it determines the heat flux due to these modes), it does not fundamentally alter the physics of the TSM (see discussion in Section~\ref{sec:TSM_R}) and therefore it does not significantly affect the TSM growth rate.

Given the relatively uninteresting dependence on the phase shift $\delta$, we will not consider finite phase shifts $\delta$ in the remainder of the simulations presented in this article, with $\eta_\parallel^P$ and $\eta_\perp^P$ chosen to be positive (negative) constants to model the parallel (anti-parallel) primary gradient cases. We note that this misses the peak of the ISM growth rate at $\delta=\pi/2$: the ISM, like the usual ITG primary mode, is destabilised by temperature gradients and stabilised by density gradients. The maximal ISM growth rate is therefore found at $\delta=\pi/2$, as there are then regions on the flux surface where the temperature gradient reaches its maximum while the density gradient vanishes.

\begin{figure}
    \centering
    \begin{subfigure}[t]{0.32\columnwidth}
        \centering
         \includegraphics[width=\textwidth, trim={1.0cm 0.5cm 3.0cm 0.5cm},clip]{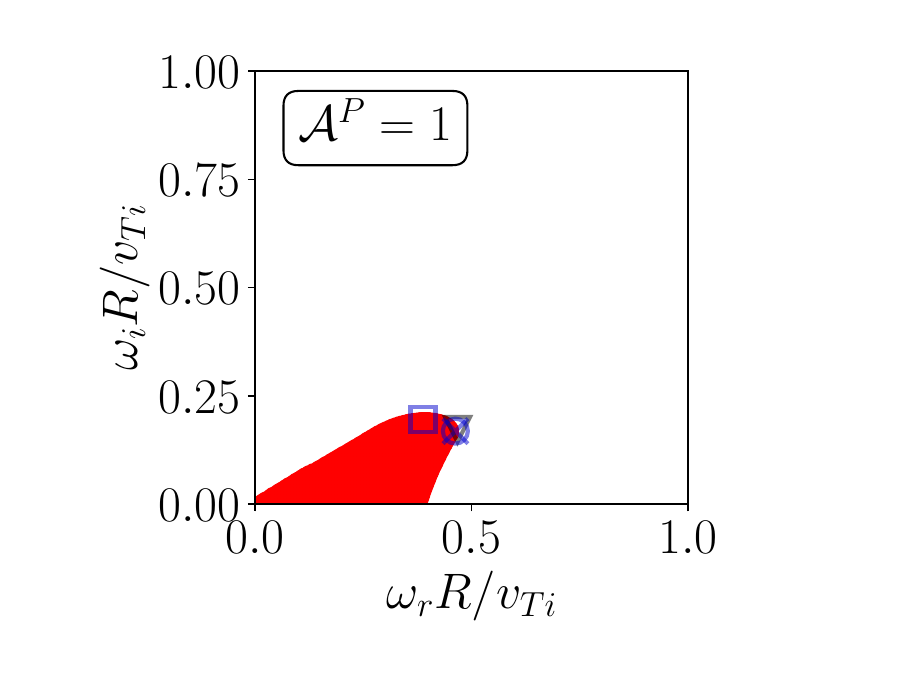}
    \end{subfigure}  
    \begin{subfigure}[t]{0.32\columnwidth}
        \centering
         \includegraphics[width=\textwidth, trim={1.0cm 0.5cm 3.0cm 0.5cm},clip]{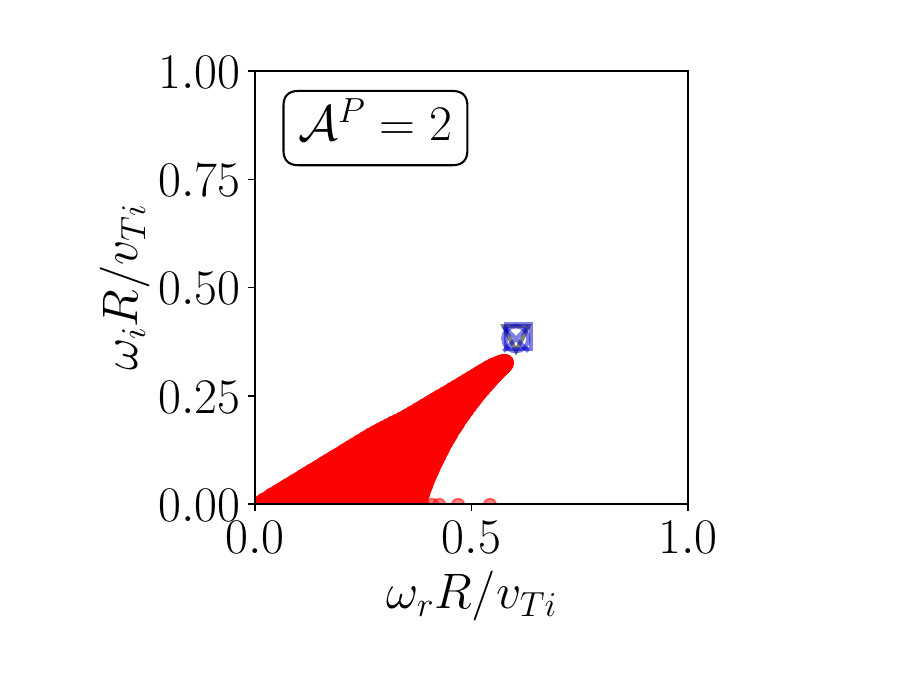}
    \end{subfigure}
    \begin{subfigure}[t]{0.32\columnwidth}
        \centering
         \includegraphics[width=\textwidth, trim={1.0cm 0.5cm 3.0cm 0.5cm},clip]{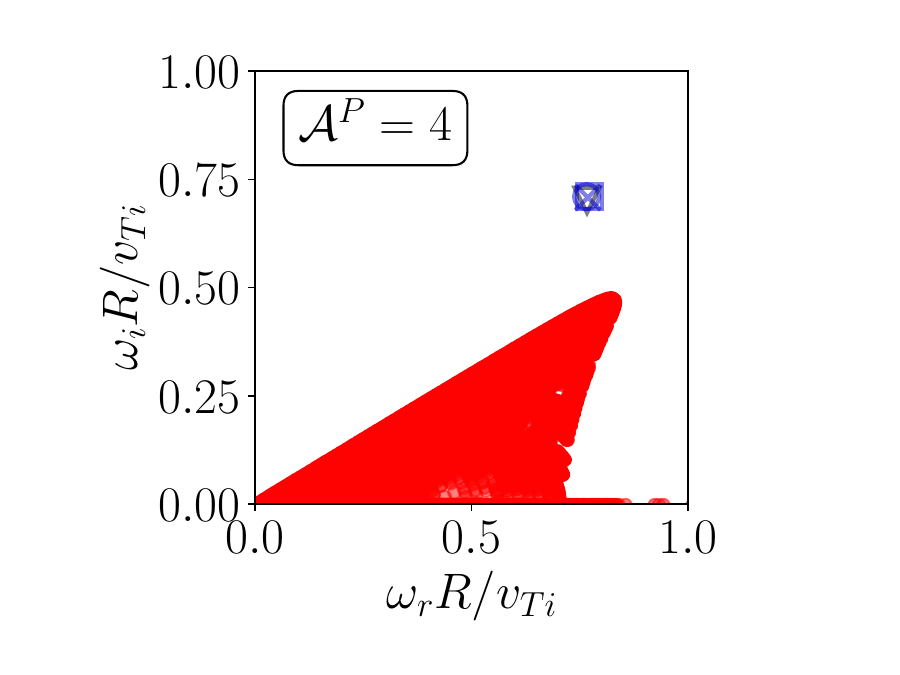}
    \end{subfigure}
   \caption{Real frequency and growth rate of ISMs for varying primary amplitudes $\mathcal{A}^P$ (left to right), with each red circle corresponding to the solution of \eqref{eq:D_ISM} at a point $(y,\theta)$ on the flux surface. The simulations correspond to the $k_x \rho_i = 0.2$ case in Figure~\ref{fig:validation_D_TSM_scan_kx}, i.e. the primary drive varies as $g_i^P \propto e^{-(\theta/\pi)^2}$ and has $\eta_\parallel^P=\eta_\perp^P =3$. The TSM solution obtained from \eqref{eq:D_TSM} is represented by the black triangle, while the mode frequencies from gyrokinetic simulations for a safety factor $q=20$ with phase shifts $\Theta=0$ and $\Theta=0.288572618$ are indicated by blue circles and crosses, respectively. The blue squares correspond to GK simulations with the parallel streaming and binormal magnetic drift set to zero.}
   \label{fig:forbidden_region}
\end{figure}

In Figure~\ref{fig:validation_D_TSM}, the toroidal secondary modes described by \eqref{eq:D_TSM} become subdominant to the local ISMs \eqref{eq:D_ISM} at small primary amplitudes $\mathcal{A}^P \lesssim 1$, at large wavenumbers $k_x \rho_i \gtrsim 1$ and at large phase shifts $\delta \sim \pi$. In these cases, the TSM complex frequency eventually enters the region of ISM complex frequencies (recall that $\mathcal{D}^\mathrm{ISM}=0$ gives a different complex frequency at each point on the flux surface because it depends on $y$ and $\theta$), as shown for a particular example in Figure~\ref{fig:forbidden_region}. For the TSM dispersion relation \eqref{eq:D_TSM} to remain analytic, the integration contours corresponding to the flux-surface average $\langle 1/\mathcal{D}^\mathrm{ISM} \rangle_\psi$ must be deformed. Similarly to the case of Landau damping, the integral then has both a principal value and a pole contribution. In practice, numerically deforming the integration contour in \eqref{eq:D_TSM} is cumbersome due to the presence of multiple poles and branch cuts. We therefore eschew such difficulties here, further motivated by the awareness that, as the TSM enters a region of ISMs, it necessarily becomes subdominant to a faster growing ISM (see also Figure~\ref{fig:forbidden_region}). We thus capture the TSMs only up to the point where they enter the ISM region, whence the solid lines in Figure~\ref{fig:validation_D_TSM_scan_kx} do not extend to the largest values of $k_x \rho_i$. Note that only the fastest growing ISMs are captured in Figure~\ref{fig:validation_D_TSM}, and the TSM will generally enter the ISM region at a smaller growth rate than the maximum ISM growth rate, see e.g. Figure~\ref{fig:forbidden_region}.

\subsection{ITG secondary modes (ISMs)} \label{sec:ISM}

We now consider the ISMs described by \eqref{eq:D_ISM}. While the conventional toroidal ITG primary modes are driven unstable by a combination of the binormal magnetic drift $(v_{My})$ \footnote{The radial magnetic drift being negligible for the usual ITG assumes a streamer primary mode, which is generally the most unstable, though there are notable exceptions \citep{parisi_toroidal_2020, parisi_three-dimensional_2022}.} and the background radial temperature gradient $(\partial_x T)$, the ISMs described by \eqref{eq:D_ISM} are driven unstable by the radial magnetic drift $(v_{Mx})$ and the primary binormal gradients $(\partial_y T_\parallel^P, \partial_y T_\perp^P)$. Indeed, the dispersion relation of a local toroidal ITG primary mode with binormal wavenumber $k_y$ is given by \eqref{eq:D_ISM} with the replacements $b_i \rightarrow k_y^2 \abs{\nabla y}^2 \rho_i^2 / 2$ and
\begin{equation}
    \mathcal{N} \rightarrow \frac{1}{n_i} \int\mathrm{d}^3 v\, J_{0i}^2 F_{Mi} \frac{k_y \tilde v_{My} - k_y \tilde v_{*i}}{\omega - k_y \tilde v_{My}}.
\end{equation}
Here, $\tilde v_{My}=\tilde{\bsy{v}}_M\cdot \nabla y$ is the binormal magnetic drift velocity and
\begin{equation}
    \tilde v_{*i} \equiv \frac{T_i}{Z_i e}\frac{\bhat \times \nabla x \cdot \nabla y}{B} \partial_x \ln n_i \left( 1 + \eta \left(\frac{v_\parallel^2 + v_\perp^2}{v_{Ti}^2} - \frac{3}{2} \right) \right) \equiv v_{*i} \left( 1 + \eta \left(\frac{v_\parallel^2 + v_\perp^2}{v_{Ti}^2} - \frac{3}{2} \right) \right)
\end{equation}
is the ion diamagnetic velocity, with $\eta = \partial_x \ln T_i / \partial_x \ln n_i$.

The instability criterion for primary ITG modes is generally expressed in terms of $\eta$ and $v_{*i} \eta / v_{My}$. Instability requires $\eta (\eta - 2/3) > 0$ and $v_{*i} \eta / v_{My}$ to be sufficiently positive \citep[see e.g.][]{biglari_toroidal_1989}. Considering the case $\eta_\parallel^P=\eta_\perp^P=\eta^P$ for simplicity, these instability criteria translate directly to the ISMs. Their instability thus requires $\eta^P (\eta^P - 2/3) > 0$ and sufficiently large $\tau v_{Ex}^P \eta / v_{Mx}$. The latter criterion is generally easily satisfied for any finite $\mathcal{A}^P$ at certain points on the flux surface, as there are always points with $v_{Mx} = 0$, e.g. at $\theta \in \{0, \pi\}$ in a circular cross-section tokamak for which $v_{Mx} \propto \sin \theta$.

In contrast, the TSM's instability threshold is at a finite $\mathcal{A}^P$, as discussed in Section~\ref{sec:TSM_R}. Therefore, the theory predicts a region of unstable ISMs and stable TSMs at weak primary drive. At sufficiently large primary drive and moderate radial wavelengths $k_x \rho_i \lesssim 1$, the TSM grows faster than the ISMs, see Figure~\ref{fig:validation_D_TSM_scan_kx}. This will be explained further in Section~\ref{sec:TSM_NR}, where the non-resonant limit of the dispersion relations \eqref{eq:D_ISM} and \eqref{eq:D_TSM} is considered. Even at large primary drive, the ISM generally remains the dominant instability at short wavelengths $k_x \rho_i \gtrsim 1$. While a full theoretical analysis is beyond the scope of this study, this result may be understood qualitatively by noting that ITG primary modes typically remain unstable even at wavelengths shorter than the ion gyroradius \citep{smolyakov_short_2002, gao_short_2005}, while the RDK secondary mode (towards which the TSM tends at large primary amplitude, see Section~\ref{sec:TSM_NR}) is generally stabilised at short wavelengths by FLR effects, see \ref{sec:RDK_SW}. Therefore, for large $\mathcal{A}^P$, the ISM should be the dominant instability at short radial wavelengths; however, its growth rate will typically be smaller than that of the RDK secondary mode at long radial wavelengths, see e.g. Figure~\ref{fig:summary_regions_GK}.

\subsection{Toroidal secondary modes (TSMs)}\label{sec:TSM_R}

\begin{figure}
    \centering
    \begin{subfigure}[t]{0.49\columnwidth}
            \centering
            \includegraphics[width=\textwidth, trim={0.5cm 0.5cm 0.4cm 0.3cm}, clip]{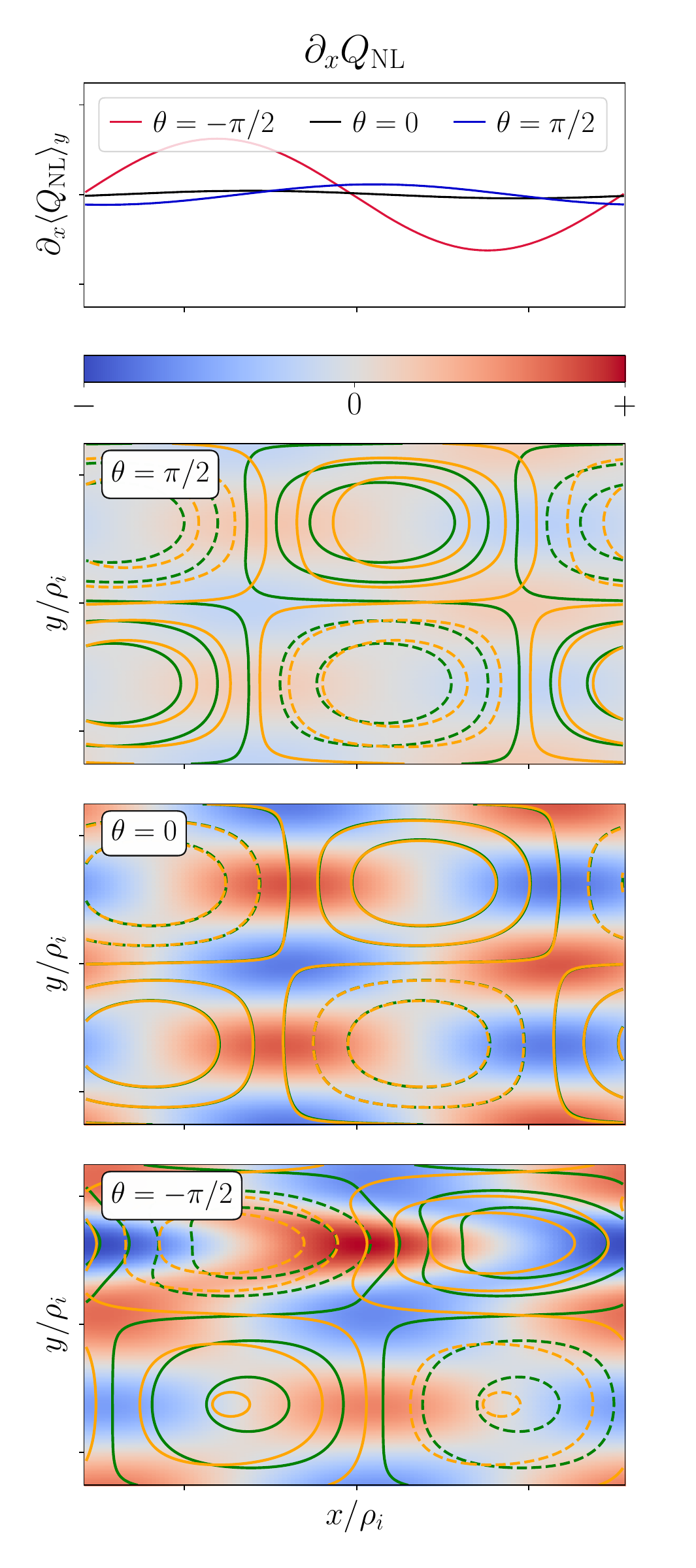}
    \caption{$\mathcal{A}^P=4$}
    \end{subfigure}
    \begin{subfigure}[t]{0.49\columnwidth}
            \centering
            \includegraphics[width=\textwidth, trim={0.5cm 0.5cm 0.4cm 0.3cm}, clip]{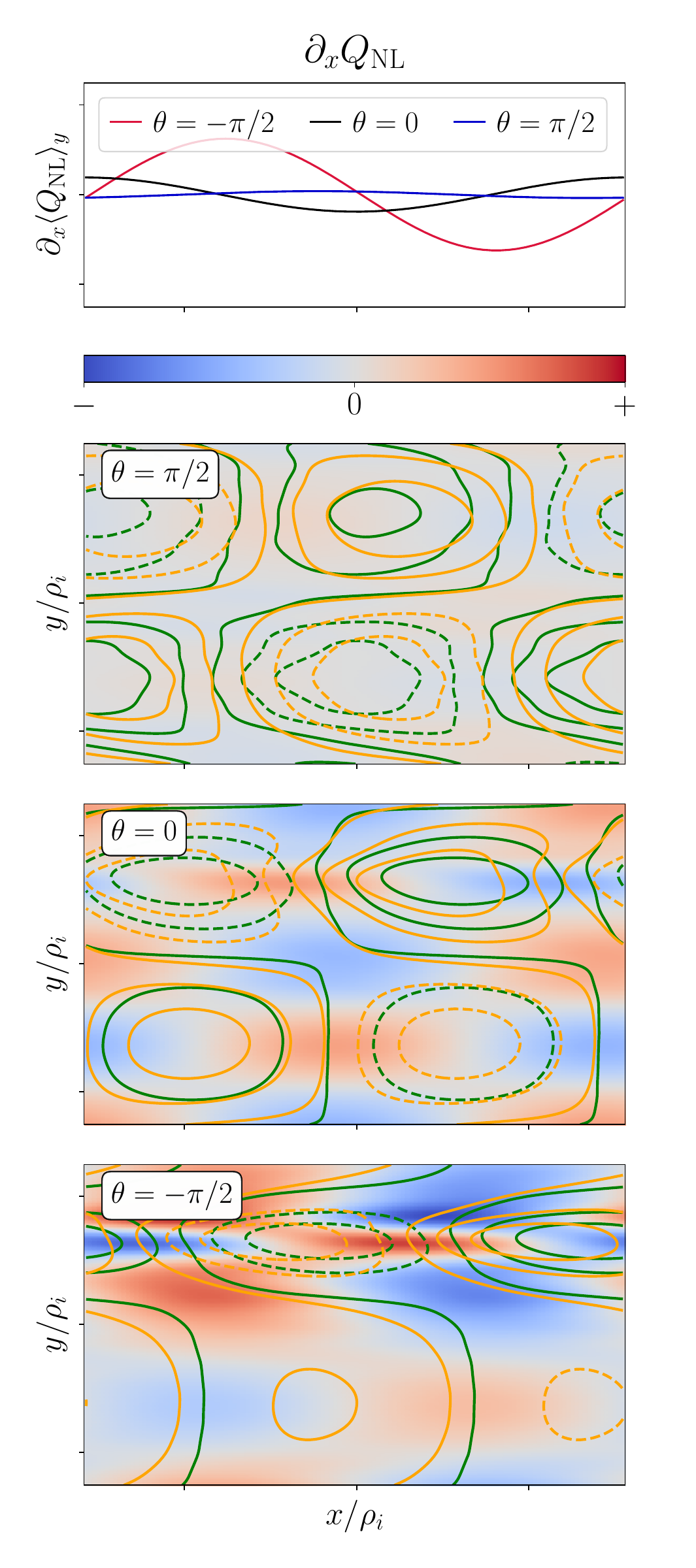}
        \caption{$\mathcal{A}^P=1$}
    \end{subfigure}
    \caption{Spatial variation of the nonlinear contribution $\partial_x Q_\mathrm{NL}$ \eqref{eq:QSW_NL} to the SW force evolution \eqref{eq:SW-force_time_der} in the toroidal secondary mode simulations presented in Figure~\ref{fig:validation_D_TSM_scan_kx} for $\mathcal{A}^P=4$ (left column) and $\mathcal{A}^P=1$ (right column) at $k_x \rho_i = 0.1$. The $y$-integrated value of $\partial_x Q_\mathrm{NL}$ is shown in the top row as a function of $x$ for three values of $\theta$, $\theta \in \{\pi/2, 0, -\pi/2\}$ (above, on, and below the tokamak midplane, respectively). The second, third, and fourth rows show the variation of $\partial_x Q_\mathrm{NL}$ in $(x,y)$ for these three $\theta$ values, as well as contours of the nonzonal pressure fluctuations $P^\mathrm{NZ}$ (orange contours) and electrostatic potential fluctuations $\varphi^\mathrm{NZ}$ (green contours), with negative values indicated by dashed lines. For $\mathcal{A}^P=4$, the $P^\mathrm{NZ}$ and $\varphi^\mathrm{NZ}$ contours overlap at $\theta=0$, and have phase shifts of opposite signs at $\theta=\pm \pi/2$ due to the radial magnetic drift advection. Moreover, the outward-propagating ($\omega/k_x > 0$) TSM shown here is strongly localised to $\theta=-\pi/2$. As a result, $\langle \partial_x Q_\mathrm{NL}\rangle_y$ has a large up-down asymmetry, see first row. In the $\mathcal{A}^P=1$ case, $\partial_x Q_\mathrm{NL}$ exhibits contributions from higher $y$-harmonics, but its $y$-averaged value still has a similar up-down asymmetry.}
    \label{fig:TSM_phase}
\end{figure}

\begin{figure}
    \centering
    \begin{subfigure}[t]{0.99\columnwidth}
            \centering
            \includegraphics[width=\textwidth, trim={6cm 2.2cm 0cm 0.6cm}, clip]{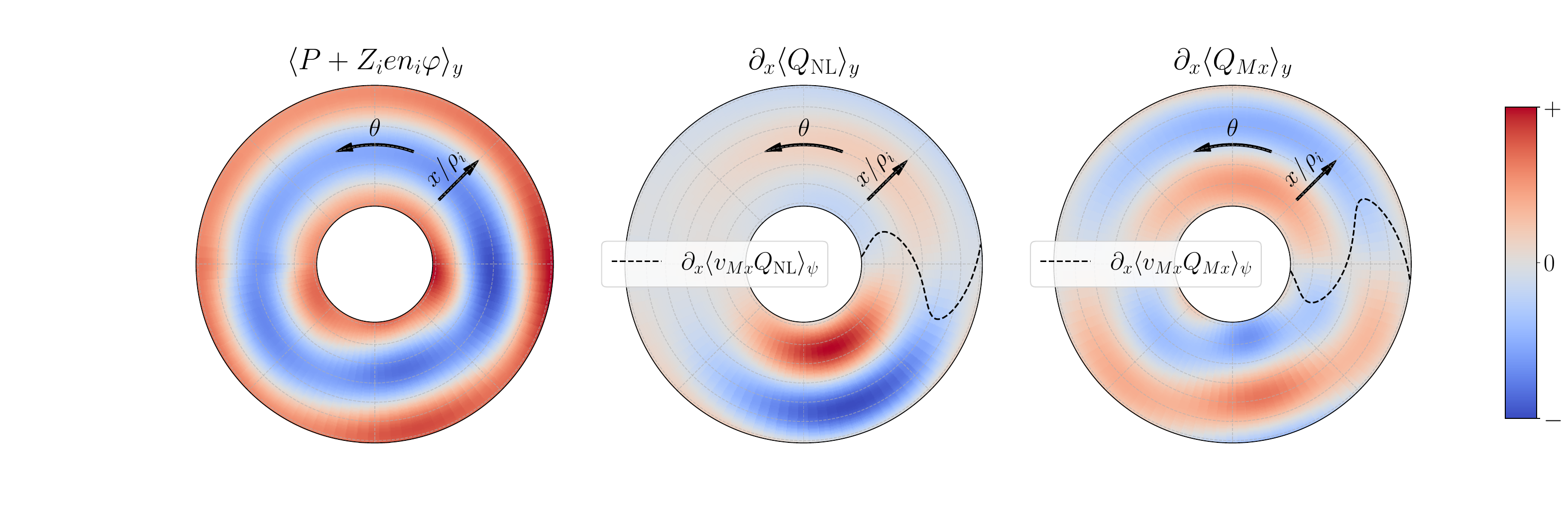}
    \caption{$\mathcal{A}^P=4$}
    \end{subfigure}
    \begin{subfigure}[t]{0.99\columnwidth}
            \centering
            \includegraphics[width=\textwidth, trim={6cm 2.2cm 0cm 0.6cm}, clip]{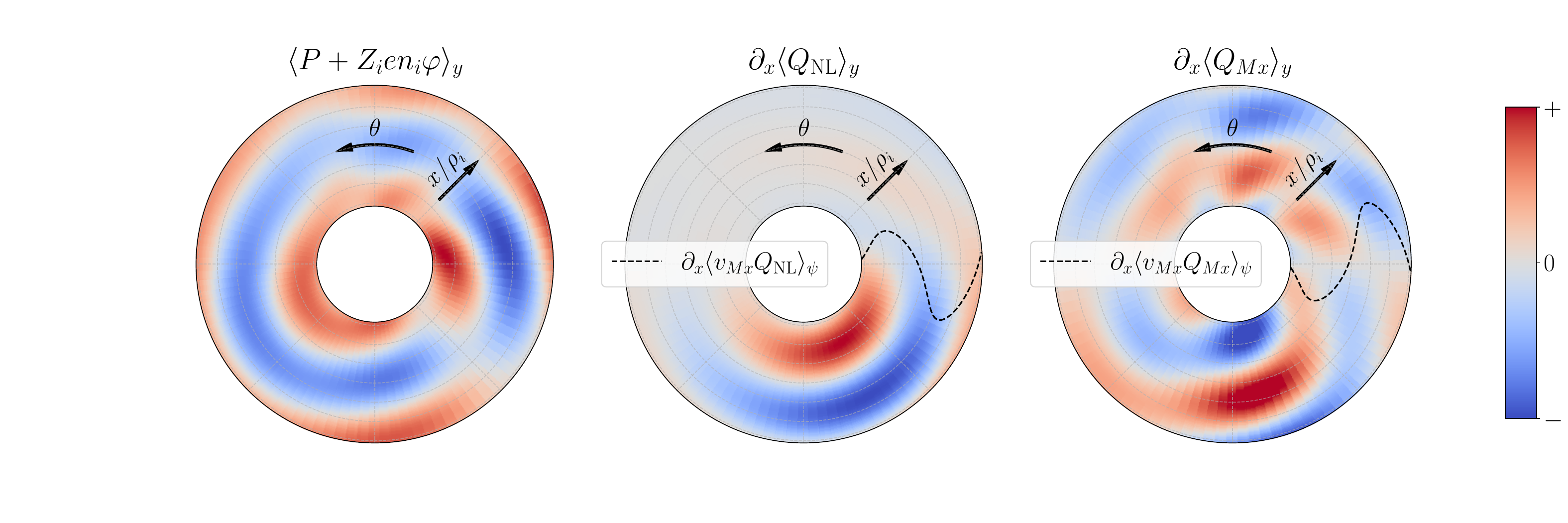}
        \caption{$\mathcal{A}^P=1$}
    \end{subfigure}
    \caption{Contour plots in $x$ and $\theta$ of integrand in SW force (left plots), whose time evolution \eqref{eq:SW-force_time_der} results from the nonlinear term \eqref{eq:QSW_NL} (middle plots) and the magnetic drift term \eqref{eq:QSW_Mx} (right plots). The contours correspond to the simulations presented in Figure~\ref{fig:validation_D_TSM_scan_kx} for $k_x \rho_i = 0.1$, $\mathcal{A}^P=4$ (top row) and $\mathcal{A}^P=1$ (bottom row). The contributions to the time evolution of the SW force on each flux surface, $\partial_x \langle v_{Mx} Q_\mathrm{NL} \rangle_\psi (x)$ and $\partial_x \langle v_{Mx} Q_{Mx} \rangle_\psi (x)$, are indicated by black dashed lines on the outboard midplane in the middle and right plots, respectively. These two contributions cancel out, such that $\langle P+Z_i e n_i \varphi \rangle_y$ has no up-down asymmetric component (left plots), i.e. $\langle v_{Mx} (P+Z_i e n_i \varphi) \rangle_\psi \approx 0$.}
    \label{fig:TSM_mechanism}
\end{figure}

We proceed to discuss the physics of the toroidal secondary modes in more detail. We first investigate whether the basic physical picture of the TSM put forward in \cite{nies_saturation_2024} is consistent with the GK simulations. The main argument is reproduced here, considering for simplicity a tokamak with circular flux surfaces, where the radial magnetic drift velocity $\tilde v_{Mx} \propto \sin\theta$ is up-down asymmetric. It will prove convenient to define the pressure-like quantity
\begin{equation} \label{eq:pressure_moment}
    P = T_i \int\mathrm{d}^3 v\,  \frac{v_\parallel^2 + v_\perp^2/2}{v_{Ti}^2} g_i.
\end{equation}
Note the factor of two between the parallel and perpendicular velocity contributions, which reflects the velocity dependence of the radial magnetic drift \eqref{eq:v_Mx}. The TSM relies on the Stringer-Winsor (SW) force in the vorticity equation \eqref{eq:vorticity}, which arises due to up-down asymmetry in $\langle P + Z_i e n_i \varphi \rangle_y$. By multiplying the GK equation \eqref{eq:gyrokinetic_eq_realspace} by $\tilde v_{Mx}$, integrating in velocity space, and flux-surface averaging, we find that, at long perpendicular wavelengths, the pressure contribution to the SW force evolves in time as
\begin{equation}\label{eq:SW-force_time_der_P}
    \partial_t \langle v_{Mx} P \rangle_\psi = T_i  \left\langle \int\mathrm{d}^3 v\, \tilde{v}_{Mx} \, \partial_t g_i \right\rangle_\psi  = - \partial_x  \left\langle v_{Mx} \left( Q_\mathrm{NL} + Q_{Mx,P} \right) \right\rangle_\psi.
\end{equation}
Here, the contribution from the nonlinear heat flux is given by
\begin{equation} \label{eq:QSW_NL}
    Q_\mathrm{NL} =  T_i v_{Ex} \int\mathrm{d}^3 v\,  \frac{v_\parallel^2 + v_\perp^2/2}{v_{Ti}^2}  g_i,
\end{equation}
and the contribution from the radial magnetic drift is
\begin{equation}
    \label{eq:QSW_P_Mx}
    Q_{Mx,P} = v_{Mx} \left( \frac{7}{4} n_i Z_i e \varphi + T_i \int\mathrm{d}^3 v\,  \left(\frac{v_\parallel^2 + v_\perp^2/2}{v_{Ti}^2}\right)^2   g_i \right) .
\end{equation}
To obtain \eqref{eq:SW-force_time_der_P}, we have neglected the contributions from binormal derivatives and parallel streaming, as befits the assumptions made in deriving the dispersion relation of toroidal secondary modes \eqref{eq:D_ISM}. We have also neglected FLR effects to simplify the discussion. The derived time evolution equation does not include the effect of the $\theta$-varying potential on the SW force. To calculate this piece, we evaluate the time derivative of the quasineutrality equation \eqref{eq:quasineutrality}, which after multiplication by $v_{Mx}$ and flux-surface averaging gives
\begin{equation}\label{eq:SW-force_time_der_phi}
    \partial_t \langle v_{Mx} Z_i e n_i \varphi \rangle_\psi = \frac{T_i}{\tau} \left\langle v_{Mx} \int\mathrm{d}^3 v\, \partial_t g_i \right\rangle_\psi  = - \partial_x  \left\langle v_{Mx}\,Q_{Mx,\varphi}  \right\rangle_\psi,
\end{equation}
with
\begin{equation}\label{eq:QSW_phi_Mx}
    Q_{Mx,\varphi} = \frac{v_{Mx}}{\tau} \left( n_i Z_i e \varphi + T_i \int\mathrm{d}^3v\, \frac{v_\parallel^2 + v_\perp^2/2}{v_{Ti}^2} g_i  \right).
\end{equation}
In deriving \eqref{eq:SW-force_time_der_phi}, we have again neglected the effects of parallel streaming and FLR effects. Combining \eqref{eq:SW-force_time_der_P} and \eqref{eq:SW-force_time_der_phi}, the time derivative of the SW force is derived to be
\begin{equation}\label{eq:SW-force_time_der}
    \partial_t \langle v_{Mx} \left( P + Z_i e n_i \varphi \right) \rangle_\psi = -\partial_x  \left\langle v_{Mx} \left( Q_\mathrm{NL} + Q_{Mx} \right) \right\rangle_\psi,
\end{equation}
with $Q_\mathrm{NL}$ given by \eqref{eq:QSW_NL} and $Q_{Mx}$ by
\begin{equation}\label{eq:QSW_Mx}
    Q_{Mx} = Q_{Mx, P} + Q_{Mx, \varphi}.
\end{equation}

The physical mechanism behind the TSM may be understood qualitatively as follows. The zonal flow shears the (primary) background turbulence to small radial scales. The radial magnetic drift's velocity dependence ($\tilde v_{Mx} \propto v_\parallel^2 + v_\perp^2/2$) then causes a phase shift between the sheared potential and pressure fluctuations, as they are advected at different rates in the $x$-direction. This phase shift leads to an up-down asymmetric $\langle Q_\mathrm{NL} \rangle_y \propto \langle P \partial_y \varphi \rangle_y$ (as the phase shift is caused by $\tilde v_{Mx} \propto \sin \theta$) and thence a contribution to the SW force. The up-down asymmetric phase shift between potential and pressure fluctuations in GK simulations of the TSM is shown in Figure~\ref{fig:TSM_phase}, alongside the resulting $Q_\mathrm{NL}$.

If the ZF inertia in \eqref{eq:vorticity} is negligible (as is the case at long radial wavelengths $k_x \rho_i \lesssim 0.5$), the ZF amplitude quickly adjusts to ensure the $Q_{Mx}$ contribution to the SW force evolution \eqref{eq:SW-force_time_der} cancels that from $Q_\mathrm{NL}$. The cancellation between $Q_\mathrm{NL}$ and $Q_{Mx}$ is indeed observed in GK simulations of the toroidal secondary mode at long radial wavelengths, as shown in Figure~\ref{fig:TSM_mechanism}. 

The cancellation is predicated on the smallness of the inertia and nonlinear stress contributions to the vorticity equation \eqref{eq:vorticity}, which need not be small for $k_x \rho_i \sim 0.5$, the scale at which the toroidal secondary mode amplitude peaks in turbulence simulations, see Figure~\ref{fig:ZF_real_Fourier}. The Stringer-Winsor force can then synergise or compete with the ZF inertia and the nonlinear stresses -- such effects are captured by the dispersion relation \eqref{eq:D_TSM} and will be discussed below when considering its non-resonant limit \eqref{eq:D_TSM_NR}.

The TSM propagates radially with a speed approximately equal to the magnetic drift velocity $\omega/k_x \sim v_{Mx}(\theta = \pm \pi/2)$ due to the role of the radial magnetic drift in generating the up-down asymmetric $Q_\mathrm{NL}$. The TSM is thus subject to kinetic damping by the magnetic drifts and requires a sufficiently large primary drive to become unstable, typically $\mathcal{A}^P \gtrsim 1$, depending on other parameters such as $\eta_{\parallel}^P$ and $\eta_\perp^P$.

The outward-propagating TSMs ($\omega / k_x > 0$) considered in Figures~\ref{fig:TSM_phase} and \ref{fig:TSM_mechanism} have $\partial_x  \langle Q_\mathrm{NL} \rangle_y$ strongly localised below the midplane ($\theta = -\pi/2$). This localisation follows from the denominator in \eqref{eq:gS_TSM}, with particle resonances being dominant when the signs of the real frequency $\omega_r$ and of the magnetic drift frequency $k_x v_{Mx}$ are identical. For $\mathcal{A}^P=4$, the radial magnetic drift contribution $\partial_x \langle Q_{Mx} \rangle_y$ dominantly varies as $\sin \theta$, which follows from the first terms in \eqref{eq:QSW_P_Mx} and \eqref{eq:QSW_phi_Mx} being dominant, and $\langle \varphi \rangle_y$ being approximately constant in $\theta$ in this case (see also the non-resonant limit in Section~\ref{sec:TSM_NR}). For $\mathcal{A}^P=1$, $\partial_x \langle Q_{Mx} \rangle_y$ still has a dominant $\sin\theta$-variation but also exhibits higher $\theta$-harmonics.

Finally, we note the physical mechanism presented above does not strongly depend on the phase shift between the primary potential and pressure perturbations. This is confirmed by the simulations in Figure~\ref{fig:validation_D_TSM_scan_phase}. While this phase shift strongly affects the total ($x$-integrated) nonlinear heat flux $Q_\mathrm{NL}$, it does not contribute to the SW force which requires a radial modulation of $\partial_x Q_\mathrm{NL}$.

\subsection{Non-resonant limit of secondary modes}\label{sec:TSM_NR}

The following study of the non-resonant limit is primarily motivated by the aspiration to identify the various modes described by \eqref{eq:D_TSM}. If the primary amplitude is sufficiently large, the effect of the radial magnetic drift should become subdominant and we expect to recover the purely growing RDK secondary mode \citep{rogers_generation_2000}. Moreover, when the primary amplitude vanishes, we expect to recover GAMs from \eqref{eq:D_TSM}. The non-resonant limit of \eqref{eq:D_TSM} will help understand under which conditions the various modes described by the dispersion relation become dominant. Furthermore, by also considering the non-resonant limit of the ISM dispersion relation, we will be able to explain why the ISM becomes subdominant to the TSM for sufficiently large $\mathcal{A}^P$ (see Figure~\ref{fig:validation_D_TSM_scan_kx}).

We consider the subsidiary ordering
\begin{equation}\label{eq:NR_ordering}
    b_i^{1/2} \sim \frac{ k_x \tilde{v}_{Mx}}{\omega} \ll \frac{k_x \tilde v_g^P}{\omega} \sim \frac{k_x v_{Ex}^P}{\omega} \sim \tau \sim 1,
\end{equation}
corresponding to a regime where the secondary mode is strongly driven and the FLR effects are weak. For convenience, we define some velocities associated with various moments of the primary distribution $g_i^P$, first that corresponding to the primary diamagnetic momentum
\begin{equation} \label{eq:vdia_P}
    v_\mathrm{dia}^P = \frac{1}{n_i}\int\mathrm{d}^3 v\, F_{Mi} \tilde{v}_g^P \frac{v_\perp^2}{v_{Ti}^2},
\end{equation}
and secondly those moments associated with powers of the magnetic drift velocity $\tilde{v}_{Mx}$ \eqref{eq:v_Mx}
\begin{equation}
    v_{j}^P = \frac{1}{n_i}\int\mathrm{d}^3 v\, F_{Mi} \tilde{v}_g^P \left( \frac{v_\parallel^2 + v_\perp^2/2}{v_{Ti}^2} \right)^{j},
\end{equation}
with $v_0^P=\tau v_{Ex}^P$ due to quasineutrality \eqref{eq:quasineutrality} in the assumed limit of large binormal scales $k_y^P \rho_i \ll 1$. 

The non-resonant limit of the generalised secondary mode dispersion relation \eqref{eq:D_TSM} corresponding to the ordering \eqref{eq:NR_ordering} may be written as

\begin{equation} \label{eq:D_TSM_NR}
    0 = \mathcal{D}^\mathrm{NR} = \frac{1}{\tau}\left\langle b_i + b_i k_x^2 \frac{v_{Ex}^P\left( v_{Ex}^P+v_\mathrm{dia}^P \right)}{\omega^2} - \left(\frac{k_x v_{Mx}}{\omega}\right)^2\left[ \frac{7}{4}+\frac{1}{\tau} + \frac{(k_x v_1^P)^2/(\tau\omega)-k_x v_2^P}{\omega-k_x v_{Ex}^P}\right] \right\rangle_\psi.
\end{equation}
This result may be obtained by expanding \eqref{eq:D_TSM}. First, we exploit the smallness of the radial magnetic drift to expand the resonant denominator in \eqref{eq:curlyN}. Then, also using $b_i \ll 1$, the denominator in the flux-surface average \eqref{eq:D_TSM} is expanded. Finally, many terms vanish due to $\langle v_{Ex}^P \rangle_y =0$ and $\langle v_\mathrm{dia}^P \rangle_y =0$, leading to \eqref{eq:D_TSM_NR}. An alternative derivation of \eqref{eq:D_TSM_NR} from the vorticity equation \eqref{eq:vorticity} may be found in \ref{sec:strongly_driven}, where the growth of secondary modes in the limit of strong drive and long perpendicular wavelengths is considered more generally.

The contributions to \eqref{eq:D_TSM_NR} may directly be associated to terms in the vorticity equation \eqref{eq:vorticity}: the first two terms originate from the ZF inertia and the nonlinear stresses, which is composed of the Reynolds stress (the $(v_{Ex}^P)^2$ term) and the diamagnetic stress (the $v_{Ex}^P v_\mathrm{dia}^P$ term). The $v_{Mx}$ contribution in \eqref{eq:D_TSM_NR} stems from the SW force, whose time evolution \eqref{eq:SW-force_time_der} has contributions from $Q_\mathrm{NL}$ and $Q_{Mx}$, given by \eqref{eq:QSW_NL} and \eqref{eq:QSW_Mx}, respectively. With the ordering \eqref{eq:NR_ordering}, only the first terms in \eqref{eq:QSW_P_Mx} and \eqref{eq:QSW_phi_Mx} contribute to $Q_{Mx}$, giving the linear term ($\propto 7/4+1/\tau$)  in \eqref{eq:D_TSM_NR}. Finally, up-down asymmetries of $v_{Ex}$ and of $g_i$ in \eqref{eq:QSW_NL} cause the $v_1^P$ and $v_2^P$ terms in \eqref{eq:D_TSM_NR}, respectively.

The non-resonant dispersion relation \eqref{eq:D_TSM_NR} lays bare the different modes contained in the more general dispersion relation \eqref{eq:D_TSM}. First, in the limit of vanishing primary drive, the balance of ZF inertia and the linear contribution to the SW force in \eqref{eq:D_TSM_NR} recovers the GAM \citep{winsor_geodesic_1968, conway_geodesic_2021} dispersion relation
\begin{equation} \label{eq:omega_GAM}
    \omega_\mathrm{GAM}^2 = \left(\frac{7}{4}+\frac{1}{\tau}\right) \frac{\langle (k_x v_{Mx})^2 \rangle_\psi}{\langle b_i \rangle_\psi}.
\end{equation}
Second, in the limit of vanishing radial magnetic drift, the RDK dispersion relation \citep{rogers_generation_2000, plunk_nonlinear_2017} is recovered from the balance of ZF inertia and nonlinear stresses,
\begin{equation}
    \omega_\mathrm{RDK}^2 = -k_x^2 \frac{\left\langle b_i \,v_{Ex}^P \left(v_{Ex}^P+v_\mathrm{dia}^P\right) \right\rangle_\psi}{\left\langle b_i \right\rangle_\psi}, \label{eq:omegaRDK}
\end{equation}
giving a purely growing mode when the right-hand-side is negative. We may now deduce from \eqref{eq:D_TSM_NR} that the RDK secondary mode is recovered strictly only in the limit $\omega_\mathrm{RDK}^2 \sim (k_x v_{Ex}^P)^2 \gg (k_x v_{Mx})^2/ b_i \sim \omega_\mathrm{GAM}^2$. In practice, the saturated turbulence amplitude does not reach such large values, so that the effect of the radial magnetic drift on the secondary modes must be considered. Indeed, the toroidal secondary modes are clearly observed in GK simulations of ITG turbulence, see Figure~\ref{fig:ZF_real_Fourier}. 

Finally, the non-resonant limit of the toroidal secondary mode is contained in \eqref{eq:D_TSM_NR} through the nonlinear contributions to the SW force from the $v_{1}^P$ and $v_2^P$ terms. To evaluate the non-resonant limit of the TSM, we consider a simplified primary drive that is described by a single Fourier mode in $y$ and is independent of $\theta$, such that $v_{Ex}^P = v_{Ex}^{P0} \sin(k_y^P y)$, $v_\mathrm{dia}^P = v_\mathrm{dia}^{P0} \sin(k_y^P y)$, $v_{1}^P = v_{1}^{P0} \sin(k_y^P y)$, and $v_{2}^P = v_{2}^{P0} \sin(k_y^P y)$. We thus exclude here for simplicity any variation of the primary drive along the magnetic field and the possibility of a phase shift between various moments of the primary distribution. The non-resonant dispersion relation \eqref{eq:D_TSM_NR} simplifies to
\begin{equation} \label{eq:D_TSM_NR_primary_Fourier}
    0 = 1 + \left(\frac{k_x v_{Ex}^{P0}}{\omega}\right)^2 C_\mathrm{RDK} - \frac{\omega_\mathrm{GAM}^2}{\omega^2} \left[1 - C_\mathrm{TSM} \left( 1 - \frac{1}{\sqrt{1-\left(k_x v_{Ex}^{P0} / \omega \right)^2}}  \right) \right],
\end{equation}
where we used the integral \eqref{eq:integral_sqrt}. Here, the square root is evaluated with the branch cut along negative reals, such that Re$\sqrt{x}>0$. We have defined
\begin{align} 
    C_\mathrm{RDK} &= \frac{1}{2}\left( 1 + \frac{v_\mathrm{dia}^{P0}}{v_{Ex}^{P0}} \right), \label{eq:def_CRDK}\\
    C_\mathrm{TSM} &= \left(\frac{7}{4} + \frac{1}{\tau} \right)^{-1} \left( \frac{1}{\tau}\left(\frac{v_1^{P0}}{v_{Ex}^{P0}}\right)^2 - \frac{v_2^{P0}}{v_{Ex}^{P0}}\right). \label{eq:def_CTSM}
\end{align}

We may rewrite \eqref{eq:D_TSM_NR_primary_Fourier} as an equation for $k_x v_{Ex}^{P0}/\omega$,
\begin{equation} \label{eq:D_TSM_NR_primary_Fourier_abstar}
    \frac{1}{\sqrt{1-\left(k_x v_{Ex}^{P0} / \omega \right)^2}} = \frac{C_1}{\left(k_x v_{Ex}^{P0} / \omega \right)^2} + \frac{C_2}{C_1},
\end{equation}
with real parameters 
\begin{equation} \label{eq:defNR_a}
    C_1 = \frac{1}{C_\mathrm{TSM}} \left( \frac{k_x v_{Ex}^{P0}}{\omega_\mathrm{GAM}} \right)^2
\end{equation}
and
\begin{equation} \label{eq:defNR_bstar}
   C_2 = \frac{1}{C_\mathrm{TSM}^2} \left( \frac{k_x v_{Ex}^{P0}}{\omega_\mathrm{GAM}} \right)^2 \left[ C_\mathrm{RDK} \left( \frac{k_x v_{Ex}^{P0}}{\omega_\mathrm{GAM}} \right)^2 + C_\mathrm{TSM} -1  \right].
\end{equation}

\begin{figure}[!ht]
    \centering
    \begin{subfigure}[t]{0.99\columnwidth}
            \centering
            \includegraphics[width=\textwidth, trim={0.4cm 0.8cm 0.4cm 0.4cm}, clip]{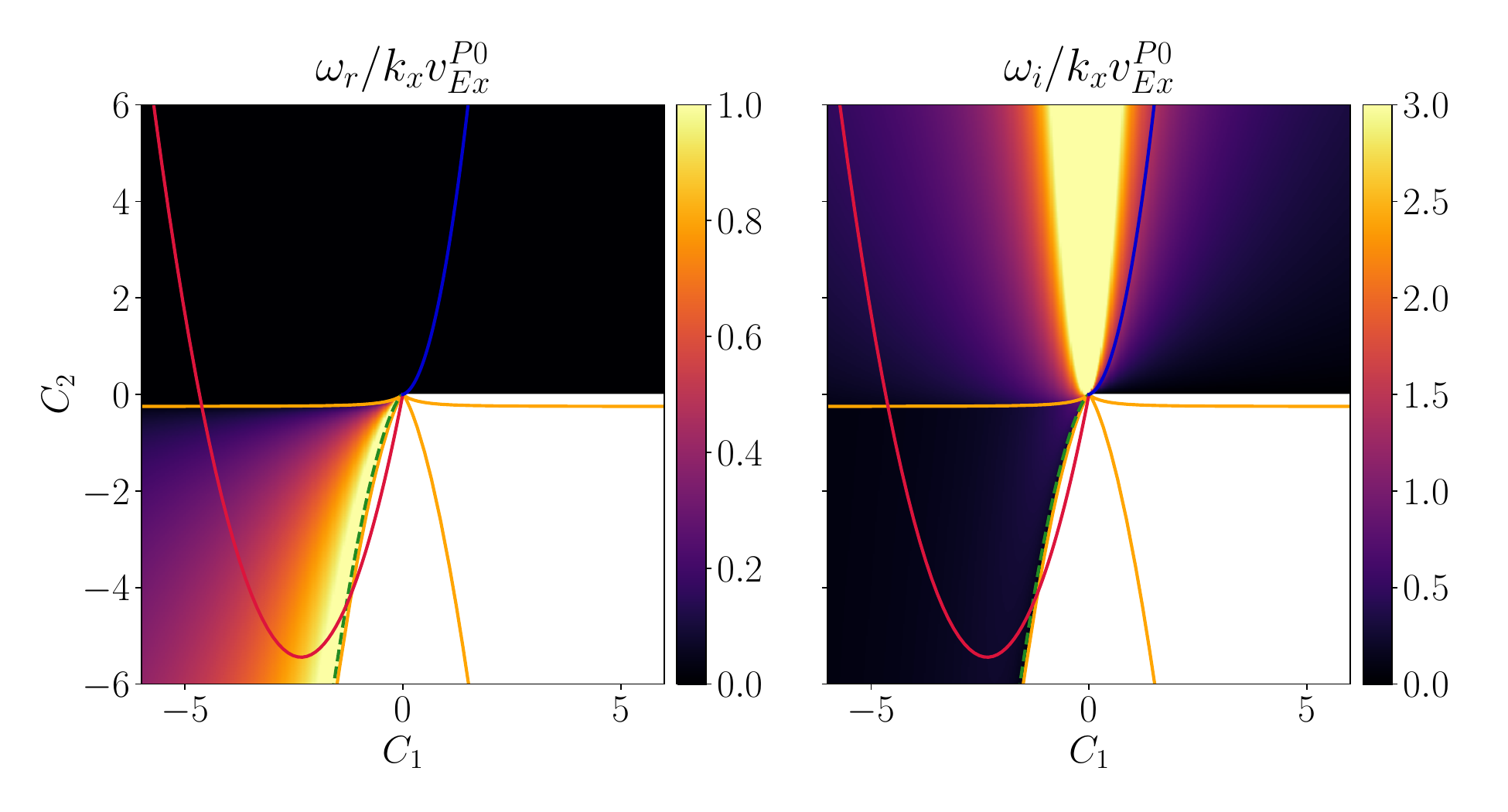}
        \caption{Real frequency (left) and growth rate (right) of fastest growing mode of the non-resonant secondary mode dispersion relation for a single Fourier mode primary drive \eqref{eq:D_TSM_NR_primary_Fourier_abstar}. The blank region has no unstable modes. The orange lines correspond to the zeroes of the discriminant \eqref{eq:discriminant_NR}. The green-dashed line is the approximation \eqref{eq:NR_stab_bdy_approx} to the curve bounding the stable region. The red and blue curves correspond to the $C_1$ and $C_2$ values in the primary amplitude scan of (b).}
        \label{fig:non_resonant_stab_bdy}
    \end{subfigure}
    \begin{subfigure}[t]{0.99\columnwidth}
            \centering
            \includegraphics[width=\textwidth, trim={0cm 0cm 0cm 0cm}, clip]{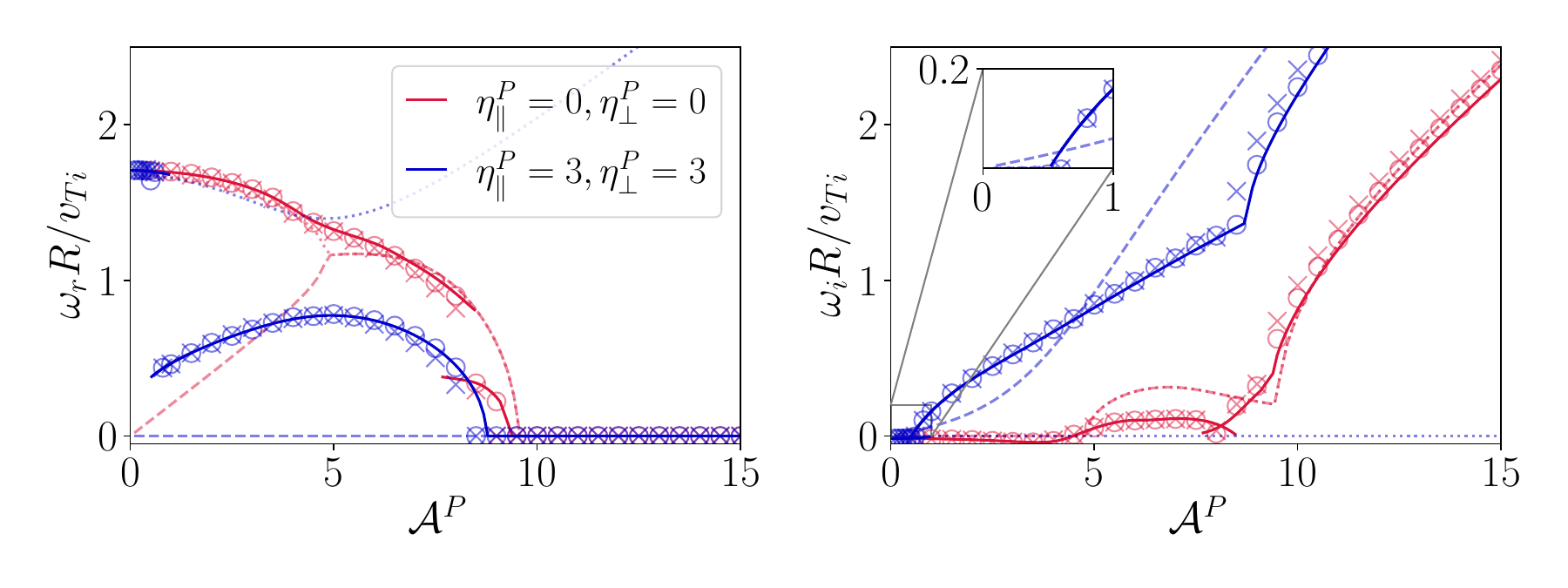}
        \caption{Real frequency (left) and growth rate (right) for $k_x \rho_i = 0.2$ as a function of the primary amplitude $\mathcal{A}^P$ obtained from the generalised secondary mode dispersion relation \eqref{eq:D_TSM} (solid lines), gyrokinetic simulations of the secondary modes (markers), and from non-resonant limit of the dispersion relation \eqref{eq:D_TSM_NR_primary_Fourier} (dashed and dotted lines). The primary drive is here taken to be constant in $\theta$ and two sets of $\eta_{\parallel,\perp}^P$ values are considered (red and blue colors). The gyrokinetic simulations of the secondary modes have a safety factor value $q=20$ and employ the phase-shift-periodic boundary condition \eqref{eq:BC_phase-shift-periodic} with $\Theta = 0$ (circles) and $\Theta = 0.28857261$ (crosses).}
        \label{fig:non_resonant_omega_AP}
    \end{subfigure}
    \caption{Solutions to the non-resonant secondary mode dispersion relation \eqref{eq:D_TSM_NR_primary_Fourier_abstar} (top) and comparison of these solutions with the solutions obtained from the general secondary mode theory and from gyrokinetic simulations (bottom).}
    \label{fig:non_resonant}
\end{figure}

The solutions to \eqref{eq:D_TSM_NR_primary_Fourier_abstar} are shown in Figure~\ref{fig:non_resonant_stab_bdy}. The stable region is bounded by two curves, the first is given by the line $\{C_2 = 0, C_1 \geq 0\}$ and the second is determined as a root of
\begin{equation}\label{eq:discriminant_NR}
    0 = \Delta = 4 C_2 (C_1^2 + C_2)^3 + C_1^2 (C_1^4 + 4 C_1^2 + 20 C_2 C_1^2 - 8 C_2^2).
\end{equation}
Here, $\Delta$ is the discriminant of the cubic equation for $\omega^2$ obtained by squaring and rearranging \eqref{eq:D_TSM_NR_primary_Fourier_abstar}. An approximate solution to the nontrivial curve bounding the stable region is given by
\begin{equation} \label{eq:NR_stab_bdy_approx}
   C_2 \approx C_1 \left( -C_1 + 3 \left(-\frac{C_1}{4}\right)^{1/3} \right).
\end{equation}
This expression is strictly valid for $-C_1 \gg 1$ but is shown in Figure~\ref{fig:non_resonant_stab_bdy} to be a good approximation for all $C_1$.

As shown in Figure~\ref{fig:non_resonant_stab_bdy}, the non-resonant secondary mode dispersion \eqref{eq:D_TSM_NR} describes both purely growing instabilities and overstable modes with $\omega_r \neq 0$ and $\omega_i >0$. The former are found approximately in the region $C_2 > 0$, which by \eqref{eq:defNR_bstar} is equivalent to $C_\mathrm{RDK}>0$ for large primary amplitudes $k_x v_{Ex}^{P0} \gg \omega_\mathrm{GAM}$ and to $C_\mathrm{TSM} > 1$ for small primary amplitudes. These correspond to the RDK secondary mode and non-resonant TSM, respectively. A necessary condition for overstability is $C_1<0$, which is equivalent by \eqref{eq:defNR_a} to $C_\mathrm{TSM} < 0$, which may be interpreted as the nonlinearly induced up-down pressure asymmetry \eqref{eq:QSW_NL} reinforcing that from the radial magnetic drift \eqref{eq:QSW_Mx}.

The non-resonant secondary mode solutions from \eqref{eq:D_TSM_NR_primary_Fourier} are shown in Figure~\ref{fig:non_resonant_omega_AP} to agree with the solutions of the resonant dispersion relation \eqref{eq:D_TSM} and with GK simulation results at large primary amplitudes $\mathcal{A}^P\gg 1$, while both qualitative and quantitative deviations are generally observed for $\mathcal{A}^P \sim 1$. When the primary mode is taken to be a pure density perturbation ($\eta_\parallel^P=\eta_\perp^P=0$), the non-resonant limit describes qualitatively how the initially stable GAM first becomes overstable at $\mathcal{A}^P\approx 5$, and then becomes a purely growing mode at $\mathcal{A}^P \approx 10$, see also trajectory in $(C_1, C_2)$-space in Figure~\ref{fig:non_resonant_stab_bdy}. However, for the $\eta_\parallel^P=\eta_\perp^P=3$ case, the non-resonant solution cannot capture the kinetic nature of the toroidal secondary mode for $\mathcal{A}^P \lesssim 10$. Indeed, it fails to reproduce the finite real frequency of the toroidal secondary mode for $\mathcal{A}^P \lesssim 10$, as $\omega_r \sim k_x v_{Mx}$ is neglected in the ordering \eqref{eq:NR_ordering}. Neither does the non-resonant dispersion relation capture the threshold in the primary amplitude for instability, as shown by the inset in Figure~\ref{fig:non_resonant_omega_AP}.

The coefficients $C_\mathrm{RDK}$ \eqref{eq:def_CRDK} and $C_\mathrm{TSM}$ \eqref{eq:def_CTSM} may be evaluated for the bi-Maxwellian primary distribution function \eqref{eq:model_gP_M} to be
\begin{align}
    C_\mathrm{RDK} & = \frac{1}{2} \left( 1+\tau(1+\eta_\perp^P) \right),\\
    C_\mathrm{TSM} & =  \tau \left( \frac{7}{4}+\frac{1}{\tau} \right)^{-1} \left( \left(1 + \frac{\eta_\parallel^P+\eta_\perp^P}{2}\right)^2 - \left( \frac{7}{4}+2\eta_\parallel^P+\frac{3 \eta_\perp^P}{2} \right) \right).
\end{align}
The RDK secondary mode requires $C_\mathrm{RDK}>0$, i.e. $\eta_\perp^P$ cannot be too negative, otherwise the diamagnetic stress opposes the Reynolds stress too strongly. The coefficient $C_\mathrm{TSM}$ is negative for $\eta_\parallel^P, \eta_\perp^P \rightarrow 0$ and positive at large $\abs{\eta_\parallel^P+\eta_\perp^P}$, explaining the differences between the pure density and $\eta_\parallel^P = \eta_\perp^P = 3$ cases in Figure~\ref{fig:non_resonant}.

\begin{figure}
    \centering
    \includegraphics[width=\textwidth, trim={0.8cm 1cm 1cm 0cm}, clip]{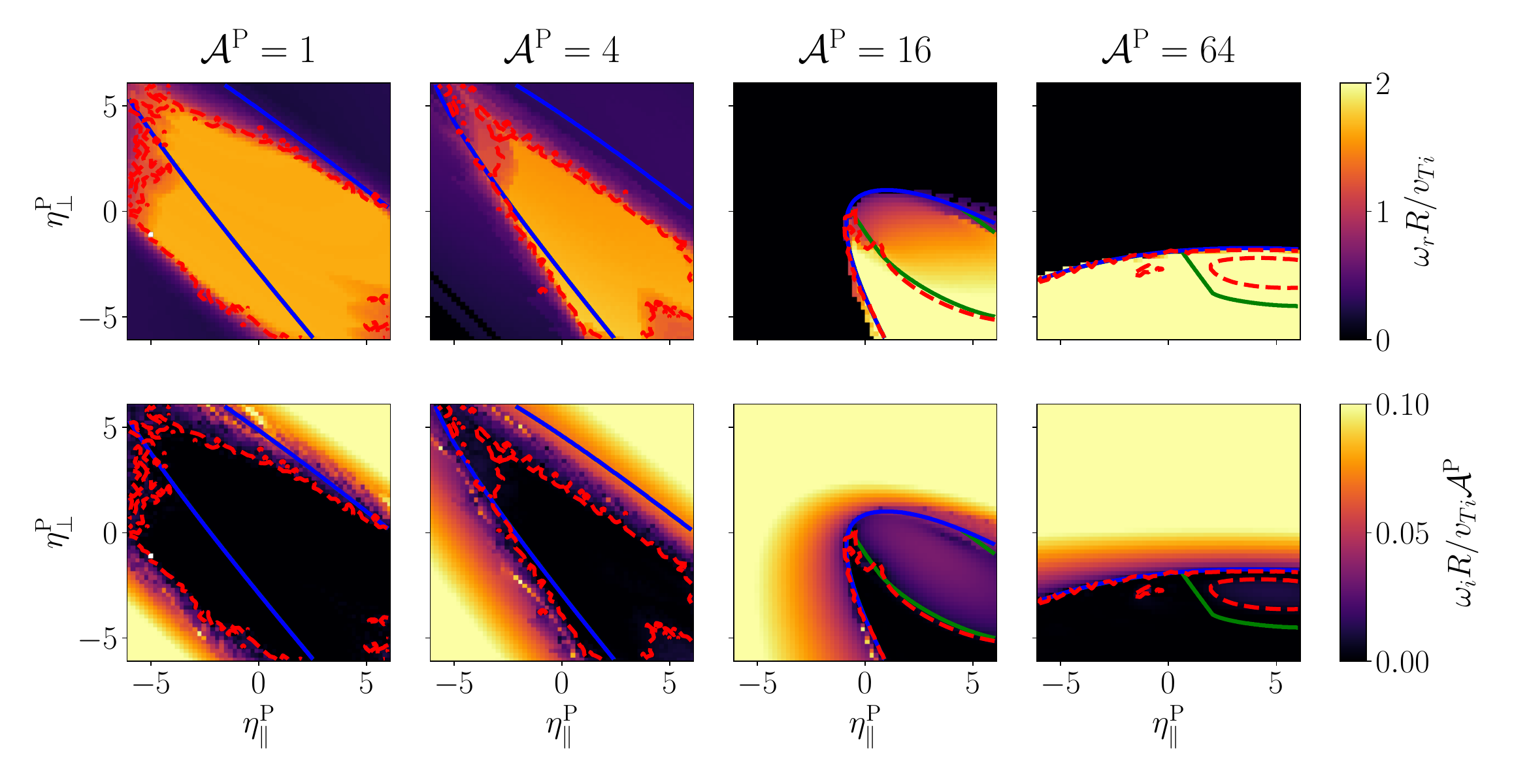}
    \caption{Real frequency (top row) and growth rate normalised by $\mathcal{A}^P$ (bottom row) at $k_x \rho_i = 0.1$ from gyrokinetic simulations for a bi-Maxwellian primary distribution function \eqref{eq:model_gP_M} constant in $\theta$, with different values of $\eta_\parallel^P$ and $\eta_\perp^P$ and four primary amplitude values $\mathcal{A}^P$ (left to right). A safety factor value $q=20$ is chosen and the phase-shift-periodic boundary condition \eqref{eq:BC_phase-shift-periodic} with $\Theta = 0$ is used. Due to computational cost, the simulations are performed with a reduced resolution $\{N_y, N_\theta, N_{v_\parallel}, N_\mu\} = \{4, 24, 32, 8\}$. The transitions from stable to purely growing modes ($\omega_r=0$) and from stable to overstable modes ($\omega_r\neq 0$) given by the non-resonant dispersion \eqref{eq:D_TSM_NR_primary_Fourier} are indicated by blue and green lines, respectively. The numerically determined instability threshold is indicated by red dashed lines.}
    \label{fig:etaprp_etapar}
\end{figure}

The stability of secondary modes as a function of $\eta_\parallel^P$ and $\eta_\perp^P$ for various primary amplitudes is studied in Figure~\ref{fig:etaprp_etapar}. The stability limits obtained from the non-resonant dispersion relation are shown therein to approximate those obtained from GK simulations. Although discrepancies are observed at small $\mathcal{A}^P\sim 1$, the non-resonant limit still qualitatively reproduces the stability boundary, i.e. instability at large $\abs{\eta_\parallel^P+\eta_\perp^P}$. The non-resonant limit also correctly captures the region of overstable solutions, see e.g. green curve for $\mathcal{A}_P=16$ in Figure~\ref{fig:etaprp_etapar}.

Finally, we investigate the conditions under which the ISM dominates over the TSM and RDK secondary mode in the non-resonant limit. Just like for the usual ITG primary mode, a non-resonant dispersion relation for the ISM may be found by considering a large driving temperature gradient compared to the density gradient, i.e. $\eta_\parallel^P, \eta_\perp^P \gg 1$. We thus consider the ordering $b_i^{1/2} \sim k_x v_{Mx} \ll \omega \sim k_x v_{Ex}^P \ll \eta_{\parallel}^P k_x v_{Ex}^P \sim \eta_\perp^P k_x v_{Ex}^P$. The ISM dispersion relation \eqref{eq:D_ISM} then simplifies to
\begin{equation}
    0 = \mathcal{D}^{\mathrm{ISM, NR}} = \frac{1}{(\omega - k_x v_{Ex}^P)^2} \left( \tau \omega (\omega - k_x v_{Ex}^P) + k_x^2 v_{Mx} v_{1}^P  \right),
\end{equation}
with roots
\begin{equation} \label{eq:omega_ISM_NR}
    \omega_\mathrm{ISM, NR} = \frac{1}{2} k_x \left( v_{Ex}^P \pm \sqrt{ ( v_{Ex}^P )^2 - 4  v_{Mx} v_{1}^P /\tau}   \right).
\end{equation}
In this limit, the ISM is unstable when the argument of the square root is negative, i.e. for large $\abs{v_1^P}$ and $v_1^P v_{Mx} > 0$ (corresponding to `bad curvature'). Its growth rate is bounded by Im$(\omega_\mathrm{ISM}) \leq k_x \sqrt{\abs{v_{Mx}v_1^P}/\tau}$, i.e. the bound on the growth rate scales with the square root of the primary temperature gradient. At large $\mathcal{A}^P$ (and/or large $\eta_{\parallel}^P, \eta_{\perp}^P$), the ISM will thus be subdominant to the RDK secondary mode and the TSM, whose growth rates increase linearly with the primary gradients.

\subsection{Secondary modes with finite binormal wavenumber and parallel streaming}
\label{sec:robustness_TSM}

The theory of secondary modes in toroidal geometry presented above assumed small binormal wavenumbers and parallel streaming, which allowed us to simplify the GK equation \eqref{eq:gyrokinetic_eq_realspace_secondary} to an algebraic equation \eqref{eq:gyrokinetic_eq_fourierspace_secondary_no_streaming}, leading to the dispersion relations \eqref{eq:D_ISM} and \eqref{eq:D_TSM}. In this section, the limits of validity of these assumptions are studied. First, the effects of finite binormal wavenumbers are studied numerically in Section~\ref{sec:vy_shat}. Then, the effects of parallel streaming on the TSM and RDK secondary mode are considered in Section~\ref{sec:streaming_TSM}. 

We note that the GK simulations of the secondary mode still enforce the primary drive to be constant in time, an assumption which may become unjustified when the binormal drifts and parallel streaming are sufficiently large. For example, the results presented here may overemphasise the role of binormal drifts and parallel streaming as the simulations will exaggerate the frequency mismatch between the primary drive and the secondary mode, which has previously been found to stabilise secondary instabilities \citep{chen_excitation_2000}.

\subsubsection{Binormal drifts and magnetic shear effects}\label{sec:vy_shat}

~\\\vspace{-0.2cm}

\begin{figure}
    \centering
    \includegraphics[width=\textwidth, trim={0.8cm 0.8cm 0.8cm 0cm}, clip]{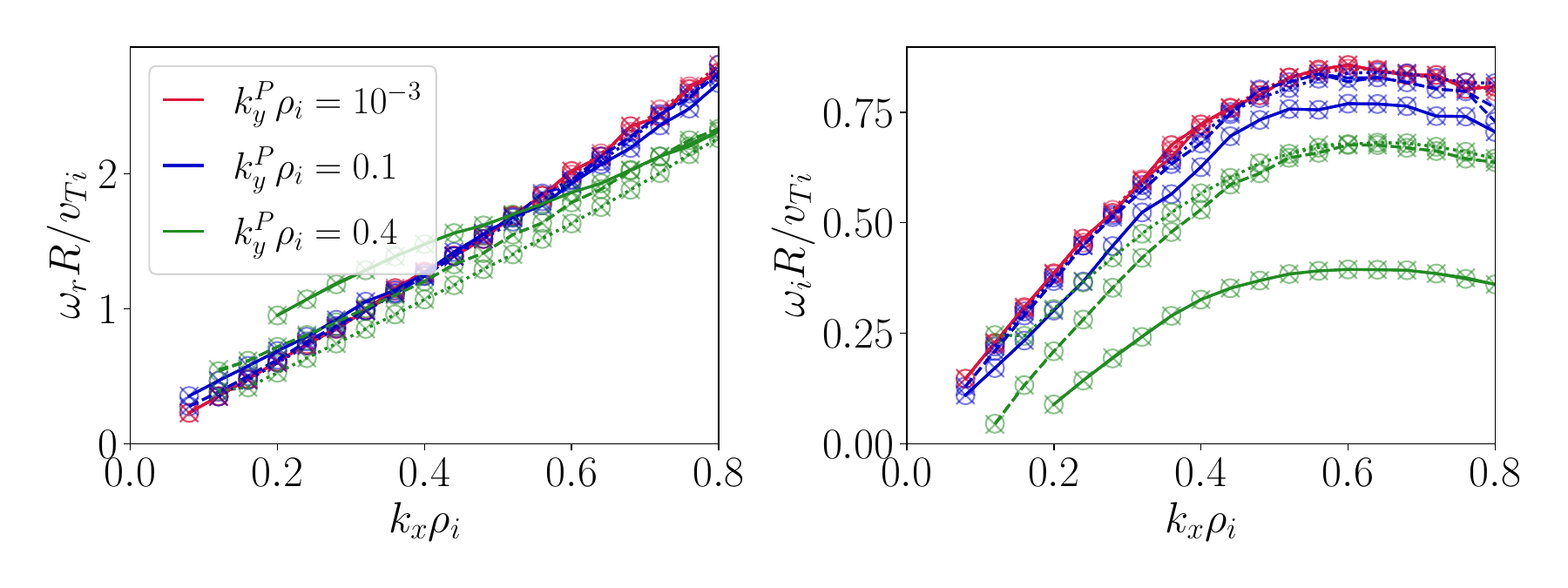}
    \caption{Secondary mode frequency (left) and growth rate (right) as a function of $k_x \rho_i$ from gyrokinetic simulations of secondary modes for varying primary binormal wavenumbers $k_y^P$ (colors) and magnetic shear values $\hat s =0.8$ (solid lines) and $\hat s=0.01$ (dashed lines). A case with $\hat s =0.8$ and with binormal magnetic drifts artificially removed in the simulations is also included (dotted lines). The phase-shift factors $\Theta = 0$ (circles) and $\Theta = 0.288572618$ (crosses) are considered. The primary distribution has the form \eqref{eq:model_gP_M} with $\eta_\perp^P = \eta_\parallel^P=3$, amplitude $\mathcal{A}^P=2$, and it varies along the magnetic field as $g_i^P \propto e^{-(\theta/\pi)^2}$. A safety factor value $q=20$ is used.}
    \label{fig:scan_kyP}
\end{figure}

\noindent In Figure~\ref{fig:scan_kyP}, gyrokinetic simulations of secondary modes are shown for various binormal wavenumbers of the primary mode. Both the usual magnetic shear value $\hat s = 0.8$ and a smaller value $\hat s = 0.01$ are considered. A case with artificially removed binormal drifts is also included to separate the potential stabilisation due to FLR effects from the stabilisation due to the binormal component of the magnetic drift. 

We remind the reader that the magnetic shear $\hat s$ enters $\nabla y$ and therefore affects both the binormal magnetic drift $\bsy{\tilde v}_M \cdot \nabla y$ and the strength of the FLR effects through $k_\perp^2 = k_x^2 \abs{\nabla x}^2 + 2 k_x k_y \nabla x \cdot \nabla y + k_y^2 \abs{\nabla y}^2$. In a large aspect ratio tokamak with circular flux surfaces, the binormal magnetic drift's variation along the magnetic field is approximately $\bsy{\tilde v}_M \cdot \nabla y \propto \cos\theta + \hat s \theta \sin\theta$. Therefore, for larger $\hat s$, the strength of the binormal magnetic drift at $\theta \sim \pm \pi/2$ is substantially increased, and it should be expected to affect the TSM more strongly.

Indeed, Figure~\ref{fig:scan_kyP} shows that increasing the primary binormal wavenumber affects the TSM growth rate most strongly through the binormal component of the magnetic drift at sufficiently large magnetic shear (solid lines). In comparison, the reduction in the peak TSM growth rate is modest in the cases where the magnetic shear is small (dashed lines) or the binormal component of the magnetic drift is zeroed out (dotted lines). For smaller $k_x \rho_i$, the binormal magnetic drift stabilises the TSM independently of the value of the magnetic shear (dashed and solid lines), bringing about a noticeable threshold in $k_x \rho_i$ for instability. We note that the stabilisation of the TSM by binormal magnetic drifts and FLR effects in Figure~\ref{fig:scan_kyP} requires large $k_y^P \rho_i$, as it remains weak even for binormal wavenumbers as large as $k_y^P \rho_i = 0.1$. 

We did not consider here background density and temperature gradients, which could make the inclusion of binormal magnetic drifts destabilising to the TSM by bringing about a primary-type instability mechanism, i.e. the usual ITG instability. 

The effects of finite binormal magnetic drifts on the TSM in turbulence simulations should generally be expected to depend on the parameter regime considered. For instance, the binormal scale of strongly driven ITG tokamak turbulence is expected to scale as $k_y^P \rho_i \sim (q R /L_{Ti})^{-1}$ based on the `critical balance' conjecture \citep{barnes_critically_2011, ghim_experimental_2013, nies_saturation_2024}. Therefore, the TSM's stabilisation by binormal magnetic drifts might prove more important near the turbulence marginal stability threshold, i.e. for smaller temperature gradient values.

\subsubsection{Parallel streaming effects} \label{sec:streaming_TSM}

~\\\vspace{-0.2cm}

\begin{figure}
    \centering
    \includegraphics[width=\textwidth, trim={2cm 1.4cm 1.8cm 0.4cm},clip]{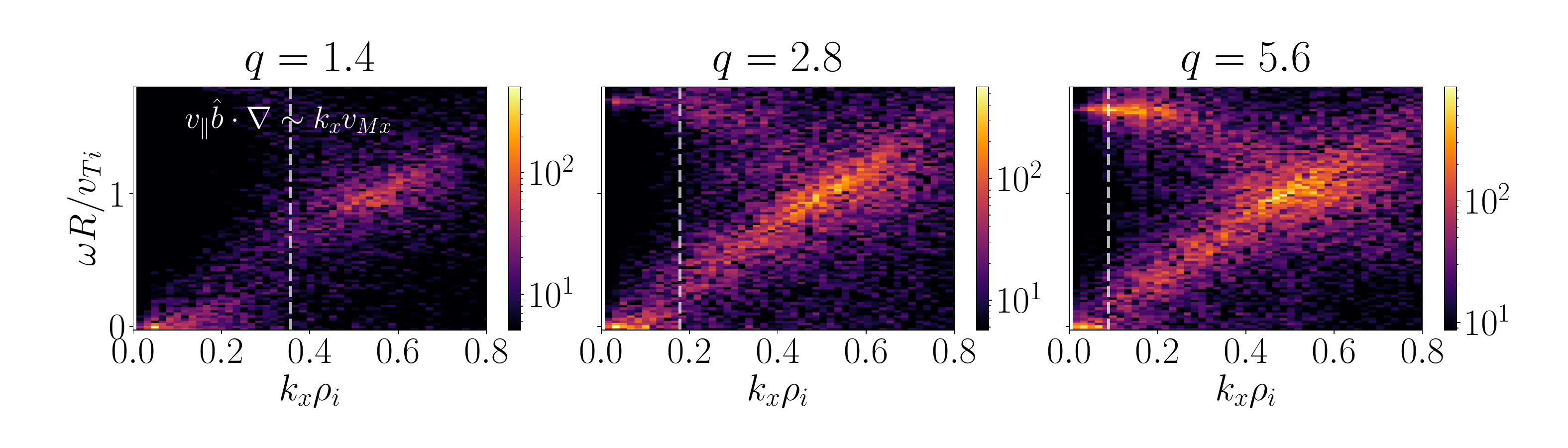}
    \caption{Zonal flow spectra $\abs{(v_E^\mathrm{ZF} R/ v_{Ti}\rho_i)_{k_x,\omega}}$ in nonlinear gyrokinetic simulations of ITG turbulence for varying safety factor $q$. Only positive frequencies are shown here for clarity as the ZF amplitude at negative frequencies is very similar to that at positive frequencies, see e.g. Figure~\ref{fig:ZF_real_Fourier}. The dashed line at $k_x \rho_i = 1/2q$ is the wavenumber where the parallel streaming rate $v_\parallel \bhat \cdot \nabla \sim v_{Ti}/qR$ and mode frequency $\omega \approx 2  k_x \rho_i v_{Ti}/R$ are comparable.}
    \label{fig:NL_spectra_scan_q}
\end{figure}

\noindent The assumption of negligible parallel streaming underlying the theory of the toroidal secondary mode requires the (complex) frequency of the secondary mode to be much larger than the parallel streaming rate. In a tokamak, $v_\parallel \bhat \cdot \nabla \sim v_{Ti}/qR$, so all simulations shown in Section~\ref{sec:secondary_modes_toroidal_geo} considered a large safety factor $q = 20$ to satisfy the assumption of small parallel streaming. However, this assumption always breaks at sufficiently long radial wavelengths, as the secondary mode frequency becomes vanishingly small in this limit. In particular, for the toroidal secondary mode near marginal stability, $\omega \sim k_x v_{Mx}$ becomes comparable to the parallel streaming rate for $k_x \rho_i \sim 1/q$ (neglecting an order unity coefficient that may be estimated given data for the mode frequency, see Figure~\ref{fig:NL_spectra_scan_q}). The toroidal secondary modes only survive above this threshold, as evidenced by the zonal flow spectra from fully nonlinear GK simulations of the turbulence at varying safety factor values, see Figures~\ref{fig:NL_spectra_scan_q} and \ref{fig:ZF_real_Fourier}. 

\begin{figure}
    \centering
    \includegraphics[width=\textwidth, trim={0.8cm 0.8cm 0.8cm 0cm}, clip]{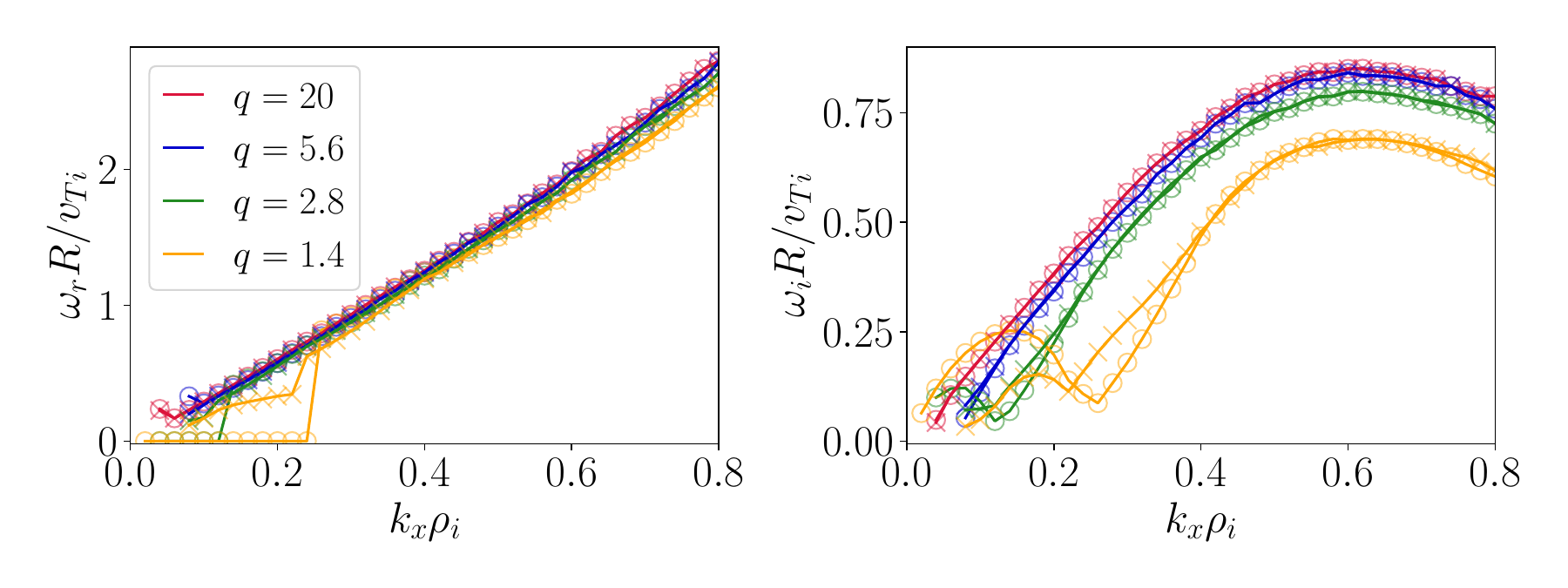}
    \caption{Secondary mode frequency (left) and growth rate (right) from gyrokinetic simulations of secondary modes for varying safety factor $q$ and phase-shift factors $\Theta = 0$ (circles) and $\Theta = 0.288572618$ (crosses). The primary distribution has the form \eqref{eq:model_gP_M} with $\eta_\perp^P = \eta_\parallel^P=3$, amplitude $\mathcal{A}^P=2$, and it varies along the magnetic field as $g_i^P \propto e^{-(\theta/\pi)^2}$.}
    \label{fig:scan_qinp}
\end{figure}

As the wavenumber threshold is approached from above, the toroidal secondary modes become Landau damped, as shown in Figure~\ref{fig:scan_qinp}. This damping by parallel streaming may further be interpreted as the short-circuiting of the regions below and above the midplane via parallel fluxes of particles and energy that reduce the up-down pressure asymmetry generating the TSM. Landau damping is also known to damp GAMs, with a typical scaling of the damping rate Im$(\omega_\mathrm{GAM}) \propto -e^{-q^2}$ in tokamaks, such that GAMs are more often observed in the edge (where $q$ is large) than in the core of tokamak devices. While the Landau damping of TSMs is expected to be stronger than that of GAMs due to the TSM's smaller frequency (see e.g. Figure~\ref{fig:ZF_real_Fourier}), the TSM turbulence drive is also typically stronger. For this reason, the GAM amplitude is small in the turbulence simulations at low safety factor $q=1.4$ and $q=2.8$ (see Figure~\ref{fig:NL_spectra_scan_q}), while the TSM amplitude remains sizeable at short radial wavelengths $k_x \rho_i \sim 0.5$.

The choice of boundary conditions, e.g. the choice of $\Theta$ in \eqref{eq:BC_phase-shift-periodic}, becomes important as parallel streaming becomes relevant to the mode dynamics. Indeed, Figure~\ref{fig:scan_qinp} shows that the choice of $\Theta$ affects the damping of toroidal secondary modes and, more importantly, $\Theta$ determines whether growing modes exist at long radial wavelengths.

In the simulations with periodic boundary conditions $(\Theta = 0)$, long-wavelength purely growing RDK secondary modes are observed for weak primary drive $k_x \tilde{v}_{g}^\mathrm{P} \lesssim v_\parallel \bhat \cdot \nabla$, see Figure~\ref{fig:scan_qinp}. In this case, the parallel streaming spreads the secondary mode along the field line, reducing the nonlinear interaction with the primary mode. Therefore, unlike the TSM, the RDK secondary mode in this case has a reduced growth rate but it is not fully stabilised. For different boundary conditions at the ends of the flux tube, e.g. for different $\Theta$, the RDK secondary mode is affected differently, as may be seen in Figure~\ref{fig:scan_qinp}.

Moreover, while we found in Section~\ref{sec:TSM_NR} that the RDK secondary mode mechanism is dominant at short wavelengths ($k_x \tilde v_{Mx} \gtrsim v_\parallel \bhat \cdot \nabla$) only for large primary amplitudes $k_x \tilde{v}_{g}^{P} \gtrsim \omega_\mathrm{GAM}$, this requirement may be relaxed to $k_x \tilde{v}_{g}^{P} \gtrsim v_\parallel \bhat \cdot \nabla$ at long wavelengths $k_x \tilde v_{Mx} \lesssim v_\parallel \bhat \cdot \nabla$ because in this regime parallel flows are very efficient in damping up-down asymmetries.

Generally, one must be cautious to apply results of the secondary model in the regime of weak primary drive $k_x \tilde{v}_{g}^P \lesssim v_\parallel \bhat \cdot \nabla$ and long radial wavelengths $k_x \tilde v_{Mx} \lesssim v_\parallel \bhat \cdot \nabla$. Indeed, the separation of scales assumed in the secondary model can easily break, e.g. the binormal magnetic drift and the diamagnetic drift may become important, or the frozen streamer primary drive assumption may become unjustified because the primary drive will change significantly in a time comparable to the inverse of the secondary mode frequency. It may therefore be necessary to include more physics or to employ other models for the growth of zonal flows in this regime.

Finally, we note that the ISMs studied in this work are the secondary mode equivalents of the toroidal branch of the ITG. Introducing parallel streaming dynamics brings about the slab branch of the ITG mode \citep{cowley_considerations_1991} and should therefore also give rise to a slab branch of the ISM. Such modes have been considered previously, see e.g. \cite{drake_streamer_1988, cowley_considerations_1991, rath_core_1992, ivanov_dimits_2022}, so we will not study them here.

\section{Summary and discussion}\label{sec:conclusions}

We have shown in this study how toroidicity affects the nonlinear physics of zonal flows. We refer back to Figure~\ref{fig:summary_regions} for an overview of the modes discussed in this study and their respective regions of validity.

The secondary model was used to derive the dispersion relation \eqref{eq:D_TSM}, notably describing the toroidal secondary mode (TSM), a branch of growing and radially propagating zonal flows introduced in \cite{nies_saturation_2024}. The theory also led to the dispersion relation \eqref{eq:D_ISM} for ITG secondary modes (ISMs), secondary modes localised around particular points on the flux surface and driven unstable by the primary temperature gradient, see Section~\ref{sec:ISM}. The physical mechanism that drives the TSM, based on the Stringer-Winsor force and the generation of up-down pressure asymmetry through a combination of zonal flow shearing and advection by the radial magnetic drift, was studied in Section~\ref{sec:TSM_R}.

By considering the non-resonant limit of the secondary modes, we investigated in Section~\ref{sec:TSM_NR} the regions in parameter space where the RDK secondary modes, TSMs, and ISMs grow. The RDK secondary mode limit was found to require a large primary drive $k_x v_{Ex}^P \gg \omega_\mathrm{GAM} \sim v_{Ti}/R$, though this requirement might be relaxed at long wavelengths due to parallel streaming, as discussed in Section~\ref{sec:streaming_TSM}. The TSM was shown in Figure~\ref{fig:etaprp_etapar} to be driven unstable for sufficiently large primary temperature gradient drive, similar to the ISM, though the TSM is a global mode on the flux surface and is dominant over the ISM at sufficiently large primary drive (see \eqref{eq:omega_ISM_NR} and surrounding discussion).

The effects of finite binormal wavenumbers and parallel streaming were considered in Section~\ref{sec:robustness_TSM}. Importantly, these effects were shown in Figures~\ref{fig:scan_kyP}~and~\ref{fig:scan_qinp} to introduce a $k_x$-threshold for the TSM. In GK simulations of strongly driven tokamak ITG turbulence, the threshold is set by the parallel streaming to $k_x\rho_i \approx 1/2q$, as demonstrated by Figures~\ref{fig:ZF_real_Fourier} and \ref{fig:NL_spectra_scan_q}.

The simulations presented in this study considered a circular cross-section tokamak and up-down symmetric primary modes for simplicity. As a result, inward-propagating and outward-propagating TSMs are driven equally strongly. Making the magnetic geometry or the driving primary modes up-down asymmetric would instead drive the TSM preferentially above or below the midplane, leading to a preferred direction of propagation. We leave the study of such scenarios for future work, noting that the TSM dispersion relation \eqref{eq:D_TSM} is valid for arbitrary magnetic geometries and variation of the primary mode along the magnetic field. Additional physical effects not considered in this work that might affect the TSM include background flow shear, non-adiabatic electron physics, collisions, and electromagnetic fluctuations. Future work could extend the secondary model of Section~\ref{sec:theory_ISM_TSM} to include these effects. 

The TSM is very prominent in gyrokinetic simulations of ITG turbulence (see Figures~\ref{fig:ZF_real_Fourier} and \ref{fig:NL_spectra_scan_q}) and was shown in \cite{nies_saturation_2024} to contribute to turbulence saturation. The TSM could also potentially explain the avalanches widely reported in gyrokinetic simulations. Across our simulations (see Figures~\ref{fig:ZF_real_Fourier}~and~\ref{fig:NL_spectra_scan_q}), the TSMs are observed predominantly at radial wavelengths $k_x \rho_i \approx 0.5$ and frequencies $\omega \approx v_{Ti}/R$, such that their associated propagation velocity is $v_\mathrm{prop} \approx 2 v_{Ti} \rho_i/R$. A preliminary analysis suggests the TSM scale and propagation speed are consistent with the avalanches reported in \cite{mcmillan_avalanchelike_2009, idomura_study_2009, gorler_nonlocal_2010, villard_turbulence_2014, rath_comparison_2016, wang_understanding_2017}. 

Moreover, zonal oscillations at $\omega \approx \omega_\mathrm{GAM}/2 \approx v_{Ti}/R$, close to the dominant TSM frequency in turbulence simulations (see Figures~\ref{fig:ZF_real_Fourier}~and~\ref{fig:NL_spectra_scan_q}), have been reported in tokamak I-mode experiments \citep{feng_i-mode_2019, mccarthy_low_2022, bielajew_edge_2022, liu_characteristics_2023}. The edge temperature ring oscillations observed in the EAST tokamak \citep{feng_i-mode_2019, liu_characteristics_2023} could potentially be explained by the Stringer-Winsor force having to vanish, similar to the TSM (see also left column in Figure~\ref{fig:TSM_mechanism}). One might thus speculate whether the TSM explains these low-frequency zonal modes. We note that it has been suggested that these oscillations are instead explained by Q-GAMs \citep{lee_destabilization_2024}, which share some similarities with the TSM. Both the Q-GAM and the TSM are driven by the SW force mechanism, though the required up-down pressure asymmetry for the Q-GAM originates from background parallel fluxes, unlike the TSM for which it is generated by a combination of ZF shearing and radial magnetic drift advection (see Section~\ref{sec:TSM_R} and \cite{nies_saturation_2024}).

\section*{Data availability statement}

The data supporting the findings of this article are available at \url{https://datacommons.princeton.edu/discovery/catalog/doi-10-34770-envh-7358}.

\ack R.N. thanks Thomas Foster, Xu Chu, and Jungpyo Lee for helpful discussions. This work was supported by U.S. DOE DE-AC02-09CH11466 through PPPL's Laboratory Directed Research and Development (LDRD). The simulations presented in this article were performed on computational resources managed and supported by Princeton Research Computing, a consortium of groups including the Princeton Institute for Computational Science and Engineering (PICSciE) and the Office of Information Technology's High Performance Computing Center and Visualization Laboratory at Princeton University.

\appendix

\section{Strongly-driven secondary modes at long perpendicular wavelengths} \label{sec:strongly_driven}

In this appendix, we consider the growth of secondary modes in the limit of large primary drive and long perpendicular wavelengths. Unlike the theory presented in the main text and \ref{sec:RDK_SW}, we do not assume the primary drive to be a streamer, i.e. $\partial_x g_i^P$ need not be zero. The primary drive is still assumed to be purely nonzonal, i.e. $\langle g_i^P \rangle_y = 0$.

We consider two expansion parameters $\epsilon_\mathrm{SD} \ll 1$ and $\epsilon_\mathrm{LW} \ll 1$, with
\begin{equation}
    v_\parallel \bhat \cdot \nabla g_i^S \sim \bsy{\tilde v}_{Mi}\cdot \nabla g_i^S  \sim \langle \bsy{v}_E^S \rangle_{\bsy{R}_i} \cdot \nabla F_{Mi} \sim \epsilon_\mathrm{SD} \partial_t g_i^S \sim \epsilon_\mathrm{SD} \langle \bsy{v}_E^S \rangle_{\bsy{R}_i} \cdot \nabla g_i^P \sim \epsilon_\mathrm{SD}  \langle \bsy{v}_E^P \rangle_{\bsy{R}_i} \cdot \nabla g_i^S,
\end{equation}
corresponding to a strong primary drive in the gyrokinetic equation for the secondary \eqref{eq:gyrokinetic_eq_realspace_secondary}, and
\begin{equation}
    \rho_i \nabla_\perp \ln g_i^S \sim \rho_i \nabla_\perp \ln g_i^P \sim \epsilon_\mathrm{LW},
\end{equation}
corresponding to long perpendicular wavelengths. We note that $g_i^S/F_{Mi} \sim Z_i e \varphi^S / T_i$ is assumed throughout.

For long perpendicular wavelengths $\epsilon_\mathrm{LW} \ll 1$, gyro-averages may be approximated as
\begin{equation} \label{eq:gyroavg_expansion}
    \langle f(\bsy{r}) \rangle_{\bsy{R}_i} \approx \left( 1 + \frac{v_\perp^2}{4 v_{Ti}^2}\rho_i^2 \nabla_\perp^2  \right) f(\bsy{R}_i), \qquad \langle f(\bsy{R}_i) \rangle_{\bsy{r}} \approx \left( 1 + \frac{v_\perp^2}{4 v_{Ti}^2}\rho_i^2 \nabla_\perp^2  \right) f(\bsy{r}).
\end{equation}
The vorticity equation \eqref{eq:vorticity} may then be shown to simplify to
\begin{align} \label{eq:vorticity_LW}
    -\partial_t \left\langle \frac{\rho_i^2 \abs{\nabla x}^2 \partial_x^2 \varphi^S}{2} \right\rangle_\psi  = & -\partial_x^2 \left\langle \frac{\rho_i^2}{2} \left[ \nabla x \cdot \nabla \varphi^P (v_{Ex}^S + v_\mathrm{dia}^S) + \nabla x \cdot \nabla \varphi^S (v_{Ex}^P + v_\mathrm{dia}^P) \right] \right\rangle_\psi   \nonumber\\
    &  - \frac{T_i}{Z_i e n_i} \left\langle \int\mathrm{d}^3 v\, \tilde v_{Mx} \partial_x \left( g_i^S + \frac{Z_i e \varphi^S}{T_i} F_{Mi} \right) \right\rangle_\psi,
\end{align}
with the primary $\bsy{E}\times\bsy{B}$ and diamagnetic velocities defined in \eqref{eq:vEx_P} and \eqref{eq:vdia_P}, respectively. The definition of $v_{Ex}^S$ and $v_\mathrm{dia}^S$ is identical to that for the primary quantities, except for the replacement of the primary electrostatic potential and distribution function with the secondary ones. The three terms in \eqref{eq:vorticity_LW} correspond to the ZF inertia, the nonlinear Reynolds and diamagnetic stresses, and the Stringer-Winsor (SW) force (see also discussion around \eqref{eq:vorticity}).

It is convenient to define the deviation of the fluctuating electrostatic potential from its flux-surface averaged value,
\begin{equation}
    \delta \varphi \equiv \varphi - \langle \varphi \rangle_\psi,
\end{equation}
with $\langle \delta \varphi \rangle_\psi =0 $ by definition. The smallness of $\epsilon_\mathrm{SD}$ and $\epsilon_\mathrm{LW}$ may be used to expand $g_i^S$ and $\delta \varphi^S$ as
\begin{equation}
    g_i^S = g_{i0}^S + g_{i1}^S + \mathcal{O}(\epsilon_\mathrm{SD}^2 g_i^S)+ \mathcal{O}(\epsilon_\mathrm{LW}^2 g_i^S) , \qquad \delta\varphi^S = \delta\varphi_0^S + \delta\varphi_1^S  + \mathcal{O}(\epsilon_\mathrm{SD}^2 \delta\varphi^S)+ \mathcal{O}(\epsilon_\mathrm{LW}^2 \delta\varphi^S),
\end{equation}
with $g_{i1}^S \sim \epsilon_\mathrm{SD}g_i^S $ and $\delta\varphi_{1}^S  \sim \epsilon_\mathrm{SD} \delta\varphi^S$. There are no contributions to $\mathcal{O}(\epsilon_\mathrm{LW} g_i^S)$ and  $\mathcal{O}(\epsilon_\mathrm{LW} \delta\varphi^S)$ as demonstrated by the fact that \eqref{eq:gyroavg_expansion} only contains corrections quadratic in $\epsilon_\mathrm{LW}$. Various moments of $g_i^S$ used below will follow the same notation for the expansion as $\delta\varphi^S$, e.g. $P^S = P_0^S + P_1^S + ...$ for the pressure-like quantity defined in \eqref{eq:pressure_moment}.

To leading order, the GK equation \eqref{eq:gyrokinetic_eq_realspace_secondary} gives
\begin{equation} \label{eq:gi0_S}
    \partial_t g_{i0}^S = -\{ \varphi^P, g_{i0}^S \} - \{ \langle \varphi^S \rangle_\psi + \delta\varphi_0^S, g_i^P \},
\end{equation}
where we have written the advection by the $\bsy{E}\times\bsy{B}$ flow using the Poisson bracket
\begin{equation}
    \{ f, g \} = \frac{\bhat \cdot\nabla x \times \nabla y}{B} \left(\partial_x f \partial_y g - \partial_y f \partial_x g\right)
\end{equation}
such that $\bsy{v}_E\cdot \nabla g_i = \{ \varphi, g_i\}$. Again to leading order, taking a time derivative of the quasineutrality equation \eqref{eq:quasineutrality_secondary} and inserting \eqref{eq:gi0_S} leads to
\begin{equation} \label{eq:phiS_RDK}
    \partial_t \delta\varphi_0^S = - \{ \langle \varphi^S \rangle_\psi, \varphi^P \}.
\end{equation}
The nonzonal secondary potential thus evolves due to the zonal flow advecting the primary potential. Differentiating \eqref{eq:gi0_S} with respect to $t$ leads to
\begin{equation} \label{eq:dt_gi0}
    0 = \partial_t \left( \partial_t g_{i0}^S + \{ \langle \varphi^S \rangle_\psi, g_{i}^P \} \right) + \{ \varphi^P, \partial_t g_{i0}^S + \{ \langle \varphi^S \rangle_\psi, g_{i}^P \} \} - \{ \langle \varphi^S \rangle_\psi, \{ \varphi^P, g_i^P \} \},
\end{equation}
where we used the Leibniz identity for Poisson brackets
\begin{equation}
    \{f, \{g, h\}\} + \{g, \{h, f\}\} + \{h, \{f, g\}\} = 0.
\end{equation}

We assume the primary drive to not be self-interacting, i.e. $\{\varphi^P, g_i^P\}=0$, such that the last term in \eqref{eq:dt_gi0} vanishes. A nonlinearly self-interacting primary drive would lead to a `forced' growth of the ZF, different from the modulational instability mechanism studied in this work. The self-interaction can be important in the case of ZFs driven by Alfv\'en primary modes due to their global character, see e.g. \cite{qiu_effects_2016}. In the case of ZF drive by turbulent fluctuations, the self-interaction becomes important when the primary modes are elongated enough along the magnetic field to `bite their own tail' \citep{c_j_how_2020}. We do not consider such cases here. Then, the quantity 
\begin{equation}
    G^S =  \partial_t g_{i0}^S + \{ \langle \varphi^S \rangle_\psi, g_i^P \}
\end{equation}
is simply advected by the primary $\bsy{E}\times\bsy{B}$ flow,
\begin{equation}
    \partial_t G^S + \{ \varphi^P, G^S\}=0.
\end{equation}
This advection does not lead to exponential growth, so we limit ourselves to the case $G^S = 0$. We therefore obtain
\begin{equation} \label{eq:gS_RDK}
    \partial_t g_{i0}^S = - \{ \langle \varphi^S \rangle_\psi, g_i^P \}.
\end{equation}
To leading order, the secondary distribution thus results purely from the zonal flow advecting the primary distribution function and is therefore nonzonal, $\langle g_{i0}^S \rangle_\psi = 0$. This is not in contradiction with the zonal flow being finite at lowest order because of the factor $(1-\Gamma_{0i}) \sim \epsilon_\mathrm{LW}^2$ in the quasineutrality equation \eqref{eq:quasineutrality}, such that the zonal flow is contained in the higher order contribution $\mathcal{O}(\epsilon_\mathrm{LW}^2)$ to $g_i^S$.

The leading order expressions \eqref{eq:phiS_RDK} and \eqref{eq:gS_RDK} are sufficient to derive a dispersion relation for the RDK secondary modes \citep{rogers_generation_2000}. Using
\begin{align}
    & -\partial_t \left( \nabla x \cdot \nabla \varphi^P (v_{Ex,0}^S + v_\mathrm{dia,0}^S) + \nabla x \cdot \nabla \delta\varphi_0^S (v_{Ex}^P + v_\mathrm{dia}^P) \right) \nonumber\\
    & = \nabla x \cdot \nabla \varphi^P \{ \langle \varphi^S \rangle_\psi,  \left( v_{Ex}^P + v_\mathrm{dia}^P \right) \} +  \nabla x \cdot \nabla \{ \langle \varphi^S \rangle_\psi, \varphi^P \}\left( v_{Ex}^P + v_\mathrm{dia}^P \right) \nonumber\\
    & = \abs{\nabla x}^2 \partial_x^2 \langle \varphi^S \rangle_\psi v_{Ex}^P \left(v_{Ex}^P + v_\mathrm{dia}^P \right) + \left\{ \langle \varphi^S \rangle_\psi, \nabla x \cdot \nabla \varphi^P  \left(v_{Ex}^P + v_\mathrm{dia}^P \right)\right\},
\end{align}
the time derivative of the vorticity equation \eqref{eq:vorticity_LW} simplifies to
\begin{align} \label{eq:RDK_disp_rel}
    - \left\langle \frac{\rho_i^2 \abs{\nabla x}^2}{2} \right\rangle_\psi  \partial_t^2 \partial_x^2 \langle \varphi^S \rangle_\psi = &  \partial_x^2 \left( \partial_x^2 \langle \varphi^S \rangle_\psi \left\langle \frac{\rho_i^2 \abs{\nabla x}^2 }{2}  v_{Ex}^P \left(v_{Ex}^P + v_\mathrm{dia}^P \right) \right\rangle_\psi \right) \nonumber \\
    & -  \left\langle v_{Mx} \partial_x  \partial_t \left( \frac{P_1^S}{Z_i e n_i} + \delta\varphi_1^S \right) \right\rangle_\psi.
\end{align}
Here, we have used the definition of the pressure-like quantity $P$ in \eqref{eq:pressure_moment} and $\langle \{ \langle \varphi^S \rangle_\psi, f \} \rangle_\psi = 0$. This latter identity, combined with $\langle v_{Mx} \rangle_\psi = 0$, causes the SW force contribution to vanish at $\mathcal{O}(\epsilon_\mathrm{SD})$, and the higher order contributions $P_1^S$ and $\delta\varphi_1^S$ must therefore be derived to evaluate it. 

In the limit $\epsilon_\mathrm{SD} \ll \epsilon_\mathrm{LW}$, the SW force contribution may be ignored entirely ($P_1^S, \delta\varphi_1^S \rightarrow 0$) and \eqref{eq:RDK_disp_rel} describes the RDK secondary mode \citep{rogers_generation_2000}. The equation thus obtained is slightly more general than similar equations in previous work which assumed the primary drive to have large radial scales compared to the secondary mode, by either assuming the primary mode to be a streamer ($\partial_x g_i^P =0$) \citep{rogers_generation_2000, plunk_gyrokinetic_2007} or by employing a WKB approximation \citep{plunk_nonlinear_2017}.  It is worth noting that equations \eqref{eq:phiS_RDK} and \eqref{eq:gS_RDK} justify the use of 4MT (four mode truncation) models to describe the RDK secondary mode: for a primary drive described by the Fourier mode $(k_x = 0, k_y = k_y^P)$, the secondary mode will be fully described by a zonal mode $(k_x = k_x^S, k_y = 0)$ and the sidebands $(k_x = k_x^S, k_y = \pm k_y^P)$. This is not the case for electron-scale secondary modes \citep{plunk_gyrokinetic_2007} or for secondary modes with finite radial magnetic drift \eqref{eq:gS_TSM}. In these cases, the secondary distribution and potential generally consist of many Fourier modes in $k_y$ even when the primary drive is described by a single Fourier mode. For an $x$-independent primary drive, Fourier-Laplace transforming \eqref{eq:RDK_disp_rel} in $x$ and $t$ leads to the RDK secondary mode dispersion relation \eqref{eq:omegaRDK}. We note that a dispersion relation for RDK secondary modes for arbitrary radial wavelengths of the secondary mode ($\epsilon_\mathrm{LW} \sim 1$) may be derived assuming a streamer primary mode $(\partial_x g_i^P=0)$, as shown in \ref{sec:RDK_SW}.

We now consider $\epsilon_\mathrm{SD} \sim \epsilon_\mathrm{LW}$, in which case we must compute the corrections $g_{i1}^S$ and $\delta\varphi_1^S$ to the secondary distribution and electrostatic potential so as to be able to evaluate the SW force contribution to the vorticity equation. First, we define various velocity-space moments that will prove useful, i.e. the parallel flow
\begin{equation}
    u_{\parallel} = \frac{1}{n_i} \int\mathrm{d}^3 v\, v_\parallel g_i,
\end{equation}
the parallel flux of parallel energy
\begin{equation}
    q_{\parallel} = \int\mathrm{d}^3 v\, v_\parallel \frac{m_i v_\parallel^2}{2} g_i,
\end{equation}
the parallel flux of perpendicular energy
\begin{equation}
    q_{\perp} = \int\mathrm{d}^3 v\, v_\parallel \frac{m_i v_\perp^2}{2} g_i,
\end{equation}
and finally the fourth-order moment (with units of pressure)
\begin{equation}
    \chi = T_i \int\mathrm{d}^3 v\, \left( \frac{v_\parallel^2 + v_\perp^2/2}{v_{Ti}^2} \right)^2 g_i.
\end{equation}

The GK equation \eqref{eq:gyrokinetic_eq_realspace_secondary} at $\mathcal{O}(\epsilon_\mathrm{SD}\partial_t g_i^S)$ is given by
\begin{equation} \label{eq:dt_gi1S}
    \partial_t g_{i1}^S + \{ \varphi^P, g_{i1}^S \} + \{\delta \varphi_1^S, g_i^P\} = - \left(v_\parallel \bhat + \tilde{\bsy{v}}_M \right) \cdot \nabla \left( g_{i0}^S + \frac{Z_i e (\langle \varphi^S \rangle_\psi + \delta\varphi_0^S)}{T_i} F_{Mi} \right) - \{ \delta\varphi_0^S, F_{Mi} \}.
\end{equation}
To evaluate the SW force contribution, we must evaluate the time derivatives of $P_1^S$ and $\delta\varphi_1^S$. The latter is obtained by considering the time derivative of the quasineutrality equation \eqref{eq:quasineutrality_secondary} to $\mathcal{O}(\epsilon_\mathrm{SD}\varphi^S)$,
\begin{equation}
    \tau \partial_t \delta\varphi_1^S = \frac{T_i}{Z_i e n_i} \int\mathrm{d}^3 v\, \partial_t g_{i1}^S.
\end{equation}
Using \eqref{eq:dt_gi1S}, we derive
\begin{equation} \label{eq:dt_phi1S}
    \tau \frac{ Z_i e}{T_i}  \partial_t \delta\varphi_1^S = - \bsy{B} \cdot \nabla \left( \frac{u_{\parallel,0}^S}{B} \right) - \bsy{v}_M \cdot \nabla \left( \frac{P_0^S}{n_i T_i} + \frac{Z_i e (\langle \varphi^S \rangle_\psi + \delta\varphi_0^S)}{T_i} \right) - v_{Ex,0}^S \partial_x \ln n_i.
\end{equation}
Here, we have used 
\begin{equation}
    \mathrm{d}^3 v = \frac{ 2\pi  B}{m_i^2 \abs{v_\parallel}} \sum_{\mathrm{sgn}(v_\parallel)} \mathrm{d}E_i \mathrm{d}\mu_i,
\end{equation}
with the ion energy $E_i = m_i v^2/2$ and magnetic moment $\mu_i = m_i v_\perp^2 / 2B$, which are held constant as the spatial derivatives are evaluated. Furthermore, we assumed for simplicity that the pressure gradient contribution to the magnetic drift in \eqref{eq:v_M} is negligible, which is satisfied e.g. at low plasma beta, $\beta \equiv 2\mu_0 p/B^2 \ll 1$. The magnetic drift velocity then reduces to
\begin{equation}
    \tilde{\bsy{v}}_M = \bsy{v}_M \frac{v_\parallel^2 + v_\perp^2/2}{v_{Ti}^2},
\end{equation}
similar to the expression for the radial component of the magnetic drift \eqref{eq:v_Mx}. If needed, the pressure gradient contribution to the curvature drift could be kept, at the price of another moment of the distribution function to carry through in the derivation. We note that \eqref{eq:dt_phi1S} satisfies the consistency condition $\langle \delta \varphi_1^S \rangle_\psi = 0$, as $\langle v_{Mx} \rangle_\psi = 0$ and the leading order distribution function $g_{i0}^S$ is nonzonal (see \eqref{eq:gS_RDK}).

The time derivative of the pressure moment $P_1^S$ is derived by evaluating the appropriate moment of \eqref{eq:dt_gi1S},
\begin{align}
    \partial_t P_1^S + & \{ \varphi^P, P_1^S\} + \{ \delta\varphi_1^S, P^P\} =  - \bsy{B} \cdot \nabla \left( \frac{q_{\parallel,0}^S}{B} \right) - \frac{\bhat \cdot \nabla q_{\perp,0}^S}{2} \nonumber\\  \label{eq:dt_P1S}
    & - \bsy{v}_M \cdot \nabla \left( \chi_0^S + \frac{7}{4} Z_i e n_i (\langle \varphi^S \rangle_\psi + \delta \varphi_0^S) \right) - v_{Ex,0}^S \partial_x (n_i T_i).
 \end{align}

The SW force contribution due to $\delta \varphi_1^S$ in \eqref{eq:RDK_disp_rel} is derived from \eqref{eq:dt_phi1S} to have the simple form
\begin{equation} \label{eq:phi1S_SW_contribution}
    \left\langle v_{Mx} \partial_x \partial_t \delta \varphi_1^S \right\rangle_\psi = - \frac{1}{\tau} \left\langle v_{Mx}^2  \right\rangle_\psi \partial_x^2 \langle \varphi^S \rangle_\psi.
\end{equation}
Note that many terms in \eqref{eq:dt_phi1S} have not contributed to \eqref{eq:phi1S_SW_contribution} because the average over $y$ of $g_{i0}^S$ vanishes due to \eqref{eq:gS_RDK}.

Evaluating the contribution of $P_1^S$ to the SW force is more involved due to the presence of the primary $\bsy{E}\times\bsy{B}$-advection term in \eqref{eq:dt_P1S}. One may simply consider the closed system of equations \eqref{eq:phiS_RDK}, \eqref{eq:gS_RDK}, \eqref{eq:RDK_disp_rel}, \eqref{eq:dt_phi1S}, and \eqref{eq:dt_P1S} to describe the evolution of strongly driven secondary modes at long perpendicular wavelengths. This system of equations describes how the leading order secondary distribution function $g_{i0}^S$ and nonzonal potential fluctuation $\delta \varphi_0^S$, which due to \eqref{eq:gS_RDK} and \eqref{eq:phiS_RDK} result from the zonal flow shearing of the primary distribution and potential, source higher order pressure and potential fluctuations $P_1^S$ and $\delta\varphi_1^S$ through parallel streaming, magnetic drifts, and the $\bsy{E}\times\bsy{B}$-mixing of the background Maxwellian. Then, $P_1^S$ and $\delta \varphi_1^S$  feed back on the zonal flow generation through the SW force, with the contribution from $\delta\varphi_1^S$ evaluated in \eqref{eq:phi1S_SW_contribution}.

To derive a more explicit dispersion relation, we Laplace-transform the secondary distribution in time and consider exponentially growing solutions $\varphi^S = \hat\varphi^S e^{-i \omega t}$ (similar notation will be used for $g_i^S$ and its velocity-space moments) with Im$(\omega) > 0$. Furthermore we define the differential operator $\omega_E^P$ such that
\begin{equation} \label{eq:omegaE_operator}
    i \omega_E^P f = \{ \varphi^P, f \},
\end{equation}
which represents the advection by the primary $\bsy{E}\times\bsy{B}$ flow. The leading order expressions \eqref{eq:phiS_RDK} and \eqref{eq:gS_RDK} may now be expressed as
\begin{equation}
    i \omega \delta\hat \varphi_{0}^S = \{\langle \hat \varphi^S\rangle_\psi, \varphi^P \}, \qquad i \omega \hat g_{i0}^S = \{\langle \hat \varphi^S\rangle_\psi, g_i^P \}.
\end{equation}
Furthermore, we obtain
\begin{align}
    -i\omega \, & \tau \frac{Z_i e \delta\hat\varphi_1^S}{T_i} = - v_{Mx} \partial_x \left( \frac{Z_i e \langle \hat\varphi^S \rangle_\psi}{T_i} \right) - \frac{1}{i\omega} \left\{ v_{Mx}\partial_x \langle \hat\varphi^S\rangle_\psi, \frac{P^P}{n_i T_i} + \frac{Z_i e \varphi^P}{T_i} \right\} \nonumber \\
    - & \frac{1}{i\omega} \left\{ \langle \hat\varphi^S \rangle_\psi, \bsy{B} \cdot \nabla \left( \frac{u_{\parallel}^P}{B} \right) + \bsy{v}_M \cdot \nabla \left( \frac{P^P}{n_i T_i} + \frac{Z_i e \varphi^P}{T_i} \right) + v_{Ex}^P \partial_x \ln n_i \right\} \label{eq:phi1S_omega}
\end{align}
from \eqref{eq:phiS_RDK}, \eqref{eq:gS_RDK}, and \eqref{eq:dt_phi1S}, as well as 
\begin{align}
    -i & (\omega - \omega_E^P)  \hat P_1^S = - \{ \delta\varphi_1^S, P^P\} - \frac{7}{4} Z_i e n_i v_{Mx} \partial_x \langle \hat\varphi^S \rangle_\psi - \frac{1}{i\omega} \left\{ v_{Mx}\partial_x \langle \hat\varphi^S\rangle_\psi, \chi^P + \frac{7}{4} Z_i e n_i \varphi^P \right\} \nonumber \\
    - & \frac{1}{i\omega} \left\{ \langle \hat\varphi^S \rangle_\psi, \bsy{B} \cdot \nabla \left( \frac{q_{\parallel}^P}{B} \right) + \frac{\bhat \cdot \nabla q_{\perp}^P}{2}  + \bsy{v}_M \cdot \nabla \left( \chi^P + \frac{7}{4} Z_i e n_i \varphi^P \right) + v_{Ex}^P \partial_x (n_i T_i)\right\} 
\end{align}
from \eqref{eq:gS_RDK}, \eqref{eq:dt_P1S}, and \eqref{eq:omegaE_operator}. The latter expression may also be written as
\begin{align}
    -i & \omega \hat P_1^S =  -\frac{7}{4} Z_i e n_i v_{Mx} \partial_x \langle \hat\varphi^S \rangle_\psi + i (\omega - \omega_E^P)^{-1} \Bigg[ i\omega \{ \delta\hat\varphi_1^S, P^P\} + \left\{ v_{Mx}\partial_x \langle \hat\varphi^S\rangle_\psi, \chi^P \right\} \nonumber\\
    & + \left\{ \langle \hat\varphi^S \rangle_\psi, \bsy{B} \cdot \nabla \left( \frac{q_{\parallel}^P}{B} \right) + \frac{\bhat \cdot \nabla q_{\perp}^P}{2}  + \bsy{v}_M \cdot \nabla \left( \chi^P + \frac{7}{4} Z_i e n_i \varphi^P \right) + v_{Ex}^P \partial_x (n_i T_i) \right\} \Bigg].\label{eq:P1S_omega}
\end{align}
The expressions \eqref{eq:phi1S_omega} and \eqref{eq:P1S_omega} can now be inserted into \eqref{eq:RDK_disp_rel} to derive a dispersion relation for secondary modes. We do not write this expression out due to its unwieldiness. Note that calculating the frequency would require the evaluation of the operator $(\omega- \omega_E^P)^{-1}$.

For simplicity, we now consider an up-down symmetric tokamak and assume the even moments of $g_i^P$ in $v_\parallel$ (e.g. the gyrocentre density and pressure) to be symmetric in $\theta$ while the odd moments of $g_i^P$ in $v_\parallel$ (e.g. the parallel flow and parallel fluxes) are antisymmetric in $\theta$. For a primary drive representing a linearly unstable mode, these assumptions are justified based on the symmetry of the linear GK equation \citep{peeters_linear_2005}. Then, $\varphi^P, P^P, \chi^P, \bhat \cdot \nabla u_\parallel^P, \bhat \cdot \nabla q_\parallel^P, \bhat \cdot \nabla q_\perp^P$ are all taken to be up-down symmetric. Furthermore, the binormal component of the magnetic drift is also up-down symmetric. For example, for a large aspect ratio tokamak with circular flux surfaces, the radial and binormal magnetic drift components vary along the field line as $\bsy{v}_M \cdot \nabla x \propto \sin\theta$ and $\bsy{v}_M \cdot \nabla y \propto \cos\theta + \hat s \theta \sin\theta$, respectively. In that case, the only non-vanishing contributions to the SW force from $P_1^S$ and $\delta\varphi_1^S$ stem from the radial magnetic drift contributions in \eqref{eq:phi1S_omega} and \eqref{eq:P1S_omega}. Therefore, for an up-down symmetric tokamak with up-down symmetric primary drive, equation \eqref{eq:RDK_disp_rel} gives
\begin{alignat}{2} 
    & \omega^2  \left\langle \frac{\rho_i^2 \abs{\nabla x}^2}{2} \right\rangle_\psi  \partial_x^2 \langle \hat\varphi^S \rangle_\psi && = \partial_x^2 \left( \partial_x^2 \langle \hat\varphi^S \rangle_\psi \left\langle \frac{\rho_i^2 \abs{\nabla x}^2 }{2}  v_{Ex}^P \left(v_{Ex}^P + v_\mathrm{dia}^P \right) \right\rangle_\psi \right) + \partial_x^2 \langle \hat \varphi^S \rangle_\psi \langle v_{Mx}^2 \rangle_\psi \left( \frac{7}{4}+\frac{1}{\tau} \right) \nonumber \\
    & -  \partial_x \Bigg\langle v_{Mx} i (\omega-\omega_E^P)^{-1} \bigg[ && \frac{1}{\tau} \left\{  v_{Mx} \partial_x \langle \hat\varphi^S \rangle_\psi + \frac{1}{i\omega}  v_{Mx} \partial_x  \left\{ \langle\hat\varphi^S\rangle_\psi, \frac{P^P}{Z_i e n_i} + \varphi^P \right\}, \frac{P^P}{Z_i e n_i} \right\} \nonumber\\
    & &&+ v_{Mx} \partial_x \left\{ \langle \hat\varphi^S \rangle_\psi, \frac{\chi^P}{Z_i e n_i} \right\} + \frac{7}{4}\left\{ \langle \hat \varphi^S \rangle_\psi, v_{Mx} \partial_x  \varphi^P \right\} \bigg] \Bigg\rangle_\psi.  \label{eq:RDK_disp_rel_symm_tokamak}
\end{alignat}

The theory in Section~\ref{sec:secondary_modes_toroidal_geo} assumes a streamer primary ($\partial_x g_i^P=0$), in which case the secondary mode may be Fourier expanded in $x$. The frequency of the secondary mode with wavenumber $k_x$ may then be derived from the dispersion relation \eqref{eq:RDK_disp_rel_symm_tokamak} using the fact that the operator \eqref{eq:omegaE_operator} simplifies to
\begin{equation}
    \omega_E^P \, e^{i k_x x} = k_x v_{Ex}^P e^{i k_x x}.
\end{equation}
The resulting dispersion relation is equation \eqref{eq:D_TSM_NR} in the main text and is discussed there in more detail.

\section{Rogers-Dorland-Kotschenreuther (RDK) secondary mode for arbitrary radial wavelengths}
\label{sec:RDK_SW}

In \ref{sec:strongly_driven} and Section~\ref{sec:TSM_NR}, we studied the RDK secondary mode in the limit of long perpendicular wavelengths. Here, we discuss the RDK secondary mode for arbitrary radial wavelengths. To this end, we assume a frozen streamer primary, like in the theory of the generalised secondary mode of Section~\ref{sec:theory_ISM_TSM}, and we consider the limit of vanishing linear terms (magnetic drifts, parallel streaming, and diamagnetic drift) in the GK equation. A dispersion relation may then be derived from \eqref{eq:D_TSM} with $\mathcal{N}$ evaluated in the limit of $\tilde v_{Mx} \rightarrow 0$. If we further assume the primary drive to be given by the bi-Maxwellian \eqref{eq:model_gP_M}, equation \eqref{eq:curlyN} for $\mathcal{N}$ simplifies to
\begin{equation}
    \mathcal{N} = - \frac{\tau k_x v_{Ex}^P}{\omega - k_x v_{Ex}^P} \left( \Gamma_{0i}(b_i) + b_i \Gamma_{0i}'(b_i) \eta_\perp^P  \right),
\end{equation}
with $\Gamma_{0i}'(b_i) = \mathrm{d}\Gamma_{0i}(b_i) / \mathrm{d}b_i$. The dispersion relation \eqref{eq:D_TSM} becomes
\begin{equation} \label{eq:RDK_SW_FSA}
    0 = 1 - \tau \left\langle \frac{\omega - k_x v_{Ex}^P}{\omega f_\omega - k_x v_{Ex}^P f_E }\right\rangle_\psi,
\end{equation}
which may be shown to be equivalent to equation (4) in \cite{rogers_generation_2000}. Here, we defined
\begin{equation}\label{eq:fE_fomega}
    f_\omega \equiv 1-\Gamma_{0i}+\tau , \qquad  f_E \equiv (1-\Gamma_{0i})(1+\tau) - \tau \eta_\perp^P b_i \Gamma_{0i}' .
\end{equation}

We consider a primary drive sinusoidally varying in $y$ and constant along the magnetic field, $v_{Ex}^P = v_{Ex}^{P0} \sin(k_y^P y)$, and $\eta_\perp^P$ is taken to be constant. For simplicity, we further assume that the variation of $b_i$ along the magnetic field is negligible, as is appropriate for e.g. a large aspect ratio tokamak with circular flux surfaces. The flux-surface average in \eqref{eq:RDK_SW_FSA} then requires evaluating the integral
\begin{equation} \label{eq:integral_sqrt}
    \frac{1}{2\pi} \int_0^{2\pi}  \frac{\mathrm{d}\phi}{1-A \sin \phi} = \frac{1}{\sqrt{ 1 - A^2 }},
\end{equation}
which may be obtained e.g. by changing variables to $z=e^{i\phi}$ and using the method of residues. Here, the square root is evaluated with the branch cut along negative reals such that Re$\sqrt{x} \geq 0$.

\begin{figure}
    \centering
    \includegraphics[width=0.65\textwidth, trim={0.8cm 0.8cm 0.8cm 0cm}, clip]{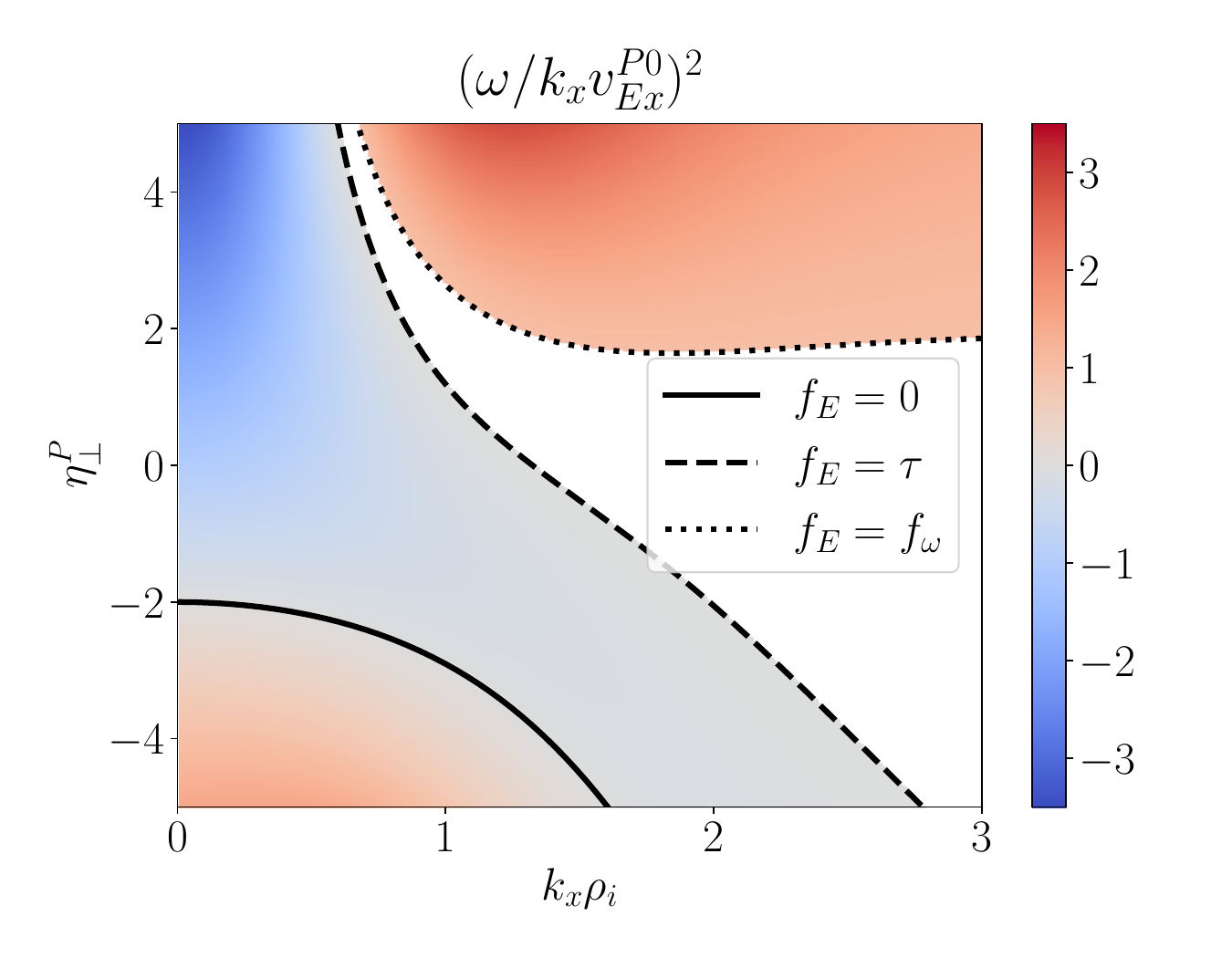}
    \caption{Rogers-Dorland-Kotschenreuther (RDK) secondary mode frequency from \eqref{eq:RDK_any_b} as a function of $\eta_\perp^P$ and $k_x \rho_i$ (with $b_i = k_x^2 \rho_i^2/2$). The primary drive is given by \eqref{eq:model_gP_M}, with $v_{Ex}^P=v_{Ex}^{P0} \sin(k_y^P y)$ and no variation along the magnetic field. There are no modes in the blank region ($\tau \leq f_E \leq f_\omega$), see \eqref{eq:RDK_SW_mode_existence}. Purely growing RDK secondary modes ($\omega^2 < 0$) are found for $0 < f_E < \tau$, see \eqref{eq:RDK_SW_instab_condition_fE}.}
    \label{fig:RDK_SW}
\end{figure}

The dispersion relation \eqref{eq:RDK_SW_FSA} then simplifies to
\begin{equation} \label{eq:RDK_SW_disp_sqrt}
    0 = 1 - \frac{\tau}{f_E} - \tau \left( \frac{1}{f_\omega} - \frac{1}{f_E} \right) \Bigg/ \sqrt{ 1 - \left( \frac{k_x v_{Ex}^{P0} f_E}{\omega f_\omega} \right)^2 },
\end{equation}
As the real part of the square root must be positive, a necessary condition for the existence of solutions to \eqref{eq:RDK_SW_disp_sqrt} is 
\begin{equation} \label{eq:RDK_SW_mode_existence}
    \frac{f_E - f_\omega}{f_E - \tau} \geq 0.
\end{equation}
When this condition is satisfied, rearranging and squaring \eqref{eq:RDK_SW_disp_sqrt} leads to the secondary mode frequency
\begin{equation} \label{eq:RDK_any_b_fE}
    \left(\frac{\omega}{k_x v_{Ex}^{P0}}\right)^2 = \frac{(f_E-\tau)^2}{f_\omega-\tau} \frac{f_E}{f_E(f_\omega+\tau) - 2\tau f_\omega}.
\end{equation}
Noting that $f_\omega - \tau > 0$ by \eqref{eq:fE_fomega}, the mode is unstable when the second fraction in \eqref{eq:RDK_any_b_fE} is negative, i.e. $0 < f_E <  2\tau f_\omega / (\tau + f_\omega)$. As the condition for mode existence \eqref{eq:RDK_SW_mode_existence} sets a stricter upper bound on $f_E$, the condition for the existence of unstable RDK secondary modes may be succinctly written as
\begin{equation} \label{eq:RDK_SW_instab_condition_fE}
    0 < f_E \leq \tau .
\end{equation}

Using \eqref{eq:fE_fomega}, the secondary mode frequency \eqref{eq:RDK_any_b_fE} may also be written as
\begin{equation} \label{eq:RDK_any_b}
    \left(\frac{\omega}{k_x v_{Ex}^{P0}}\right)^2 = \frac{\left( 1- \Gamma_{0i} - \tau\Gamma_{0i} -  \tau \eta_\perp^P b_i \Gamma_{0i}' \right)^2}{1-\Gamma_{0i}} \frac{ (1+\tau)(1-\Gamma_{0i}) -  \tau \eta_\perp^P b_i \Gamma_{0i}'}{(1-\Gamma_{0i})^2 (1+\tau) - 2 \tau^2 \Gamma_{0i} - \tau \eta_\perp^P b_i \Gamma_{0i}' \left( 1-\Gamma_{0i} + 2 \tau \right)  }.
\end{equation}
which recovers equation (5) of \cite{rogers_generation_2000} in the limit $\eta_\perp^P = 0$. Condition \eqref{eq:RDK_SW_instab_condition_fE} for instability may be written as
\begin{equation} \label{eq:RDK_SW_instab_condition}
    0 < (1-\Gamma_{0i})(1+\tau) - \tau \eta_\perp^P b_i \Gamma_{0i}' \leq \tau.
\end{equation}
For small $b_i \ll 1$, the instability criterion simplifies to $\eta_\perp^P > -(1+\tau^{-1})$, as expected from the RDK dispersion relation for long perpendicular wavelengths \eqref{eq:omegaRDK}. For large $b_i \gg 1$, for which $\Gamma_{0i} \approx (2\pi b_i)^{-1/2}$, the second inequality in \eqref{eq:RDK_SW_instab_condition} is satisfied only for large negative $\eta_\perp^P$ values. Generally (for $\eta_\perp^P \sim 1$), there will therefore be no unstable RDK secondary mode for large $b_i \gg 1$, as shown in Figure~\ref{fig:RDK_SW}.

\bibliographystyle{plainnat}
\bibliography{references}

\end{document}